\documentclass[preprint2]{aastex62}

\usepackage{graphicx}
\usepackage{subfigure}
\usepackage{longtable}
\usepackage{wrapfig}
\usepackage[figuresleft]{rotating}\usepackage{epstopdf}
\usepackage{amsmath}
\usepackage{multirow}
\usepackage[mathscr]{eucal}

\newcommand{\kms}{\mbox{${\rm km~s^{-1}}$} }

\newcommand{\kmsMpc}{km s$^{-1}$ Mpc$^{-1}$ }
\newcommand{\hi}{H{\sc\,i} }
\newcommand{\dg}{$^{\circ}$ }

\newcommand{\hnut}{$H_0$}

\graphicspath{{./}{}}

\received{2020 February 4}
\revised{2020 April 9}
\accepted{2020 April 29}

\submitjournal{ApJ}

\shorttitle{Multi-wavelength calibration of the Tully-Fisher Relations}
\shortauthors{Kourkchi et al.}

\begin{document}

\title{Cosmicflows 4: The Calibration of Optical and Infrared Tully-Fisher Relations}

\email{ehsan@ifa.hawaii.edu}

\author[0000-0002-5514-3354]{Ehsan Kourkchi}
\affil{Institute for Astronomy, University of Hawaii, 2680 Woodlawn Drive, Honolulu, HI 96822, USA}
\author[0000-0002-9291-1981]{R. Brent Tully}
\email{tully@ifa.hawaii.edu}
\affil{Institute for Astronomy, University of Hawaii, 2680 Woodlawn Drive, Honolulu, HI 96822, USA}
\author[0000-0002-5259-2314]{Gagandeep S. Anand}
\affil{Institute for Astronomy, University of Hawaii, 2680 Woodlawn Drive, Honolulu, HI 96822, USA}
\author[0000-0003-0509-1776]{H\'el\`ene M. Courtois}
\affil{University of Lyon, UCB Lyon 1, CNRS/IN2P3, IUF, IP2I Lyon,  France}
\author[0000-0001-9005-2792]{Alexandra Dupuy}
\affil{University of Lyon, UCB Lyon 1, CNRS/IN2P3, IUF, IP2I Lyon, France}
\author[0000-0002-0466-1119]{James D. Neill}
\affil{California Institute of Technology, 1200 East California Boulevard, MC 278-17, Pasadena, CA 91125, USA}
\author[0000-0003-0882-2327]{Luca Rizzi}
\affil{W.M. Keck Observatory, 65-1120 Mamalahoa Highway, Kamuela, HI 96743, USA}
\author[0000-0002-1143-5515]{Mark Seibert}
\affil{The Observatories of the Carnegie Institute of Washington, 813 Santa Barbara Street, Pasadena, CA 91101, USA}

\begin{abstract}

This study is a part of the {\it Cosmicflows-4} project with the aim of measuring the distances of more than $\sim$10,000 spiral galaxies in the local universe up to $\sim$15,000~\kms. New \hi line width information has come primarily from the Arecibo Legacy Fast ALFA Survey.  Photometry of our sample galaxies has been carried out in optical (SDSS {\it u}, {\it g}, {\it r}, {\it i}, and {\it z}) and infrared (WISE {\it W1} and {\it W2}) bands. Inclinations have been determined using an online graphical interface accessible to a collaboration of citizen scientists. Galaxy distances are measured based on the correlation between the rotation rate of spirals and their absolute luminosity, known as the Tully-Fisher Relation (TFR). In this study, we present the calibration of the TFR using a subsample of $\sim$600 spirals located in 20 galaxy clusters. Correlations among such observables as color, surface brightness, and relative \hi content are explored in an attempt to reduce the scatter about the TFR with the goal of obtaining more accurate distances.  A preliminary determination of the Hubble constant from the distances and velocities of the calibrator clusters is $H_0=76.0\pm1.1$(stat.)$\pm2.3$(sys.)~\kmsMpc.

\end{abstract}


\section{Introduction} \label{sec:intro}

Through the {\it Cosmicflows} program, the dark matter distribution in the universe is reconstructed based on the peculiar motions of galaxies in response to the underlying gravitational field. The peculiar velocities of galaxies are calculated by subtracting the Hubble expansion rate at their position from their observed radial velocities. Therefore, the precise measurement of distances is vital in this analysis. 

To date, three editions of {\it Cosmicflows} have been published \citep{2008ApJ...676..184T, 2013AJ....146...86T, 2016AJ....152...50T}.  Throughout, the samples are heterogeneous
in that there is input involving multiple methodologies and multiple sources.  The scale foundations are set by the Cepheid Period-Luminosity Relation \citep{1912HarCi.173....1L} and the Tip of the Red Giant Branch (TRGB) standard candle \citep{1993ApJ...417..553L} supplemented by methods that have high precision but limited applicability such as Detached Eclipsing Binaries \citep{2019Natur.567..200P} and the nuclear maser in NGC~4258 \citep{2013ApJ...775...13H}.  These local calibrators constrain the scaling of methods useful at intermediate distances, such as the amplitudes of surface brightness fluctuations in early-type galaxies \citep{1988AJ.....96..807T}, the three-parameter fundamental plane correlation also applicable to early types \citep{1987ApJ...313...59D, 1987ApJ...313...42D}, and the luminosity-rotation rate relation found to hold for spiral galaxies \citep{1977A&A....54..661T}.  All of these procedures must be shown to be consistent with each other and, in turn, with the methodology that extends to large distances founded on the predictability of the peak luminosities of supernovae of type Ia \citep{1993ApJ...413L.105P}.

{\it Cosmicflows-3} (CF3), the last edition reporting galaxy distances, contains information on 17,647 systems\footnote{The CF3 catalog updated with minor corrections is available at http://edd.ifa.hawaii.edu}.  Roughly half the contributions in that compilatfion come from a single source: Fundamental Plane measures from the Six Degree Field Redshift Survey \citep{2012MNRAS.427..245M, 2014MNRAS.445.2677S}.  This contribution is restricted to $\delta \leq 0$, the celestial south.  As a consequence, the coverage in CF3 is strongly tilted toward the south.  It is a particular concern of the next addition, {\it Cosmicflows-4}, that this imbalance in sky coverage be redressed.  This new edition will contain updates across a broad spectrum of inputs, but most of all, there will be a major enhancement in the number of spiral luminosity-rotation rate measurements made available as a consequence of the large number of neutral hydrogen line profile detections resulting from the Arecibo Legacy Fast ALFA Survey \citep[ALFALFA; ]{2018ApJ...861...49H} covering the declination range $0 < \delta < +38$. Photometry for these galaxies is alternatively available from the Sloan Digital Sky Survey \citep[SDSS;][]{2015ApJS..219...12A} and the Wide-field Infrared Satellite Explorer \citep[WISE;][]{2010AJ....140.1868W}.  The present paper presents up-to-date calibrations of the luminosity-rotation rate correlations in the bands of these surveys. 

Since its inception \citep{1977A&A....54..661T, 1979ApJ...229....1A}, the luminosity-rotation rate relation (hereafter TFR) has been refined and reevaluated by many users.  If the interest is in galaxy distances, a primary concern is the control of bias \citep{1994ApJS...92....1W}.  The current analysis makes use of the correlation fit with errors taken in line widths, a procedure that has been called the ``inverse" TFR (ITFR).  This procedure was initially discussed as a general principle by \citet{1980AJ.....85..801S} and specifically with relevance to the TFR by \citet{1988Natur.334..209T}.  The issue was discussed in detail in \S3 of \citet{2000ApJ...533..744T}.  The slope of the power-law fit of the ITRFR nulls the bias to a first approximation.  Residual bias can be taken into account with a small correction to be discussed. 

It deserves emphasis that a focus on the correlation with errors taken in line width, the distance-independent parameter, is desired if the primary goal is the measurement of distances, but it is not the appropriate procedure if the interest is a physical understanding of the relationship between luminosity and rotation in spiral galaxies.  In the latter case, it is more informative to make a bivariate fit taking into account errors affecting both line widths and magnitudes \citep{1997AJ....113...53G,2006ApJ...653..861M,2017MNRAS.469.2387P,2019MNRAS.484.3267L}.

We employ the inverse relation, since our interest is the measurement of galaxy distances.
Recalibrations have been carried out in both optical and infrared bands in the course of the {\it Cosmicflows} program. \citet{2012ApJ...749...78T} calibrated the TFR at the {\it I} band using 267 spirals in 13 clusters. \citet{1979ApJ...229....1A} suggested the use of near-infrared bands to improve the TFR methodology, given the lesser dust obscuration in host galaxies. Furthermore, the light of old stars peaks at longer wave bands; therefore, near-infrared luminosities may better represent the baryonic mass of galaxies. With the motivation of reducing the scatter about the TFR and getting more accurate distances, \citet{2013ApJ...765...94S} extended the calibration of the TFR toward the infrared using {\it Spitzer} $3.6$ $\mu$m {\it Infrared Array Camera} (IRAC) imaging for the same sample of galaxies as used by Tully and Courtois. Using the available {\it I}-band magnitudes, they also observed a correlation between the optical-infrared color index and the deviation of galaxy magnitudes from the infrared TFR. They found a tighter TFR after adjusting $[3.6]$ magnitudes for the effect of an $I-[3.6]$ color term. 

\citet{2014ApJ...792..129N} followed the same procedure using the WISE \citep{2010AJ....140.1868W} and added 43 galaxies to the previous sample. They also updated the calibration of the {\it I}-band TFR using their larger sample. They studied the possibility of decreasing the scatter about the infrared TFR by taking into account the optical-infrared color terms. Moreover, they showed that adopting a quadratic form for the TFR results in slightly smaller scatter about the relation.

In our program, we have performed the multiband photometry of $\sim$20,000 {\it Cosmicflows-4} candidate galaxies in optical (SDSS {\it u, g, r, i} and {\it z}) and/or infrared WISE {\it W1} (3.4$\mu m$) and {\it W2} (4.6$\mu m$) bands. 
The purpose of the current study is to provide the calibration of the TFR at these wave bands to measure distances for our full sample of galaxies. To achieve our goal, we use a sample of 648 calibrator galaxies (roughly double that available previously) in 20 clusters (up from 13).  We explore intrinsic relations between color, morphology, and other galaxy observables with deviations from the TFR at different wave bands, toward possible ways to reduce scatter and hopefully obtain more accurate distances.

We explain our sample selection criteria and present our data product in \S \ref{sec:data}.
In \S \ref{sec:tfr_calib}, we calibrate the luminosity-line width correlations. In \S \ref{sec:tfr_addTerms}, we explore the possibility of reducing the scatter of the calibrated relations using additional observables such as color indices and surface brightnesses. 
The radial velocities of the calibrator clusters are determined in \S \ref{sec:ClusterRadial}, and their measured distance moduli are evaluated across all passbands and calibrations in \S \ref{sec:ClusterDM}. We provide estimations of the Hubble constant, \hnut, on the basis of the clusters in \S \ref{sec:HubbleParam}.
Our conclusion is presented in \S \ref{sec:conclusions}.

\section{Data} \label{sec:data}
\subsection{Calibrators} \label{sec:calibrators}

To calibrate the luminosity-line width correlation in spirals, we follow the same methodology that was originally described by \citet{2012ApJ...749...78T} and further implemented by \citet{2013ApJ...765...94S} and \citet{2014ApJ...792..129N}. In this technique, the TFR calibration is carried out using spirals that reside in galaxy clusters. In a cluster, all spirals are assumed to be at the same distance; therefore, the correlation of apparent luminosities with rotation rates is manifested.  Weighted least-squares fits assume errors in line widths (ITFR). Considering individual clusters separately, the slopes are similar but the intercepts differ because the clusters are at different distances. Upon deriving the TFR slopes for all individual clusters, the initial step toward a universal template is taken by shifting the galaxies of each cluster along the luminosity axis to a fiducial distance (for convenience, that set by the Virgo cluster sample; the nearest case and populated across the fully useful magnitude range). In this first pass, the shift for a cluster is given by the magnitude difference with respect to the fiducial at the intercept magnitude value at a log line width equal to 2.5.  The tentative universal template that has been formed, treating all of the galaxies as if belonging to the same cluster and at the same distance, has a more complete magnitude coverage and reduces the uncertainty of the calibration slope.  The tentative universal slope is now forced in a least-squares sense onto all of the individual clusters.  The intercept magnitude differentials with respect to the fiducial mandate the shifts to form a new universal relation.  The procedure is repeated iteratively, rapidly reaching convergence. 

It is to be remarked that the universal template that is constructed is decoupled from variations in the cosmic expansion.  More typically in other studies, templates are formed by  assuming the constituents, on average, participate in the Hubble flow \citep{1996ApJ...457..460W, 1997AJ....113...53G, 2007ApJS..172..599S, 2014MNRAS.445.2677S}.  We endeavor at every step to decouple the measurement of distances from velocity information, since a principal goal of our program is to identify departures from Hubble expansion.

To calibrate the zero-point of the TFR, we use a subset of nearby spirals with accurate distance measurements from independent methods, such as TRGB and/or CPLR. In case of availability of distance measurements from both methods, we use the average value. Details are provided in \S \ref{sec:tfr_zp}. The key point to mention is that the slope applied to the zero-point sample is accepted to be given by the universal cluster calibration. There is no sense of completion to any magnitude limit with the zero-point calibration sample.  The error-weighted least-squares fit to the zero-point sample transforms the universal cluster template relation to an absolute magnitude scale. 

The calibrator sample of this study is a subset of our large sample of $\sim$20,000 candidates that we compiled to measure distances using the luminosity-line width method\footnote{This catalog will be published in a separate paper.}. Galaxies of this sample are restricted to having (1) high-quality \hi measurements (see \S \ref{sec:linewidth})  (2) the morphological type Sa or later, and (3) spirals with inclinations greater than 45\dg from face-on. 

In this program, we measured the spiral inclinations by visually comparing them with a set of spirals with known inclinations. Sorting spirals based on their inclinations was executed through an online graphical tool, {\it Galaxy Inclination Zoo (GIZ)}\footnote{\url{http://edd.ifa.hawaii.edu/inclination/index.php}}, as a participative science project with citizens. Comparisons with inclinations derived from axial ratios \citep{2014ApJ...792..129N} indicate that the rms uncertainties are $\pm4^{\circ}$. In detail, the uncertainties are $\pm3^{\circ}$ more edge-on than $70^{\circ}$, degrading to $\pm5^{\circ}$ by inclinations of $50^{\circ}$. Please refer to \S 2.5 of \citet[hereafter K19]{2019ApJ...884...82K} for further details.

We note that inclinations can be determined in an independent way from the projection of disk rotation with resolved velocity field information.  The number of sources that have been given sufficient study are limited, since observations must be made with radio interferometers.  The targets must be large and \hi-rich and warrant the investment of resources.  Nonetheless, the studies that have been done provide the basis for a comparison.  We draw on intersecting samples of 30 galaxies by \citet{2017MNRAS.469.2387P} and 97 galaxies by \citet{2016ApJ...816L..14L}.  The kinematically determined inclinations in the two interferometric studies concur and are systematically more face-on than the inclinations from images by $3.3^{\circ}$ with $4\sigma$ significance.  The scatter around this offset is $\pm4.8^{\circ}$. Hence the scatter is satisfactory, suggesting statistical uncertainties at the level of $4^{\circ}$ for each method but revealing a systematic in at least one of the methods. The information necessary to provide kinematic inclinations is not available for the samples of thousands of galaxies in our study.  A systematic offset would affect our calibrators and field targets alike.  We do not detect any trend in measured distances as a function of inclination.

For our calibration, we select galaxy clusters that have substantial numbers of galaxies with the requisite line width and photometric information (median of 24 galaxies per cluster).  General properties of clusters, such as center, $R_{2t}$, $V_c$ and $\sigma_p$ are taken from a galaxy group catalog \citep{2015AJ....149..171T} that is built based on the {\it Two Micron All Sky Survey} (2MASS) Redshift Survey quasi-complete to $K_s=11.75$ \citep{2012ApJS..199...26H}, where $R_{2t}$ is the cluster projected second turnaround radius, $V_c$ is the average heliocentric radial velocity of the cluster, and $\sigma_p$ is the cluster projected velocity dispersion. A galaxy is included in a calibrator cluster if it meets one of the following conditions: (1) The galaxy projected distance from the center of the cluster, $R_p$, is less than $1.5 R_{2t}$ and its heliocentric velocity is within $3\sigma_p$ of $V_c$; (2) the projected distance is relaxed to $1.5 R_{2t} \leq R_p < 3 R_{2t}$ but the heliocentric velocity is more constrained at less than $2\sigma_p$ of $V_c$. 

Based on experience with earlier releases of $Cosmicflows$, as many as $3$\% of the galaxies selected for the TFR analysis do not follow the general trend of the TFR. The outlier galaxies could not be initially rejected based on their morphology or the quality of photometry and/or \hi measurements and/or other observables that we use in this study. We exclude them from our sample as $>3\sigma$ deviants when fitting the TFR. We label these galaxies as ``anomalous galaxies" and distinguish them in our plots. 

Our selection criteria leave us with 648 galaxies (including 13 anomalous galaxies) from our large sample as candidate members of our 20 clusters.\footnote{Galaxies with \hi line widths less than 64~\kms\ are preemptively dropped because such galaxies will not survive our final absolute magnitude cut of $M_i\leq-17$.} We use these galaxies for the slope calibration of the TFR. For the zero-point calibration, there are 94 spirals that have distance measurements from the Cepheid luminosity-period or TRGB methods.

There are a few confusing regions where clusters overlap in plane-of-sky projections. For example, in the case of the Virgo cluster we exclude galaxies that have higher chances of being members of background galaxy groups. We include a galaxy in the Virgo cluster within its virial radius if its heliocentric velocity satisfies either $V_h < 600$ \kms or $1200 < V_h < 1600$ \kms to avoid galaxies that are potentially in overlapping groups: Virgo~W and Virgo~M at $V_h>1600$~\kms\ and Virgo~W' and Virgo~W'' in the window $600<V_h<1200$~\kms. See \S6.1 of \citet{2016AJ....152...50T} for an extensive discussion of the potential confusion. In another special case, the Ursa Major entity consists of seven smaller galaxy groups identified in \citet{2017ApJ...843...16K} that are almost indistinguishably at the same distance along a filament that spans $7.5^\circ$ across the plane of the sky. 

\subsection{\hi Line Widths} \label{sec:linewidth}

One of the criteria for our sample selection is the existence of high-quality \hi data. 
We take \hi 21 cm line widths and fluxes from various sources in the order of priority listed below.

\begin{enumerate}
  
  \item The All Digital \hi catalog ({\it ADHI}) that has been compiled in the course of the $Cosmicflows$ program \citep{2009AJ....138.1938C, 2011MNRAS.414.2005C}, available at the Extragalactic Distance Database (EDD) website.\footnote{\url{http://edd.ifa.hawaii.edu}; catalog ``All Digital \hi''.} Entries in this catalog contain the \hi line width parameter $W_{mx}$, the parameter that encodes galaxy rotation rates at their negative and positive maximum intensities along the line of sight. Here $W_{mx}$ is calculated from the observed $W_{m50}$, and is adjusted for spectral resolution and redshift, and $W_{m50}$ is the line width at 50\% of the average \hi flux within the range that covers 90\% of the total \hi flux. 
  Low signal to noise ratio (S/N and confused or anomalous line profiles are cause for the rejection of potential candidates.  We only accept {\it ADHI} $W_{mx}$ values if the evaluated uncertainties are less than or equal to $20$ \kms. 
  
  \item ALFALFA ({\it ALFALFA}; \citealt{2011AJ....142..170H, 2018ApJ...861...49H}). To be compatible with the {\it ADHI} values, cataloged {\it ALFALFA} line widths are adjusted using $W_{mx}=W_{alf}-6$~\kms, a formula derived for spirals with \hi line width measurements in both catalogs. We require $S/N>10$ for line widths taken from the {\it ALFALFA}, which is fairly consistent with the adopted uncertainty condition for an {\it ADHI} estimate.
  
  \item The {\it Springob/Cornell} \hi catalog \citep{2005ApJS..160..149S}. The $W_{M50}$ values provided in this catalog are transformed using $W_{m50} = 1.015W_{m50}-11$ \kms and are then translated into $W_{mx}$ values using the {\it ADHI} standard formalism \citep{2009AJ....138.1938C}. 
  
  \item The {\it Pre Digital \hi} catalog available in EDD provides information from early analog \hi line profiles \citep{1981ApJS...47..139F, 1989gcho.book.....H}.  Recourse to this old material is needed in some cases of nearby galaxies that are much larger than the beam sizes of large radio telescopes. This catalog lists $W_{20}$, the width at 20\% of the \hi profile peak.  We convert $W_{20}$ to $W_{mx}$ based on the relation described by \citet{2009AJ....138.1938C}.
\end{enumerate}

The parameter $W_{mx}$ is taken as a robust measure of galaxy rotation after correction for the effect of inclination. We introduce a minimum threshold for our potential candidates at $W_{mx}>64$ \kms, eliminating dwarf galaxies that, at any rate, would not survive our restriction later imposed of $M_i < -17$. Input from the {\it Springob/Cornell} and {\it Pre Digital \hi} catalogs is not considered if a galaxy has high-quality \hi measurements in either of the catalogs {\it ADHI} (uncertainty of $W_{mx}$ not larger than 20 \kms) or {\it ALFALFA} ($S/N>10$). 

We take the average values of \hi flux/line width for galaxies that are in both {\it ADHI} and {\it ALFALFA} catalogs. If a galaxy is not tabulated in either of {\it ADHI} or {\it ALFALFA} catalogs, we extract \hi measurements from either {\it Springob/Cornell} or {\it Pre Digital \hi} catalogs, with {\it Springob/Cornell} having higher priority.

Considerable attention has been given in the past to studies of the resolved rotation characteristics of spirals toward the twin quests of reducing scatter in the TFR and gaining physical insight into causes of intrinsic scatter \citep{2001ApJ...563..694V,2007MNRAS.381.1463N,2016MNRAS.463.4052P,2017MNRAS.469.2387P,2016ApJ...816L..14L,2019MNRAS.484.3267L}.  These interferometric studies identify three classes: rotation curves that fall below their maximum rate at large radii, rotation curves that roughly maintain their maximum rates at large radii, and rotation curves that are continuing to rise at large radii.  Massive and relatively early types tend to manifest themselves in the first of these classes, while galaxies of low intrinsic luminosity are found in the third class.  Scatter in the TFR is minimized by characterizing rotation with the parameter $V_{flat}$, determined by the resolved rotation rate at the extremity of a galaxy \citep{2017MNRAS.469.2387P,2019MNRAS.484.3267L}.  Rotation parameters constructed from the rotation rate at interior radii, such as $V_{2.2}$, at a radius of 2.2 scale lengths cause deterioration of the TFR.  These studies are instructive, but global profiles from single-beam radio observations integrate over the \hi flux of galaxies.  It is encouraging that \citet{2019MNRAS.484.3267L} found that global profile parameters perform only slightly less well than $V_{flat}$ and significantly better than parameters representative of rotation at inner radii.  It can be gathered that much of the integrated \hi flux from a galaxy arises from relatively large radii.  

In addition to a line width, we derive an \hi 21~cm magnitude, $m_{21}$, from the \hi fluxes following 
\begin{equation}
\label{Eq:m21}
  m_{21} = -2.5 {\rm log} F_{HI} + 17.40~, 
\end{equation}
where $F_{HI}$ is the flux within the 21 cm line profile in units of Jy$\cdot$\kms\ \citep{2014A&A...570A..13M}.

\subsection{Photometry} \label{sec:photometry}

Photometry for the calibrator galaxies is carried out on optical and infrared images provided by the SDSS DR12 and WISE surveys.
The process of obtaining individual exposures, image preprocessing, and performing photometry was described in detail in \S 2.2 through \S 2.4 of K19. 
Our asymptotic magnitudes are calculated within the aperture radius at the point where the cumulative luminosity curve of growth flattens.  For comparison, we computed total magnitudes with extrapolations assuming exponential disks following \citet{1996AJ....112.2471T}, finding agreement with a scatter of 0.02 mag. Uncertainties in magnitudes are 0.05 mag, dominated by the setting of the sky background.

Out of 648 candidate calibrator galaxies, 464 spirals (eight anomalous) have SDSS photometry at the {\it u, g, r, i} and {\it z} bands. For 600 (19 anomalous) galaxies, we make use of WISE {\it W1-} and {\it W2-}band coverage. There is both optical and infrared photometry coverage for 435 galaxies.

The observed apparent magnitude at each wave band, $m^{total}_{\lambda}$, is adjusted using the following relation:
\begin{equation}
\label{Eq:bkai}
m^*_\lambda = m^{(total)}_{\lambda} - A^{\lambda}_b - A^{\lambda}_k  - A^{\lambda}_a - A^{\lambda}_i,
\end{equation}
where $A^{\lambda}_b$ is the Milky Way dust extinction, $A^{\lambda}_k$ is the correction for the redshift effect on the galaxy luminosity at each passband ($k$-correction), and $A^{\lambda}_a$ is the total flux aperture correction for the photometry of the extended objects when the photometry calibration was performed using point-source objects (stars). There is a full discussion of the calculation of $A^{\lambda}_b$, $A^{\lambda}_k$, and $A^{\lambda}_a$ in \S 2.5 of K19.

Here $A^{\lambda}_i$ is the dust attenuation within the host galaxy that depends on (1) the galaxy physical properties, modeled as a function of galaxy observables, and (2) the galaxy spatial orientation with respect to the observer (K19). At each wave band $\lambda$, $A^{\lambda}_i$ is modeled as
\begin{equation}
\label{eq:dust}
    A^{\lambda}_i=\gamma_{\lambda} \mathcal{F}_{\lambda}(\dot{\imath})~,
\end{equation}
where $\mathcal{F}_{\lambda}(i)$ is a function of galaxy inclination, {\it i},\footnote{Symbol {\it i} not to be confused with the photometric passband} and $\gamma_{\lambda}$ is a third-degree polynomial function of the main principal component, $P_{1,W2}$, that is constructed from the linear combination of galaxy observables using
\begin{equation}
\label{Eq:P1}
\begin{split}
    P_{1,W2} = 0.524({\rm log}W^i_{mx}-2.47)/0.18 \\
    + 0.601(C_{21W2}-1.63)/1.15 \\
    - 0.603(\langle \mu_2 \rangle^{(i)}_e-23.35)/1.38 ~ ,
    \end{split}
\end{equation}
where $W^i_{mx}$ is the \hi line width corrected for inclination following $W^i_{mx}=W_{mx}/{\rm sin}(i)$, 
$C_{21W2}=m_{21}-\overline{W}2$ is a pseudocolor calculated from the difference between the \hi 21 cm and {\it W2} magnitudes, and $\langle \mu_2 \rangle^{(i)}_e$ is the average surface brightness of the galaxy at the {\it W2} band within the effective radius that is corrected for the geometric effect of inclination given by 
\begin{equation}
\label{Eq:Mu50_i}
\langle \mu_2 \rangle^{(i)}_e=\langle \mu_2 \rangle_e+0.5 \log_{10} (a/b)~, 
\end{equation} 
where $a$ and $b$ are the semi-major and semi-minor axes of the photometry aperture. Given the effective radius of a galaxy, $R_{\lambda e}$, its effective surface brightness is 
\begin{equation}
\langle \mu_\lambda \rangle_e = m^{(total)}_{\lambda} +2.5 {\rm log_{10}}(2\pi R^2_{\lambda e}) ~.
\end{equation} 
There is a full discussion in K19 regarding the calculation of $A^{\lambda}_i$.

The above formalism for calculating the dust attenuation of host galaxies relies on the existence of the infrared photometry. However, there are 29 spirals in our sample of TFR calibrators for which we lack infrared photometry. We use a prediction algorithm to estimate their internal dust attenuation, $A^{\lambda}_i$, from their optical photometry and 21 cm line widths/fluxes. 

In concept, given the luminosity of a galaxy at optical wave bands and some other information about its physical properties, such as intrinsic size and/or morphology, one can predict the galaxy luminosity at longer wave bands. 
One common method is to fit the spectral energy distribution (SED) over the observed magnitudes using a set of template SEDs that represent the morphology, size, and physical properties of the sample galaxies \citep{2020AJ....159..138D}. 

In this study, we use a less complicated prediction scheme. We orchestrate a random forest algorithm together with a set of distance-independent observables to predict the missing infrared information. 
Our prediction algorithm is trained using $\sim$2,000 spirals with the full optical/infrared magnitude coverage. 
The trained algorithm is capable of predicting {\it W2} magnitudes with an rms uncertainty of $\sim 0.2$ mag. Based on the predicted {\it W2} magnitudes, the $1\sigma$ uncertainty of the predicted $\gamma_{\lambda}$ is $\sim 0.04$ mag for the optical bands and is smaller at longer wavelengths.  Then $\gamma_{\lambda}$ is multiplied by $\mathcal{F}_\lambda$ to obtain the dust attenuation $A^{\lambda}_i$. Here $\mathcal{F}_\lambda$ is a monotonic increasing function of inclination that is maximal for fully edge-on galaxies between $1.5$ and $1.75$ for the optical wave bands and $0.75$ for the {\it W1} band (K19). The overall uncertainty on our predicted $A^{\lambda}_i$ values is always not worse than $\sim 0.07$ mag. We leave further details of this procedure for a later paper (E. Kourkchi et al., 2020, in preparation).

\subsection{Data Catalog}

The measurements and the collection of parameters used in this study are collected in Table \ref{tab_data} for 648 cluster galaxies and 94 zero-point calibrators. Column contents are (1) Principal Galaxies Catalog (PGC) number; (2) common name; (3) the measured inclination angle of the galaxy in degrees; (4) the \hi line width in \kms; (5) logarithm of the inclination-corrected \hi line width, calculated from $W^i_{mx}=W_{mx}/\sin(i)$, where {\it i} is the inclination angle presented in column (3); (6) the \hi 21 cm magnitude calculated from the \hi flux, $F_{HI}$, using $m_{21}=-2.5 \log F_{HI}+17.4$, (7)-(11) SDSS {\it u,g,r,i,z} total raw magnitudes in the AB system; (12)-(13) the WISE {\it W1} and {\it W2} total raw magnitudes in the AB system,\footnote{The uncertainties on the measured {\it u}-band magnitudes are $\sim 0.1$ mag. The uncertainties on all other measured magnitudes are not worse than $0.05$ mag, which is conservatively adopted for the purpose of error propagation in this study.}; (14) the $b/a$ axial ratios of the elliptical photometry apertures used for the photometry of SDSS images, where $a$ and $b$ are the semi-major and semi-minor axes of the elliptical aperture, respectively; (15) analogous to 14 but for the photometry apertures of the WISE images; (16)-(22) the semi-major axes of apertures that enclose half the total light of galaxies at optical/infrared bands in arcmin; (23) the name of the host cluster, with the 94 zero-point calibrators identified as ``ZP calibrator'', and with an asterisk added to the end of the cluster name if a galaxy is labeled as ``anomalous," i.e., strongly deviant for no reason that is evident; (24)-(30) tabulate optical/infrared magnitudes corrected for Milky Way obscuration, redshift $k-$correction, and aperture effects, but {\bf not} the inclination dependent dust attenuation, based on the corresponding raw magnitudes listed in columns (7)-(13); (31)-(37) the dust attenuation corrections at the optical/infrared band, calculated from Eq.\ref{eq:dust}; and (38)-(44) the correspondence to columns (31)-(37) after correcting for the effect of global dust obscuration, 
$m^*_\lambda = m^{(total)}_{\lambda} - A^{\lambda}_b - A^{\lambda}_k  - A^{\lambda}_a - A^{\lambda}_i$.


\begin{turnpage}
\setlength{\tabcolsep}{0.1cm}
\begin{deluxetable*}{cc c ccc ccccccc cc ccccccc}
\tablecaption{Data Catalog\tablenotemark{*}
\label{tab_data}}
\tabletypesize{\scriptsize}
\tablehead{ \\
PGC & 
Name & 
$Inc.$ &
$W_{mx}$ & $\log (W^i_{mx})$ & $m_{21}$ & 
{\it u} & {\it g} & {\it r} & 
{\it i} & {\it z} & 
{\it W1} & {\it W2} & 
$(b/a)_S$ & $(b/a)_W$ &
$R_{eu}$ &
$R_{eg}$ &
$R_{er}$ &
$R_{ei}$ &
$R_{ez}$ &
$R_{eW1}$ &
$R_{eW2}$
\\
 &  & 
(deg) &
(\kms) &  &
(mag) & (mag) & (mag) & 
(mag) & (mag) & (mag) & 
(mag) & (mag) &
 &  & 
($'$) & ($'$) & ($'$) & 
($'$) & ($'$) & ($'$) & 
($'$) 
\\
(1) & (2) & (3) & 
(4) & (5) & (6) & 
(7) & (8) & (9) & 
(10) & (11) & (12) & 
(13) & (14) &
(15) & (16) &
(17) & (18) & (19) &
(20) & (21) &
(22) 
}
\startdata
38803 & IC 3033 & 59$\pm$5 & 111$\pm$5 & 2.112$\pm$0.030 & 15.811$\pm$0.076 & 15.89 & 14.93 & 14.55 & 14.34 & 14.31 & 14.91 & 15.21 & 0.46 & 0.53 & 0.36 & 0.35 & 0.33 & 0.33 & 0.31 & 0.38 & 0.49 \\ 
38943 & NGC 4178 & 75$\pm$3 & 249$\pm$5 & 2.410$\pm$0.011 & 12.884$\pm$0.076 & 12.43 & 11.68 & 11.21 & 10.92 & 10.83 & 11.31 & 11.88 & 0.35 & 0.34 & 1.59 & 1.38 & 1.34 & 1.34 & 1.28 & 1.21 & 1.22 \\ 
38945 & IC 3044 & 73$\pm$3 & 125$\pm$5 & 2.116$\pm$0.019 & 14.975$\pm$0.076 & 15.02 & 14.02 & 13.62 & 13.41 & 13.31 & 14.04 & 14.77 & 0.36 & 0.40 & 0.52 & 0.54 & 0.56 & 0.57 & 0.58 & 0.61 & 0.57 \\ 
39028 & NGC 4192 & 78$\pm$3 & 458$\pm$5 & 2.671$\pm$0.007 & 12.724$\pm$0.076 & 11.57 & 10.49 & 9.77 & 9.36 & 9.13 & 9.38 & 9.98 & 0.24 & 0.36 & 2.76 & 1.96 & 1.82 & 1.77 & 1.71 & 1.39 & 1.40 \\ 
39181 & IC 3066 & 83$\pm$3 & 132$\pm$8 & 2.123$\pm$0.025 & 16.958$\pm$0.076 & 16.70 & 15.38 & 14.84 & 14.57 & 14.44 & 15.14 & 15.79 & 0.33 & 0.33 & 0.22 & 0.25 & 0.25 & 0.26 & 0.26 & 0.29 & 0.29 \\ 
39224 & NGC 4212 & 53$\pm$4 & 252$\pm$5 & 2.499$\pm$0.024 & 14.734$\pm$0.076 & 12.71 & 11.51 & 10.90 & 10.57 & 10.39 & 10.67 & 11.20 & 0.70 & 0.74 & 0.59 & 0.59 & 0.58 & 0.58 & 0.56 & 0.56 & 0.56 \\ 
39246 & NGC 4216 & 84$\pm$3 & 511$\pm$5 & 2.711$\pm$0.005 & 13.752$\pm$0.076 & 12.11 & 10.48 & 9.64 & 9.15 & 8.91 & 9.04 & 9.69 & 0.24 & 0.32 & 1.76 & 1.43 & 1.27 & 1.23 & 1.13 & 1.00 & 1.04 \\ 
39256 & IC 3077 & 58$\pm$4 & 58$\pm$13 & 1.835$\pm$0.102 & 18.110$\pm$0.076 & 15.41 & 14.16 & 13.54 & 13.16 & 13.07 & 13.72 & 14.71 & 0.65 & 0.72 & 0.59 & 0.58 & 0.56 & 0.58 & 0.56 & 0.54 & 0.42 \\ 
39308 & NGC 4222 & 90$\pm$1 & 220$\pm$5 & 2.342$\pm$0.010 & 14.293$\pm$0.076 & 14.68 & 13.54 & 12.90 & 12.53 & 12.35 & 12.43 & 12.94 & 0.14 & 0.32 & 0.89 & 0.85 & 0.83 & 0.83 & 0.77 & 0.52 & 0.52 \\ 
39431 & IC 3105 & 84$\pm$3 & 83$\pm$5 & 1.922$\pm$0.026 & 15.129$\pm$0.076 & 15.21 & 14.57 & 14.35 & 14.28 & 14.15 & 15.12 & 16.00 & 0.23 & 0.42 & 0.51 & 0.51 & 0.52 & 0.52 & 0.54 & 0.44 & 0.43 \\ 
\nodata \\
\enddata
\tablenotetext{*}{The complete version of this table is available online.}
\end{deluxetable*}


\addtocounter{table}{-1}
\begin{deluxetable*}{c||cccccccccccccccccccccc}
\tablecaption{Data Catalog (continued)\tablenotemark{*}
}
\tabletypesize{\scriptsize}
\tablehead{ \\
PGC & 
Cluster & 
\colhead{$\overline{u}$} & \colhead{$\overline{g}$} & \colhead{$\overline{r}$} & 
\colhead{$\overline{i}$} & \colhead{$\overline{z}$} & 
\colhead{$\overline{W1}$} & \colhead{$\overline{W2}$} &
\colhead{$A^{(i)}_u$} & \colhead{$A^{(i)}_g$} & \colhead{$A^{(i)}_r$} & 
\colhead{$A^{(i)}_i$} &  \colhead{$A^{(i)}_z$} &  
\colhead{$A^{(i)}_{W1}$} &  \colhead{$A^{(i)}_{W2}$} & 
\colhead{$u^*$} & \colhead{$g^*$} & \colhead{$r^*$} &  
\colhead{$i^*$} &  \colhead{$z^*$} &   
\colhead{$W1^*$} &  \colhead{$W2^*$} \\
 ~ & & 
(mag) & (mag) & (mag) & 
(mag) & (mag) & 
(mag) & (mag) &
(mag) & (mag) & (mag) & 
(mag) & (mag) & 
(mag) & (mag) & 
(mag) & (mag) & (mag) & 
(mag) & (mag) & 
(mag) & (mag)  
\\
(1) & (23) & (24) & 
(25) & (26) & (27) & 
(28) & (29) & (30) & 
(31) & (32) & (33) & 
(34) & (35) &
(36) & (37) &
(38) & (39) & (40) &
(41) & (42) &
(43) & (44)
}
\startdata
38803 & Virgo & 15.75 & 14.83 & 14.48 & 14.29 & 14.27 & 14.91 & 15.21 & 0.24 & 0.16 & 0.11 & 0.09 & 0.07 & 0.01 & 0.00 & 15.51 & 14.67 & 14.37 & 14.20 & 14.20 & 14.90 & 15.21 \\ 
38943 & Virgo & 12.30 & 11.59 & 11.14 & 10.87 & 10.79 & 11.31 & 11.88 & 0.62 & 0.45 & 0.34 & 0.29 & 0.24 & 0.03 & 0.01 & 11.68 & 11.14 & 10.80 & 10.58 & 10.55 & 11.28 & 11.87 \\ 
38945 & Virgo & 14.87 & 13.90 & 13.54 & 13.35 & 13.26 & 14.03 & 14.76 & 0.43 & 0.27 & 0.19 & 0.16 & 0.12 & 0.00 & 0.00 & 14.44 & 13.63 & 13.35 & 13.19 & 13.14 & 14.03 & 14.76 \\ 
39028 & Virgo & 11.42 & 10.38 & 9.69 & 9.30 & 9.09 & 9.37 & 9.97 & 0.80 & 0.57 & 0.43 & 0.34 & 0.25 & 0.04 & 0.01 & 10.62 & 9.81 & 9.26 & 8.96 & 8.84 & 9.33 & 9.96 \\ 
39181 & Virgo & 16.56 & 15.28 & 14.77 & 14.52 & 14.40 & 15.14 & 15.79 & 0.79 & 0.53 & 0.39 & 0.32 & 0.26 & 0.02 & 0.01 & 15.77 & 14.75 & 14.38 & 14.20 & 14.14 & 15.12 & 15.78 \\ 
39224 & Virgo & 12.57 & 11.40 & 10.82 & 10.51 & 10.35 & 10.66 & 11.19 & 0.23 & 0.15 & 0.11 & 0.07 & 0.04 & 0.01 & 0.00 & 12.34 & 11.25 & 10.71 & 10.44 & 10.31 & 10.65 & 11.19 \\ 
39246 & Virgo & 11.97 & 10.37 & 9.57 & 9.10 & 8.87 & 9.03 & 9.69 & 0.81 & 0.45 & 0.29 & 0.13 & 0.00 & 0.01 & 0.01 & 11.16 & 9.92 & 9.28 & 8.97 & 8.87 & 9.02 & 9.68 \\ 
39256 & Virgo & 15.24 & 14.04 & 13.46 & 13.10 & 13.03 & 13.72 & 14.72 & 0.28 & 0.20 & 0.15 & 0.13 & 0.11 & 0.01 & 0.01 & 14.96 & 13.84 & 13.31 & 12.97 & 12.92 & 13.71 & 14.71 \\ 
39308 & Virgo & 14.54 & 13.44 & 12.83 & 12.48 & 12.31 & 12.43 & 12.94 & 1.64 & 1.26 & 1.02 & 0.88 & 0.75 & 0.05 & 0.02 & 12.90 & 12.18 & 11.81 & 11.60 & 11.56 & 12.38 & 12.92 \\ 
39431 & Virgo & 15.07 & 14.46 & 14.28 & 14.22 & 14.11 & 15.11 & 15.99 & 0.74 & 0.47 & 0.32 & 0.26 & 0.21 & 0.01 & 0.00 & 14.33 & 13.99 & 13.96 & 13.96 & 13.90 & 15.10 & 15.99 \\ 
\nodata \\
\enddata
\tablenotetext{*}{The complete version of this table is available online.}
\end{deluxetable*}
\end{turnpage}

\clearpage

\begin{table*}[ht]
\scriptsize
\tabletypesize{\tiny}
\setlength{\tabcolsep}{0.2cm}
\centering
\caption{The calibrated slope and zero-point of the ITFRs at different SDSS and WISE passbands.} 
\label{tab:allITFRs}
\begin{tabular}{c l c ccc c ccc c c}
\tablewidth{0pt}
\hline \hline
 Band & TFR & Sample & \multicolumn{3}{c}{Universal Slope} &  & \multicolumn{3}{c}{Zero-point} & & ZP-Adjustment\\
 \cline{4-6}  \cline{8-10} \cline{12-12}
  & Code & & Ngal & Slope  & rms &  & Ngal & ZP  & rms  & & C$_{zp}$\\
(1) & (2) & (3) & (4) & (5) & (6) & & (7) & (8) & (9) & & (10)\\  
\hline 
\decimals
{\it u} & TF$_{u}$ &  OP &  429 & 	-7.03$\pm$0.17 & 	0.59 & & 39 &  	-19.27$\pm$0.13 & 	0.75 & & -0.08$\pm$0.30  \\
{\it g} & TF$_{g}$ &  OP  &  430 & 	-7.37$\pm$0.13 & 	0.49 & & 39 & 	-20.15$\pm$0.11 & 	0.62 & & -0.11$\pm$0.18  \\
{\it r} & TF$_{r}$ &  OP  &  430 & 	-7.96$\pm$0.13 & 	0.49 & & 39 & 	-20.57$\pm$0.10 & 	0.59 & & -0.08$\pm$0.13  \\
{\it i} & TF$_{i}$ &  OP  &  430 & 	-8.32$\pm$0.13 & 	0.49 & & 39 & 	-20.80$\pm$0.10 & 	0.59 & & -0.04$\pm$0.10  \\
{\it z} & TF$_{z}$ &  OP  &  429 & 	-8.46$\pm$0.13 & 	0.50 & & 38 & 	-20.89$\pm$0.10 & 	0.57 & & -0.08$\pm$0.11  \\
{\it W1} & TF$_{W1}$ &  IR  &  584 & 	-9.47$\pm$0.14 & 	0.58 & & 64 & 	-20.36$\pm$0.07 & 	0.60 & & \\
{\it W2} & TF$_{W2}$ &  IR  &  584 & 	-9.66$\pm$0.15 & 	0.62 & & 64 & 	-19.76$\pm$0.08 & 	0.65 & & 0.00$\pm$0.06 \\
\hline
\end{tabular}
\end{table*}

\section{Calibration of the TFR} \label{sec:tfr_calib}

We follow the same inverse TFR (ITFR) fitting procedure as explained in \citet{2012ApJ...749...78T} to calibrate the TFR for all wave bands of interest. Accordingly, the errors and residuals in \hi line width are taken into account in the linear least-square regression.  Bias errors are reduced by a factor of 5 with respect to a direct TFR analysis \citep{1994ApJS...92....1W}, reducing the corrections required to account for bias from a serious to a minor problem.  The major bias with use of the direct relation arises because at a given line width an intrinsically brighter galaxy drawn from above the mean relation is prone to be included in a sample, while an intrinsically fainter galaxy may be lost.  With the inverse relation, galaxies of a given magnitude with wider and narrower line widths are equally likely to be included. See the discussion in \S 1 and in \citet{2000ApJ...533..744T}.

In this study, the errors of the measured magnitudes at all wave bands are much smaller than the uncertainties in the measured line widths. Thus, we only use line width errors as the input measurement error. Although the inverse relation mitigates the Malmquist selection bias, the scatter in the sample induces a residual bias, which is addressed in \S \ref{sec:bias}.

\subsection{The ITFR Template Slope} \label{sec:tfr_slope}

There is the fundamental assumption in using the luminosity-rotation rate correlation as a tool for measuring distances that galaxies in diverse environments statistically follow a common relationship.  This assumption can be evaluated by giving attention to meaningfully large and varied samples.  Drawing upon individual galaxy clusters has the useful feature that all candidates are supposed to be at the same distance.  Samples can be constructed that are quasi-complete to apparent magnitude limits, with exclusions because of face-on orientation, or confusion, or morphological anomalies that are expected to be randomly drawn from the mean correlation.  Clusters do vary in their richness and the relative proportions of early- and late-type systems.  It can be asked if spirals in clusters have the same properties as spirals outside of clusters.  There is remarkably little evidence that there are differences in ways that affect the TFR.  Plausibly, gas-rich spirals in clusters of all sorts are recent arrivals that retain their pre-entry properties.  The issue deserves continual scrutiny.  Our 20 clusters are wide-ranging in their properties, from dense and populous to diffuse and representative of the field. 
As will be shown, the 20 clusters share the commonalities of correlation slopes and scatter, supportive of the proposition of universal relations between galaxy luminosities and rotation rates.

Assuming that there is a universal TFR describable by a power-law, then a key property of ITFR fits allows us to combine multiple cluster samples to form a robust template.  The slope of the ITFR is independent of the depth of the magnitude limit, within statistical uncertainties.  By contrast, the slope of the direct TFR becomes shallower as the magnitude limit is raised.  A fit that combines errors on both axes shares the same property in a lesser fashion.  This unique property of the ITFR of the decoupling of the slope from the depth of coverage allows cluster samples to be slid on top of each other with simple shifts in magnitude.  Nearby clusters will sample to the faintest magnitudes of applicable interest.  Distant clusters will only populate the upper part of the correlation, but do so drawing from the same slope.  Of course, to be useful, each individual sample must be sufficiently numerous as to allow for an independent determination of slope to judge consistency and merge with the ensemble with meaningful statistical uncertainty. 

The algorithm to calibrate the slope of the ITFR is as follows. (1) The ITFR is fit independently to each individual cluster, given the apparent magnitudes and \hi line widths of its constituent galaxies. (2) One cluster is chosen as a reference, and and the other clusters are moved to nominally the same distance by shifting data along the magnitude axis based on the relative estimated zero-points of the individually fitted ITFRs. (3) Adopting the shifted magnitudes, the combined data of all translated clusters and the reference cluster form a tentative universal template. (4) The slope of the fitted ITFR for the combined sample of clusters (the universal slope) is fitted individually to the separate clusters. (5) We cycle back to step 2, updating iteratively following the same process until convergence. In the iterations at step 4, the newly revised universal slope is adopted, and with each cycle, the  individual zero-points quickly settle to optimal values.

\begin{figure}[ht]
\centering
\includegraphics[width=1.\linewidth]{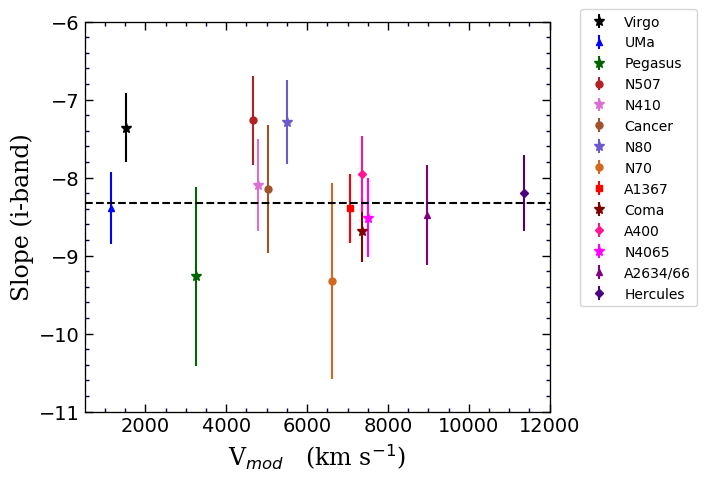}
\includegraphics[width=1.\linewidth]{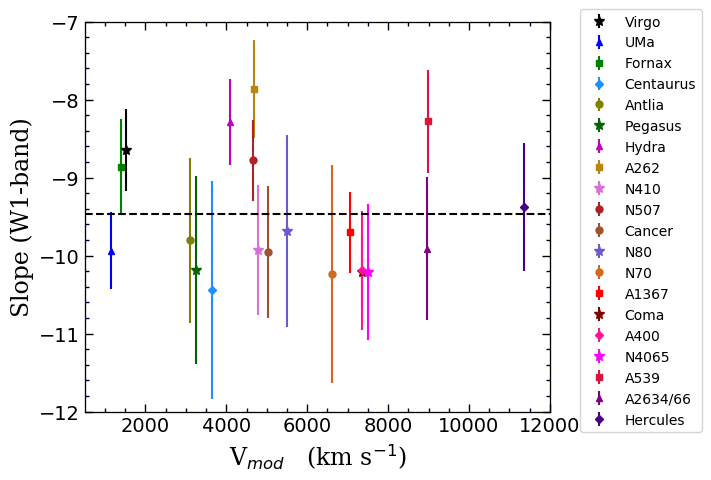}
\caption{The ITFR slopes with standard deviations for individual clusters vs. systemic velocity.  Top panel: {\it i} band; bottom panel: {\it W1} band. Horizontal dashed lines are drawn at the level of the universal slopes.
The legends on the right provide the cluster codes, listed in Tables \ref{table:i_cluster} and \ref{table:w1_cluster}.
}
\label{fig:slope_Vmod}
\end{figure}

The estimated relative distances from the preliminary individual zero-points confirm that Virgo, Ursa Major, and Fornax are the closest clusters of our sample, and therefore they have the most complete magnitude coverage. To minimize the incompleteness effect in the farther clusters, we first combine the Virgo, Ursa Major and Fornax clusters following the above algorithm. We adopt Virgo, the slightly nearer and most populated cluster, as the reference cluster. The universal slope is then updated iteratively by combining farther clusters in groups that are sorted based on their preliminary estimated relative distances. From closest to farthest, these cluster sets are (1) Centaurus, Antlia, and Pegasus; (2) Hydra, Abell 262, NGC 410, NGC 507, and Cancer; (3) NGC 80, NGC 70, Abell 1367, Coma, Abell 400; (4) Abell 2634/66 and Abell 539; and (5) Abell 2151 (Hercules). These groups of clusters are added step-by-step to form our ultimate universal ensemble. In each step, the ITFR slope of the composed ensemble in the previous cycle is adopted as a fixed parameter to derive the relative distances of the remaining clusters. It is shown that this procedure works because the slope of the ITFR does not vary with the level of magnitude truncation in each cluster \citep{2012ApJ...749...78T, 2013ApJ...765...94S, 2014ApJ...792..129N}.  This assertion is substantiated with Figure~\ref{fig:slope_Vmod}.  Closer clusters to the left in the plots sample to fainter magnitudes of the ITFR.

\begin{figure*}
\centering
\includegraphics[width=0.80\linewidth]{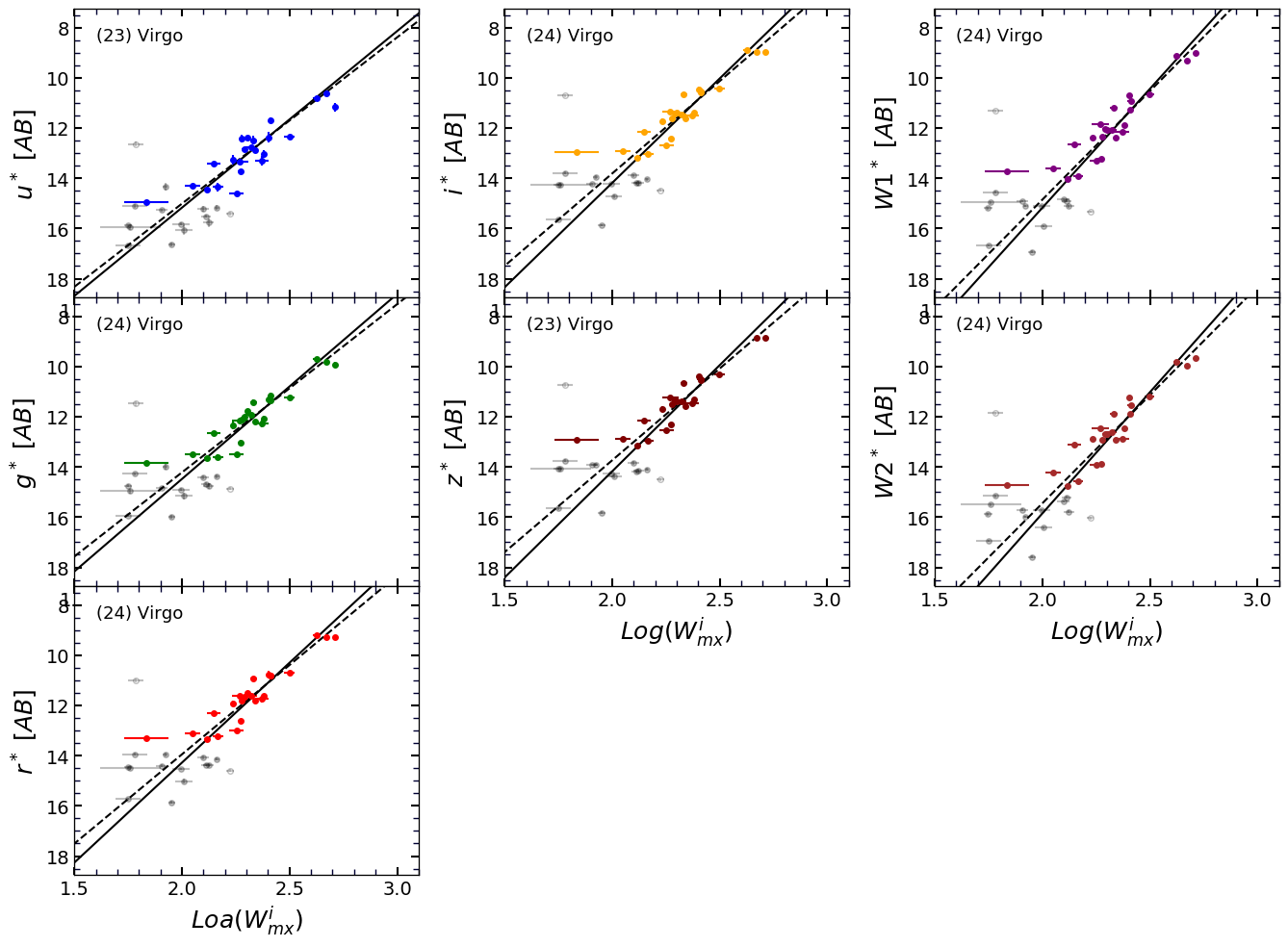}
\caption{The ITFR fit for the Virgo cluster galaxies at different optical and infrared bands. The dashed line illustrates the fitted ITFR using the filled colored points, and the solid line is the fitted ITFR using the universal slope as a fixed parameter. Gray open points represent galaxies that either are outliers or their absolute SDSS-magnitudes, after the full calibration, are fainter than $M_i=-17$ mag.}
\label{fig:TFR:Virgo}
\end{figure*}

Table \ref{tab:allITFRs} provides the universal slopes derived at different bands. Column (1) identifies the sequence of SDSS and WISE passbands. Column (2) gives the assigned codes for the resulting relations. Throughout the paper, $TF_{\lambda}$ stands for the ITFR that is constructed based on the magnitudes given by Eq. \ref{Eq:bkai} and no further corrections. Column (3) identifies the nature of the sample; {\it OP} and {\it IR} stand for the samples with optical and infrared photometry measurements, respectively.  Column (4) lists the number of galaxies that are involved in the ensemble when deriving the universal slope. Fewer galaxies with optical photometry are available than infrared because of the restricted sky coverage of SDSS.  Column (5) records the universal slope. Column (6) provides the rms of magnitude deviations from the fitted ITFR averaged over the full applicable magnitude domain.  In detail, scatter increases toward fainter magnitudes as discussed in \S \ref{sec:tfr_scatter}. Columns (7)-(9) contain the zero-point information discussed in \S \ref{sec:tfr_zp}. Column (10) lists the adjustment on the ITFR zero-point discussed in \S \ref{sec:zp_adjustment}.

\begin{table*}
\scriptsize
\setlength{\tabcolsep}{0.2cm}
\centering
\caption{The parameters of the fitted ITFRs for individual clusters in the {\it i} band, as visualized in Figure \ref{fig:i_clusters}. The cluster name is listed in the first column. The cluster codes are listed in the second column. ``Ngal" is the number of galaxies incorporated in the ITFR fit. ``Slope" and ``ZP" are the slopes and zero-points of the individually fitted ITFRs (dashed lines in Figure \ref{fig:i_clusters}). ``ZP$_0$" is the resulting individual zero-points from the linear regressions with the fixed ITFR universal slope, i.e. $-8.32\pm0.13$. The root mean square of galaxy deviations from the universal lines (solid lines in Figure \ref{fig:i_clusters}) along the magnitude axis are tabulated in the last column.} 
\label{table:i_cluster}
\begin{tabular}{lcc|cc|cc}
\hline \hline
Cluster & Code & Ngal & Slope & ZP & ZP$_0$ & $rms$ \\
\hline 
Virgo  &  V & 24 &  -7.36$\pm$0.44 &  10.13$\pm$0.09 &  10.03$\pm$0.09 &  0.69  \\
Ursa Major & U Ma &  36 &  -8.39$\pm$0.46 &  10.43$\pm$0.08 &  10.43$\pm$0.07 &  0.53  \\
Pegasus & Pe &  24 &  -9.27$\pm$1.15 &  12.24$\pm$0.23 &  12.40$\pm$0.11 &  0.59  \\
NGC507 & N5 &  20 &  -7.27$\pm$0.57 &  13.06$\pm$0.10 &  13.06$\pm$0.12 &  0.54  \\
NGC410 & N41 &  33 &  -8.10$\pm$0.58 &  13.15$\pm$0.08 &  13.14$\pm$0.08 &  0.52  \\
Cancer & Ca &  18 &  -8.15$\pm$0.82 &  13.29$\pm$0.11 &  13.29$\pm$0.11 &  0.48  \\
NGC80  & N8 &  14 &  -7.29$\pm$0.54 &  13.71$\pm$0.06 &  13.69$\pm$0.07 &  0.41  \\
NGC70  & N7 &  11 &  -9.32$\pm$1.26 &  13.78$\pm$0.15 &  13.71$\pm$0.10 &  0.35  \\
Abell 1367 & A1 &  68 &  -8.40$\pm$0.44 &  13.91$\pm$0.06 &  13.91$\pm$0.06 &  0.56  \\
Coma  & Co &  79 &  -8.68$\pm$0.40 &  13.94$\pm$0.06 &  13.95$\pm$0.06 &  0.56  \\
Abell 400  & A4 &  21 &  -7.95$\pm$0.48 &  13.99$\pm$0.06 &  14.00$\pm$0.06 &  0.58  \\
NGC4065  & N40 &  14 &  -8.51$\pm$0.50 &  14.25$\pm$0.09 &  14.26$\pm$0.08 &  0.40  \\
Abell 2634/66 & A2 &  29 &  -8.48$\pm$0.64 &  14.43$\pm$0.09 &  14.42$\pm$0.06 &  0.48  \\
Hercules & He &  39 &  -8.20$\pm$0.49 &  15.06$\pm$0.07 &  15.07$\pm$0.06 &  0.42  \\
\hline
\end{tabular}
\end{table*}

\begin{figure*}[ht]
\centering
\includegraphics[width=0.65\linewidth]{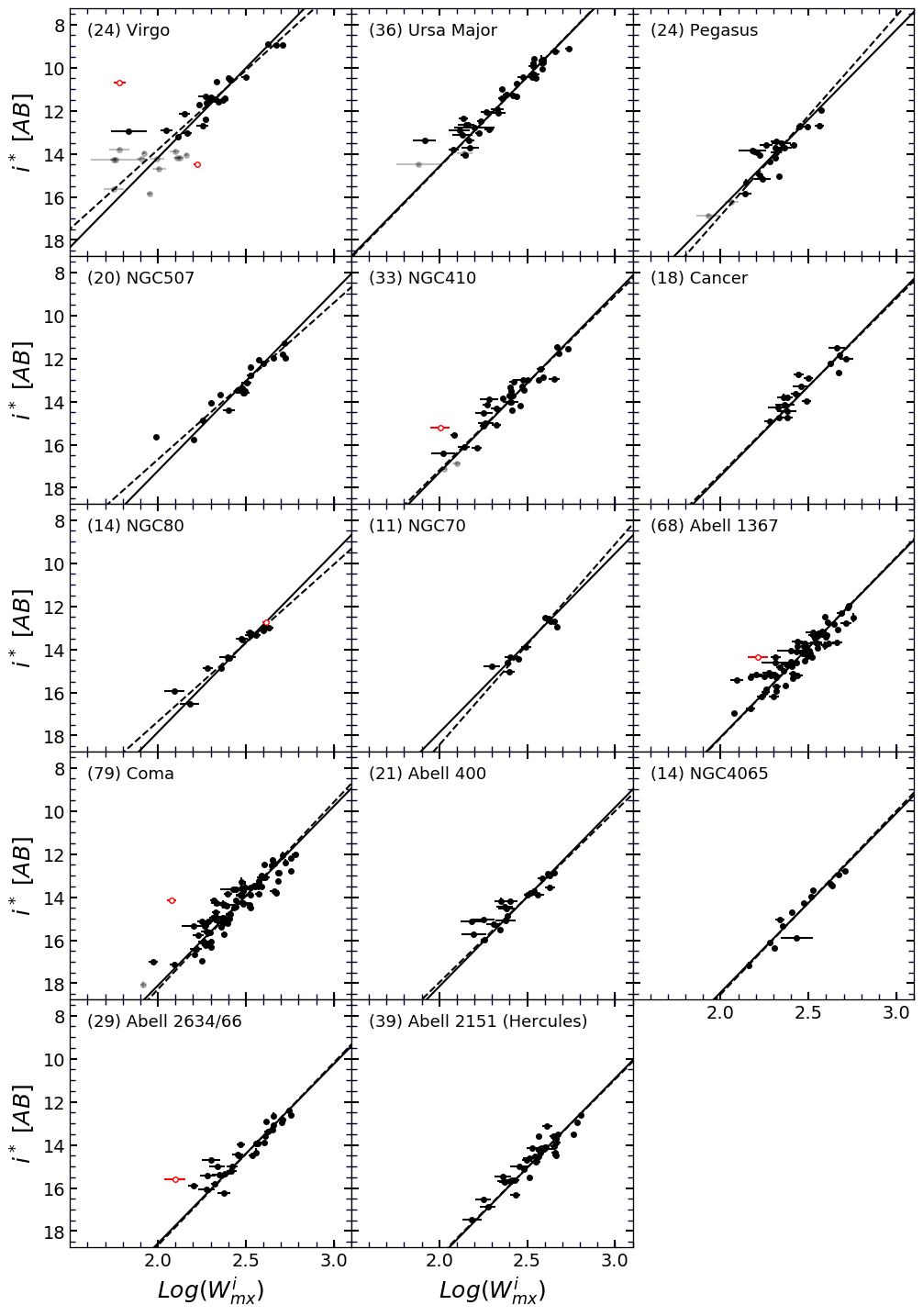}
\caption{The ITFR fits for 14 clusters with available SDSS {\it i-}band photometry. In each panel, the dashed line is the fit pertaining to the individual cluster, and the solid line is the best universal ITFR slope fit to the cluster.
Red points identify excluded outliers. Gray points are galaxies fainter than the magnitude limit of $M_i=-17$ mag. The number of contributing galaxies in the fitting process is displayed before the cluster name in the top left corner of each panel. }
\label{fig:i_clusters}
\end{figure*}

\begin{figure*}[ht]
\centering
\includegraphics[width=0.85\linewidth]{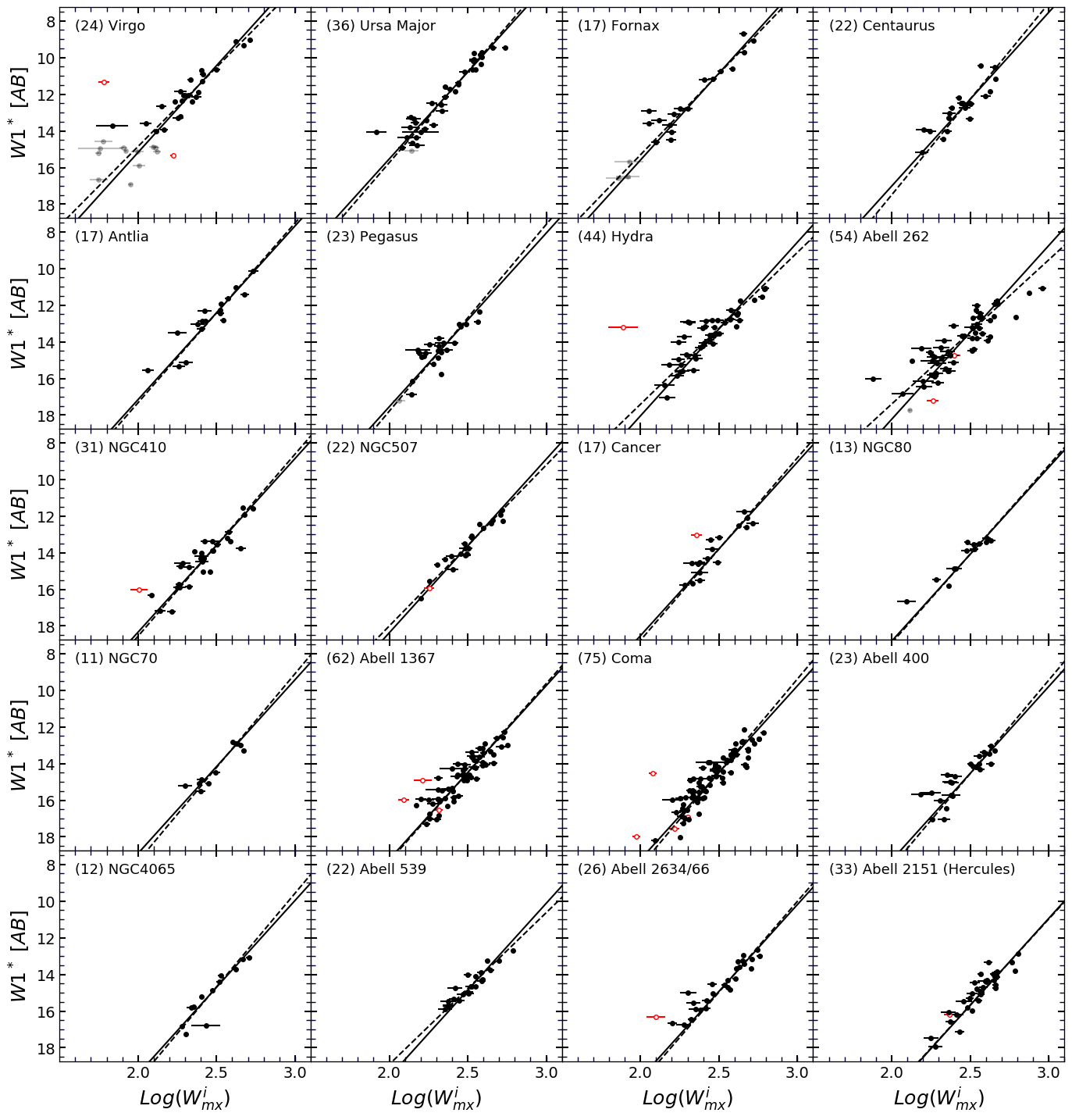}
\caption{Same as Figure \ref{fig:i_clusters} but for 20 clusters with available WISE {\it W1-}band photometry.
The faint-end magnitude cutoff is at $M_{W1}=-16.1$ mag.
}
\label{fig:w1_clusters}
\end{figure*}

As an example, Figure \ref{fig:TFR:Virgo} displays the apparent magnitudes versus line widths for galaxies in the Virgo cluster at SDSS {\it u, g, r, i, z} and WISE {\it W1} and {\it W2} passbands. 
In each panel, the dashed line is the least-squares fit to the filled colored points.  In the fitting process, we exclude the outlier galaxies NGC~4424 (PGC40809) and IC~3446 (PGC41440) and galaxies with absolute SDSS {\it i}-band magnitudes fainter than $-17$ mag (see \S \ref{sec:tfr_zp} regarding the absolute magnitude limit). In Figure~\ref{fig:TFR:Virgo}, the excluded galaxies are represented by open gray points. 
The universal ITFR slopes, fit with the zero-point as the only free parameter, are represented by the solid lines.

There are 14 clusters in our study with optical imaging coverage (the {\it OP} sample). Figure \ref{fig:i_clusters} illustrates the individual and universal ITFRs in the {\it i} band for these clusters. Galaxies that are considered anomalous are displayed in red and excluded from the least-squares fits. The slopes of the solid lines in all panels of this figure are the same and equal to the universal slope of the ITFR in the {\it i} band of $-8.32\pm0.13$. Complete information on the fitted relations is presented in Table \ref{table:i_cluster}. 

The all-sky coverage of the WISE imaging enables us to consider more clusters in infrared bands. One exceptional case is Abell 262 that is located at the edge of the SDSS survey, such that there are only seven galaxies with SDSS photometry but 55 with WISE photometry. Figure \ref{fig:w1_clusters} displays ITFRs for 20 clusters studied in the {\it W1} band, and Table \ref{table:w1_cluster} lists the parameters of the fits.

\begin{table*}
\scriptsize
\setlength{\tabcolsep}{0.2cm}
\centering
\caption{Same as Table \ref{table:i_cluster} but for the parameters of the fitted ITFRs for individual clusters in {\it W1}-band, as visualized in Figure \ref{fig:w1_clusters}. The universal slope at {\it W1}-band is $9.47\pm0.14$.}  
\label{table:w1_cluster}
\begin{tabular}{lcc|rc|cc}
\hline \hline
Cluster & Code & Ngal & Slope & ZP & ZP$_0$ & rms  \\
\hline 
Virgo  & V &  24 &  -8.64$\pm$0.53 &  10.53$\pm$0.11 &  10.44$\pm$0.10 &  0.82  \\
Ursa Major & U Ma &  36 &  -9.94$\pm$0.49 &  10.77$\pm$0.08 &  10.80$\pm$0.08 &  0.59  \\
Fornax  & F &  17 &  -8.86$\pm$0.61 &  10.88$\pm$0.13 &  10.86$\pm$0.14 &  0.78  \\
Centaurus & Ce &  22 &  -10.44$\pm$1.27 &  12.27$\pm$0.15 &  12.28$\pm$0.13 &  0.59  \\
Antlia  & An &  17 &  -9.80$\pm$1.05 &  12.43$\pm$0.12 &  12.42$\pm$0.11 &  0.61  \\
Pegasus & Pe  &  23 &  -10.19$\pm$1.21 &  12.70$\pm$0.24 &  12.81$\pm$0.12 &  0.66  \\
Hydra & Hy &  44 &  -8.19$\pm$0.55 &  13.25$\pm$0.09 &  13.29$\pm$0.10 &  0.89  \\
Abell 262 & A26 &  54 &  -7.86$\pm$0.63 &  13.47$\pm$0.11 &  13.49$\pm$0.13 &  0.98  \\
NGC 410 & N41 &  31 &  -9.93$\pm$0.83 &  13.60$\pm$0.11 &  13.61$\pm$0.10 &  0.64  \\
NGC 507 & N5 &  22 &  -8.78$\pm$0.52 &  13.61$\pm$0.08 &  13.64$\pm$0.08 &  0.38  \\
Cancer & Ca  &  17 &  -9.95$\pm$0.84 &  13.80$\pm$0.12 &  13.79$\pm$0.11 &  0.51  \\
NGC 80 & N8 &  13 &  -9.68$\pm$1.23 &  14.09$\pm$0.13 &  14.09$\pm$0.12 &  0.53  \\
NGC 70 & N7 &  11 &  -10.23$\pm$1.40 &  14.22$\pm$0.16 &  14.17$\pm$0.12 &  0.42  \\
Abell 1367 & A1 &  62 &  -9.70$\pm$0.52 &  14.49$\pm$0.06 &  14.49$\pm$0.06 &  0.59  \\
Coma  & Co & 75 &  -10.21$\pm$0.48 &  14.49$\pm$0.07 &  14.50$\pm$0.06 &  0.61  \\
Abell 400 & A4 &  23 &  -10.19$\pm$0.76 &  14.55$\pm$0.09 &  14.54$\pm$0.08 &  0.69  \\
NGC 4065 & N40 &  12 &  -10.21$\pm$0.87 &  14.68$\pm$0.12 &  14.71$\pm$0.11 &  0.53  \\
Abell 539 & A5 &  22 &  -8.28$\pm$0.66 &  14.77$\pm$0.08 &  14.86$\pm$0.08 &  0.39  \\
Abell 2634/66 & A2 &  26 &  -9.91$\pm$0.92 &  14.97$\pm$0.12 &  14.93$\pm$0.08 &  0.57  \\
Hercules & He &  33 &  -9.38$\pm$0.82 &  15.68$\pm$0.12 &  15.68$\pm$0.09 &  0.52  \\
\hline
\end{tabular}
\end{table*}

\begin{figure*}[ht]
\centering
\includegraphics[width=0.43\linewidth]{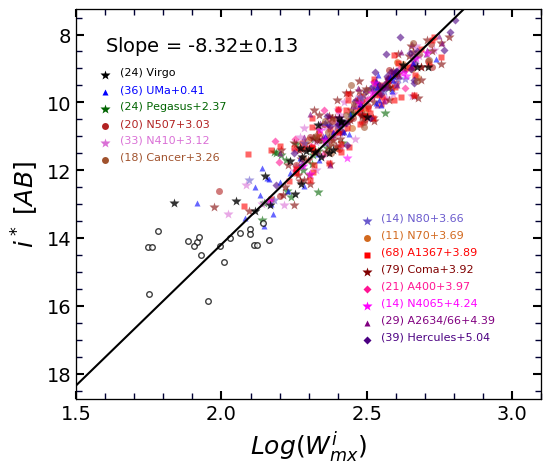}
\includegraphics[width=0.43\linewidth]{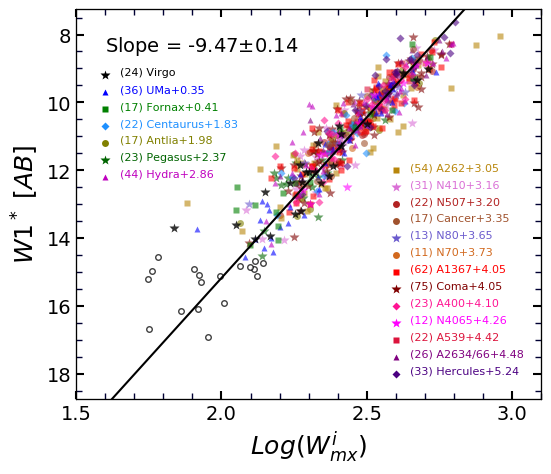}
\caption{
Universal template constructed from all clusters to find the slope of the ITFR. All clusters are shifted forward to the distance of the Virgo cluster by adjusting the apparent magnitude of their constituent galaxies. Galaxies in each cluster are displayed by unique symbol/color. Labels start with the number of contributed galaxies in each cluster, followed by the cluster name and the magnitude offset relative to Virgo. Open circles are galaxies fainter than the faint-end magnitude cutoff, i.e. $-17$ mag at the {\it i} band and $-16.1$ mag at the {\it W1} band. 
}
\label{fig:Slope_ZP_calib}
\end{figure*}

Our resulting universal ensembles in the {\it i} and {\it W1} bands are illustrated in the two panels of Figure~\ref{fig:Slope_ZP_calib}. The annotations list cluster names and the zero-point offsets relative to the Virgo cluster. In cases of clusters with both optical and infrared coverage the magnitude offsets at the {\it i} and {\it W1} bands are in reasonable agreement.

\begin{figure*}[ht]
\centering
\includegraphics[width=0.35\linewidth]{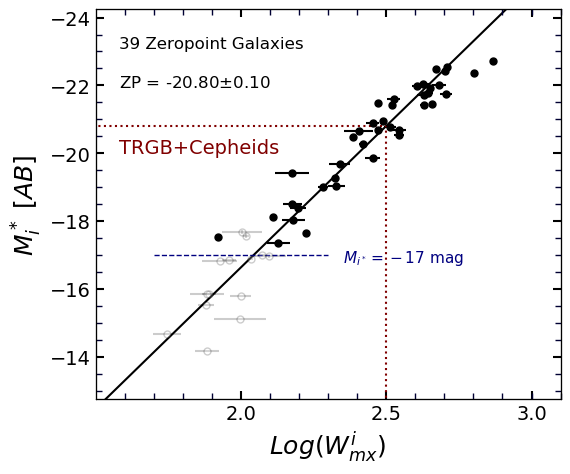}
\hspace{5mm}
\includegraphics[width=0.35\linewidth]{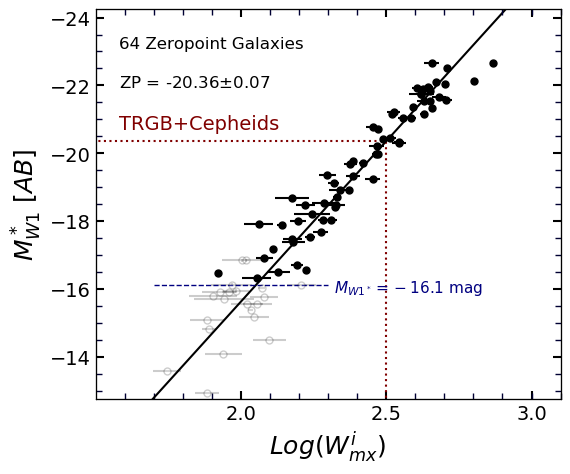}

\includegraphics[width=0.35\linewidth]{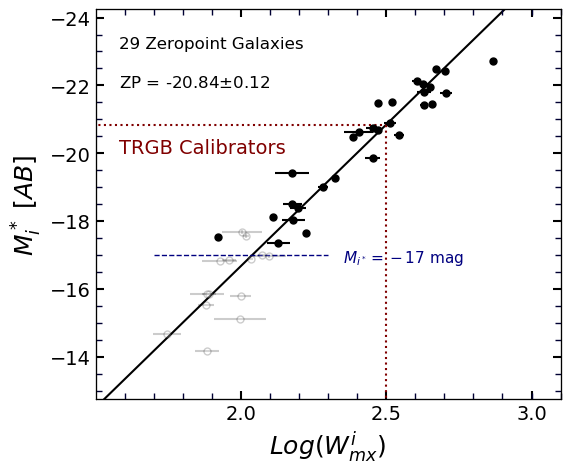}
\hspace{5mm}
\includegraphics[width=0.35\linewidth]{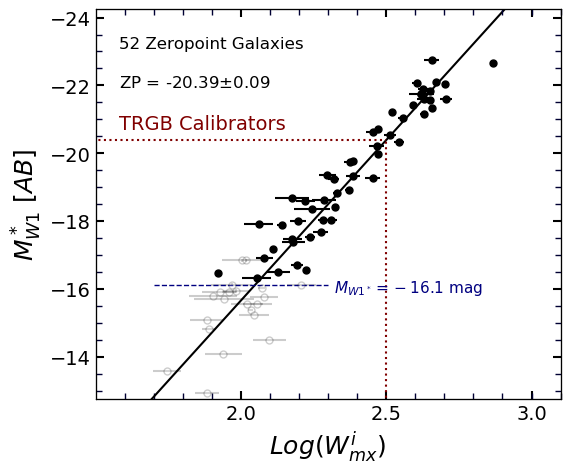}

\includegraphics[width=0.35\linewidth]{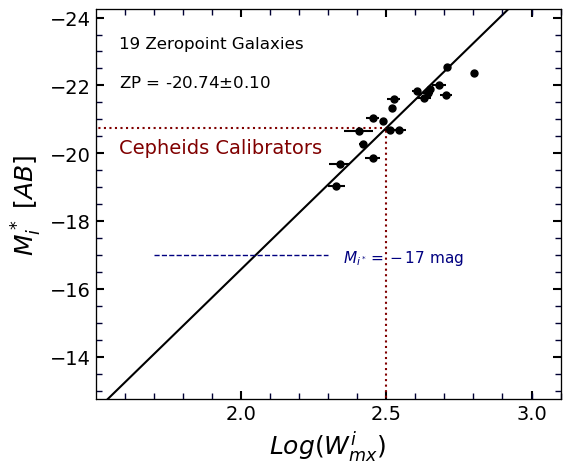}
\hspace{5mm}
\includegraphics[width=0.35\linewidth]{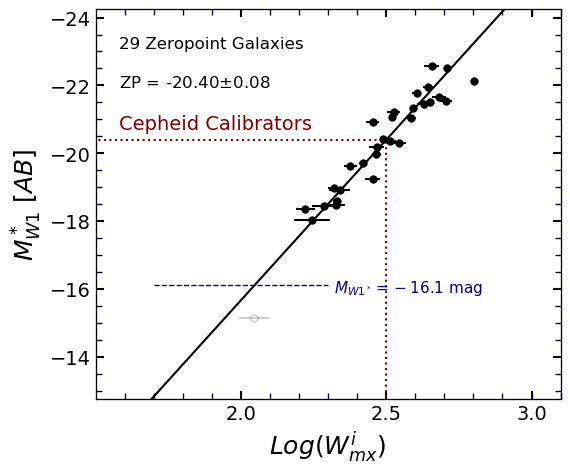}
\caption{
Absolute luminosity of the zero-point calibrators versus \hi line width at optical {\it i}-band (left) and infrared {\it W1}-band (right). The slopes of the solid lines are set by the cluster template. least-squares minimization of magnitudes set the zero-points.
Blue dashed lines mark the adopted magnitude limits. Faint and excluded zero-point calibrators are displayed in gray. Maroon dotted horizontal/vertical lines mark magnitude zero-points where log$(W^i_{mx})=2.5$.
}
\label{fig:ZP_calib}
\end{figure*}

\subsection{Zero-point Calibration and Absolute Distances} \label{sec:tfr_zp} 

We consider 94 galaxies that have distance estimates derived from either the Cepheid Period$-$Luminosity Relation (CPLR) or the Tip of the Red Giant Branch (TRGB) luminosity feature.  Not all of these are useful, particularly those that pertain to systems fainter than our ultimate $M_i = 17$ cutoff, but there are many more potential zero-point calibrators than previously available. The absolute calibration of CF3 was based on 33 galaxies: 25 CPLR, 19 TRGB, with 11 in common.  Here, as in the earlier study, the CPLR and TRGB calibration distances are found through completely separate pathways.  The two resultant scales can be compared, evaluated, and combined as deemed appropriate. 

In CF3, the CPLR calibration assumed a Large Magellanic Cloud (LMC) distance modulus of 18.50.  Here we adopt a small change.
Our zero-point scale is set to be consistent with that of \citet{2019ApJ...876...85R}, which combines linkages with the LMC assumed at a modulus of $\mu_{LMC}=18.477$ \citep{2019Natur.567..200P}, the maser galaxy NGC~4258 at a modulus of $\mu_{N4258}=29.40$ \citep{2013ApJ...775...13H}, and a Milky Way-based calibration.  That paper argues for a small modification (a decrease in moduli of $-0.024$~mag) of the CPLR distance moduli reported by \citet{2016ApJ...826...56R}.  In addition, three other sources of distances from the CPLR are incorporated.  Ground-based observations provide measurements for very nearby galaxies.  We average over two sources: those by \citet{2016AJ....151...88B} and by the Araucaria collaboration \citep{2013ApJ...773...69G, 2017ApJ...847...88Z} and related references.  The two sources agree to within 0.02 mag ($1\sigma$ with scatter $\pm 0.06$~mag) and are given equal weight in averaging. Zero-points tied to the LMC are rescaled to agree with \citet{2019ApJ...876...85R}. The other Cepheid contribution is \citet{2001ApJ...553...47F}.  The overlap with \citet{2016ApJ...826...56R} is only four objects, but the agreement is within 0.02~mag at the same LMC fiducial distance.  Each of these inputs is rescaled to be consistent with \citet{2019ApJ...876...85R}.     

Our primary source for TRGB distances is the color-magnitude diagram/TRGB catalog in the Extragalactic Distance Database (EDD)\footnote{http://edd.ifa.hawaii.edu} \citep{2009AJ....138..332J}. The zero-point in this compilation was established by \citet{2007ApJ...661..815R} based on horizontal branch/RR Lyrae distances to nearby dwarf spheroidals and other minor galaxies in the Local Group.  The sample is expanded by the addition of three objects studied by \citet{2015ApJ...807..133J, 2017ApJ...836...74J}.

As with the cluster samples, the zero-point calibrators are subject to faint-limit luminosity cuts.  The cut at the optical bands is set at the {\it i} band such that galaxies fainter than $M_i = -17$ are dropped.  The infrared cut is the roughly equivalent $M_{W1} = -16.1$.  There are four galaxies brighter than these limits that deviate by more than $3\sigma$ from the TFR and are considered anomalous cases and rejected: PGC 5896 (NGC 625), PGC 44536 (NGC 4861), PGC 48334 (NGC 5253), and PGC 68535 (NGC 7250). The optical bands are left with 39 calibrators: 19 CPLR, 29 TRGB, with 9 in common.  There are 64 calibrators in the infrared: 29 CPLR, 52 TRGB, with 17 in common.  There are fewer optical calibrators because of the restricted domain of the SDSS photometry.  The WISE photometry is available across the full sky.

Zero-point calibrations are shown for the representative {\it i} and {\it W1} bands in Figure~\ref{fig:ZP_calib}.  The linear fits obey the formulation
\begin{equation}
\label{Eq:TFR}
\mathcal{M}^*_{\lambda}=Slope \big({\rm log}W^i_{mx}-2.5 \big)+ZP ~, 
\end{equation}
where zero-point, $ZP$, is the absolute luminosity of a galaxy at log$W^i_{mx}=2.5$. 
We adopt the convention that the ITFR predictions of absolute magnitudes are given in script. The superscript asterisks indicate that magnitudes are corrected for the effects explained following Eq.~\ref{Eq:bkai}. In Figure~\ref{fig:ZP_calib}, the TRGB and CPLR calibrators are shown separately in the middle and bottom panels, respectively, and the combined samples are shown in the top panels, with straight averaging between TRGB and CPLR when both are available.  

Slopes are set by the values determined from the cluster templates, while zero-points are established by least-squares regression with the zero-point as the only free parameter.
In Figure~\ref{fig:ZP_calib} the maroon dotted lines indicate the location of the $ZP$ on the absolute magnitude axis. In these examples, 
\begin{subequations}
	\begin{align}
	\begin{split}
		\mathcal{M}^*_{i} = &- (20.80 \pm 0.10) \\
			&- (8.32 \pm 0.13)({\rm log}W^i_{mx} - 2.5), 
	\end{split} \\
	\begin{split}
		\mathcal{M}^*_{W1} = &- (20.36 \pm 0.07) \\
			&- (9.47 \pm 0.14)({\rm log}W^i_{mx} - 2.5).
	\end{split}
	\end{align}
\end{subequations}
The zero-points for all wave bands are recorded in column (8) of Table \ref{tab:allITFRs}. Column (7) of this table gives the number of zero-points calibrators used at each band, and column (9) gives the rms scatter of calibrators about the fitted lines along the absolute luminosity axis.

The distance modulus, $\mu_{\lambda}$, of a given galaxy with the apparent magnitude $m^*_{\lambda}$ is given as
\begin{equation}
\label{Eq:mu_lambda}
    \mu_{\lambda} = m^*_{\lambda} - \mathcal{M}^*_{\lambda}~.
\end{equation}
For example, the magnitude offsets between apparent magnitudes at log$W^i_{mx}=2.5$ in Figure~\ref{fig:Slope_ZP_calib} and absolute magnitudes at the same line width value in Figure~\ref{fig:ZP_calib} give measures for the distance modulus of the Virgo Cluster.  Moduli for the other clusters can be determined from the magnitude offsets derived in the construction of the universal clusters templates, recorded in the labeling of Figure~\ref{fig:Slope_ZP_calib} for the cases of the {\it i} and {\it W1} passbands.

\subsection{Alternate Zero-point Calibration: Possible Serious Systematic}

\citet{2019ApJ...882...34F}, hereafter F19, have recently provided a recalibration of the TRGB zero-point that led them to derive a value for the Hubble constant $H_0$ of $69.8\pm1.9$, significantly lower than the \citet{2019ApJ...876...85R} value of $74.0\pm1.4$ and, anticipating the results of this study, significantly lower than we find with the calibration derived in \S\ref{sec:tfr_zp}.  Given the varying degrees of tension of these values with the value of $67.4\pm0.5$ derived from early universe conditions \citep{Planck2018} and the implications for physics beyond the standard model, we are called upon to justify why we prefer our TRGB calibration to that of F19. 

We note two significant differences between the TRGB zero-point calibration by F19 and our own.  First, F19 based their calibration on the Large Magellanic Cloud (LMC) at a distance given by the combination of detached eclipsing binaries \citep{2019Natur.567..200P} and mid-infrared CPLR \citep{2011ApJ...743...76S, 2012ApJ...759..146M}.  The LMC is too large with the red giant branch stars too bright to be easily studied with the Hubble Space Telescope.  F19 drew on ground-based observations of the OGLE Collaboration \citep{2012AcA....62..247U} with photometry in the {\it V} and {\it I} bands.  Our calibration is tied to horizontal branch and RR Lyrae distances to the Local Group galaxies M33, IC~1613, NGC~185, Sculptor, and Fornax {\citep{2007ApJ...661..815R}.  The stellar photometry isolating the TRGB was with our standard filters with the Hubble Space Telescope using both the WFPC2 and ACS detectors.

Second, F19 assumed that the TRGB has a constant absolute magnitude of $M_I=-4.05$ for all manner of galaxies.\footnote{\citet{2019ApJ...886...61Y} argued that the F19 TRGB absolute magnitude should be $M_I=-3.97$ and the F19 calibration of $H_0$ should be increased 3.7\% to 72.4~\kmsMpc, primarily because, they contended, reddening is lower in the direction of the LMC fields than accepted by F19.} The \citet{2007ApJ...661..815R} calibration that we assume incorporates a weak color dependency, in response primarily to metallicity variations and secondarily to age differences.  The LMC has a relatively low metal abundance, and consequently the TRGB is expected in our formulation to be somewhat brighter than is the case with most of the galaxies in the TFR zero-point calibration. 

These alternate procedures result in differing values for the Hubble constant, so which better approximates reality?  F19 compared their TRGB and \citet{2019ApJ...876...85R} CPLR distance moduli and for 10 SN Ia systems relevant to their study, they found $<\mu_{TRGB}-\mu_{CPLR}>=+0.06\pm0.17$ with $1\sigma$ uncertainty of $\pm0.05$; evidence for good agreement.  We made a similar comparison of our TRGB and \citet{2019ApJ...876...85R} CPLR moduli with 25 galaxies and got the disturbing difference $<\mu_{TRGB}-\mu_{CPLR}>=0.13\pm0.10$ and a $1\sigma$ uncertainty of $\pm0.02$.

It could be argued that we should abandon our separate TRGB and CPLR calibrations.  Instead, we should force agreement averaged over instances where both measurements are available. We could choose one or the other of the TRGB or CPLR as a baseline or take some average of the two. However, any such approach fails.  Consider the panels of Figure~\ref{fig:ZP_calib}.  Retaining the separation of TRGB and CPLR calibrations, we can reach compatible inferred $H_0$ values for the more distant clusters in our study, as will be described culminating in \S \ref{sec:HubbleParam}.  However, if one or the other of the TRGB or CPLR scales is changed to force agreement, on average, between individual targets, then a difference at the level of 7\% is created if the zero-point calibration is based solely on TRGB or on CPLR samples. 

We review this unsatisfactory situation. If the F19 calibration of the TRGB is accepted, then there is statistical agreement between the TRGB measurements and the CPLR moduli of \citet{2019ApJ...876...85R} but there is a disconnect between the TRGB and CPLR calibrations of the TFR that leads to inconsistent determinations of $H_0$.  If, instead, the TRGB calibration of \citet{2007ApJ...661..815R} is accepted and the \citet{2019ApJ...876...85R} calibration of the CPLR are accepted then the TRGB and CPLR calibrations of the TFR are consistent in terms of values of $H_0$ found using our cluster sample.  

It is to be appreciated that the primary goal of the {\it Cosmicflows} program is to measure the peculiar velocities of galaxies, not $H_0$.  With distances $d_i$, we obtain peculiar velocities $V_{pec,i} \sim cz_i - H_0 d_i$ where $H_0 \sim \langle cz_i/d_i \rangle$.  Hence, it is important that we use a value of $H_0$ compatible with the distance measurements, not that we use the ``correct" $H_0$.  Important details of the zero-point calibration remain to be resolved.  The discussion in this section highlights the possibility of substantial systematics.

\begin{figure*}[ht]
\centering
\includegraphics[width=0.95\linewidth]{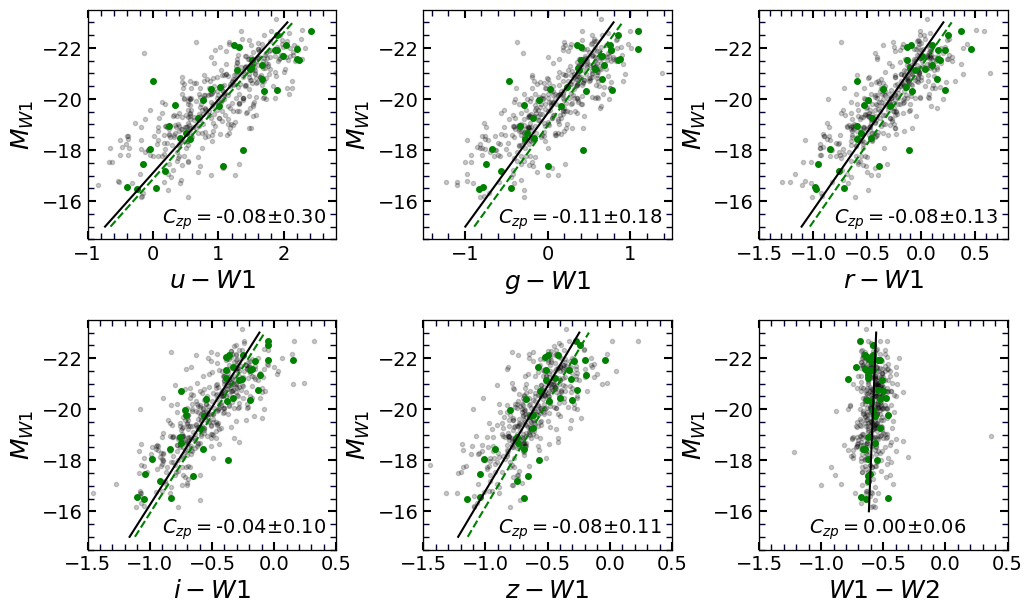}
\caption{Absolute magnitudes at {\it W1}-bands versus color indices. Gray points represent cluster galaxies used for ITFR slope calibration. Green points are zero-point calibrators. The fitted lines are the results of a least-squares regression procedure that minimized the residuals along the color axis. Black solid and green dashed lines are fitted over gray and green points, respectively. The slope of the green line is the same as the black line and is fixed in the fitting process. In each panel, C$_{zp}$ is the difference in the zero-point of the fitted lines along the horizontal axis.}
\label{fig:zp_adjustments}
\end{figure*}

\begin{figure*}[ht]
\centering
\includegraphics[width=0.90\linewidth]{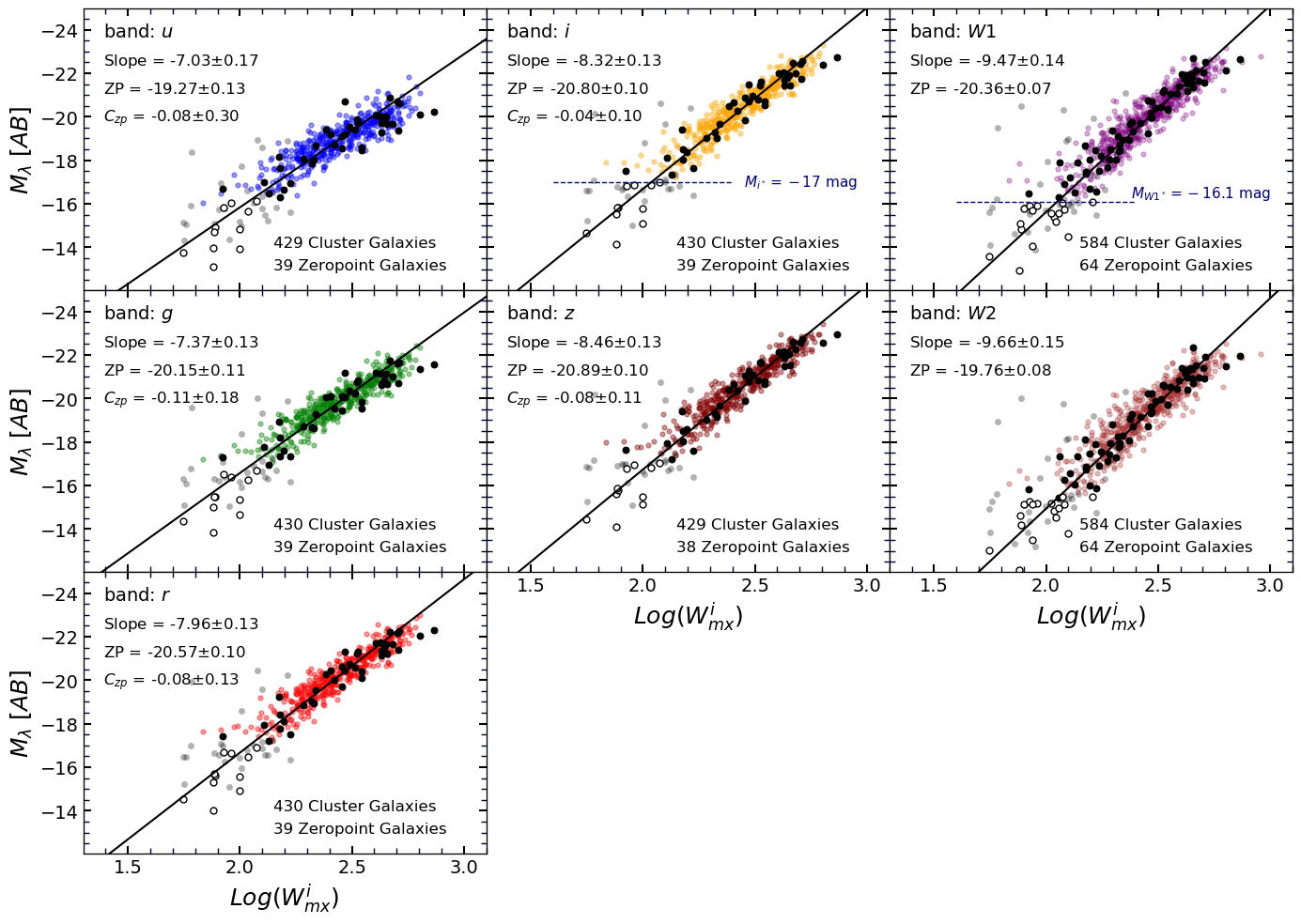}
\caption{Absolute luminosity vs. \hi line width for different wave bands. Solid straight lines illustrate the fitted ITFRs with no bandpass adjustments. Colored points show cluster galaxies that are used for slope calibration.  The magnitude limit $M_i=-17$ mag is imposed in the fitting process, illustrated by the blue dashed horizontal line in the {\it i}-band panel. 
The blue dashed line in the {\it W1}-band panel shows the equivalent magnitude limit of $-16.1$ in the {\it W1} band.
Gray points are the faint cluster galaxies or the outliers we excluded from the ensemble in the fitting process.
Black filled points are the spirals used for zero-point calibration. 
Open circles are zero-point calibrators that are fainter than the adopted magnitude limit.
}
\label{fig:TF_all}
\end{figure*}

\subsection{Zero-point Adjustments} \label{sec:zp_adjustment} 

In principle, distance moduli to a given cluster should be the same across all passbands.  However, we have detected a trend in the cluster calibrator samples whereby moduli are slightly reduced at blueward bands.  We trace this effect to a slight color offset between the ensemble cluster template sample and the sample with CPLR or TRGB distances that gives the zero-point calibration. 

The problem is illustrated in Figure~\ref{fig:zp_adjustments}.  Colors with respect to the infrared band {\it W1} are shown on the abscissa in each panel, with the clusters sample plotted as gray points and the zero-point sample in green.  The ordinate positions of the points are given by absolute magnitudes in the {\it W1} band following from the calibration described in the previous subsection.  The solid black lines in these plots are fits to the cluster template sample obeying the formulation $\lambda-W1 = m M_{W1} + b $. The intercept $b$ is the average $\lambda-W1$ color of galaxies with $M_{W1}=0$.  The least-squares minimization is carried out in the color terms to counter Malmquist-type bias.

The green dashed lines in these panels are the fits to solely to the zero-point calibrators, preserving the black solid line slopes but giving freedom to the scaling parameter $b$.  Offsets between the black and green lines are given by $C_{zp}=b_{clusters}-b_{zp}$.  In every case between an optical bandpass and the infrared bandpass {\it W1} the green dashed lines are offset redward of the black solid lines.  Never are the offsets statistically significant.  However, making an adjustment for the offsets statistically eliminates the differences in distance moduli measurements as a function of passband.  We make the adjustment
\begin{equation}
\label{Eq:zp_adjust}
     \overline{ZP} = ZP + C_{zp}   ~.
\end{equation}
which gives consistency with moduli determined in the {\it W1} band.  There is no assurance that reporting to the {\it W1} band is correct.  We take the amplitude of shifts at the level of 3\% in distances as a quantitative estimator of systematic uncertainty.  

The final linear calibrations are given in Figure~\ref{fig:TF_all} for the five optical and two infrared bands.  The points in colors represent galaxies from the cluster template brighter than our magnitude thresholds and not rejected as an outlier.  The CPLR and TRGB zero-point galaxies are represented by black points and establish the absolute magnitude scales.  Information in each panel includes the number of contributing cluster template and zero-point galaxies, the slopes derived from the cluster template, the zero-points derived directly from the galaxies with CPLR or TRGB distances, and the adjustment parameters, $C_{zp}$, required to null the color offsets demonstrated in Figure~\ref{fig:zp_adjustments}.  Neither of the infrared band correlations receives an adjustment.

To avoid confusion, the nomenclature for the ITFR with adjusted zero-points is  
\begin{equation}
\label{Eq:adjustedTFR}
\overline{\mathcal{M}^*_{\lambda}}=Slope \big({\rm log}W^i_{mx}-2.5 \big)+\overline{ZP} ~. 
\end{equation}

\subsection{Scatters about the ITFR} \label{sec:tfr_scatter}

We use the rms scatter of data points about the fitted ITFRs as a measurement of the statistical uncertainty of individual distance estimates. Absolute magnitude residuals from the ITFR are expressed as
\begin{equation}
\label{Eq:TFR_residual}
    \Delta M^i_{\lambda} = M^i_{\lambda} - \mathcal{M}^*_{\lambda}~,
\end{equation}
where $M^i_{\lambda}$ is the absolute magnitude of a target galaxy and $\mathcal{M}^*_{\lambda}$ is the estimated absolute magnitude at the target's line width from the fitted linear relation. The rms scatter of the ensemble of galaxies in Figure~\ref{fig:TF_all} (colored cluster template sample) and zero-point calibrators (black points) are listed in columns (6) and (9) of Table \ref{tab:allITFRs}, respectively. 
Residuals $\Delta M^i_{\lambda}$ follow an approximately Gaussian distribution; therefore, the rms value can be taken as the $1\sigma$ error in estimates of target distances. 

\begin{figure}[t]
\centering
\includegraphics[width=0.9\linewidth]{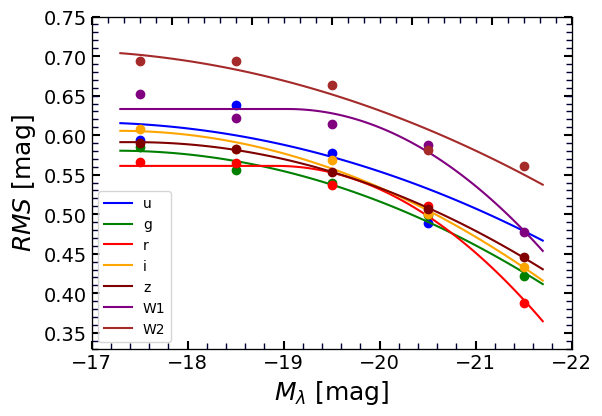}
\caption{Binned rms scatter of the template cluster galaxies about the linear ITFR at different passbands. Curves are fitted parabolic functions that are flattened after they peak at the faint-end.}
\label{fig:RMScurves}
\end{figure}

The averaged {\it u}-band scatter of $0.60$ mag is greater than at longer optical wavelengths. The {\it u} band is sensitive to young stellar populations and dust obscuration. 
Moreover, the quality of photometry is worst at the {\it u} band, where the uncertainties of measured magnitudes are $\sim0.10$ mag, about double that at other SDSS bands. At the {\it g, r, i} and {\it z} bands, the averaged rms scatter is relatively the same and equal to $\sim 0.50$ mag. Using 267 cluster galaxies, \citet{2012ApJ...749...78T} reported an rms scatter of $0.41$ mag for the {\it I}-band (Vega) calibration of the ITFR. With an additional 34 galaxies, \citet{2014ApJ...792..129N} found a slightly larger scatter of $0.46$ mag at the same wave band.  Our averaged rms scatter about the {\it i}-band relation, based on 430 galaxies, is $0.49$ mag.  Scatter tends to increase toward the fainter end of the TFR, a domain more completely probed by our more extensive samples. The increased scatter toward fainter magnitudes is demonstrated in Figure \ref{fig:RMScurves}.  Scatter increases from $\sim 0.40$ for the most luminous galaxies to $\sim 0.58$ for the faintest galaxies. It can be noted that the rms scatter for individual clusters recorded in Tables~\ref{table:i_cluster} and \ref{table:w1_cluster} tends to decrease for the more distant clusters.  Progressively with increasing distance, only the bright end of the TFR is probed.

Based on 584 galaxies in our sample with infrared photometry, we find that the dispersion about the {\it W1} and {\it W2} calibrations is $0.58$ mag and $0.62$ mag, respectively, in agreement with previous studies  \citep{2013ApJ...765...94S, 2014ApJ...792..129N}. The latter reported a scatter of $0.54$ mag for dispersion about their {\it W1} ITFR fit using 310 galaxies. 

Overall, scatter tends to decrease toward the near infrared wavelengths where with the effects of young stellar population and metallicity are minimized, as anticipated by \citet{1979ApJ...229....1A}.  The subsequent substantial increase in scatter at the WISE bandpasses is associated with a color term that is discussed in \S \ref{sec:tfr_addTerms}.

\begin{figure*}[ht]
\centering
\includegraphics[width=0.9\linewidth]{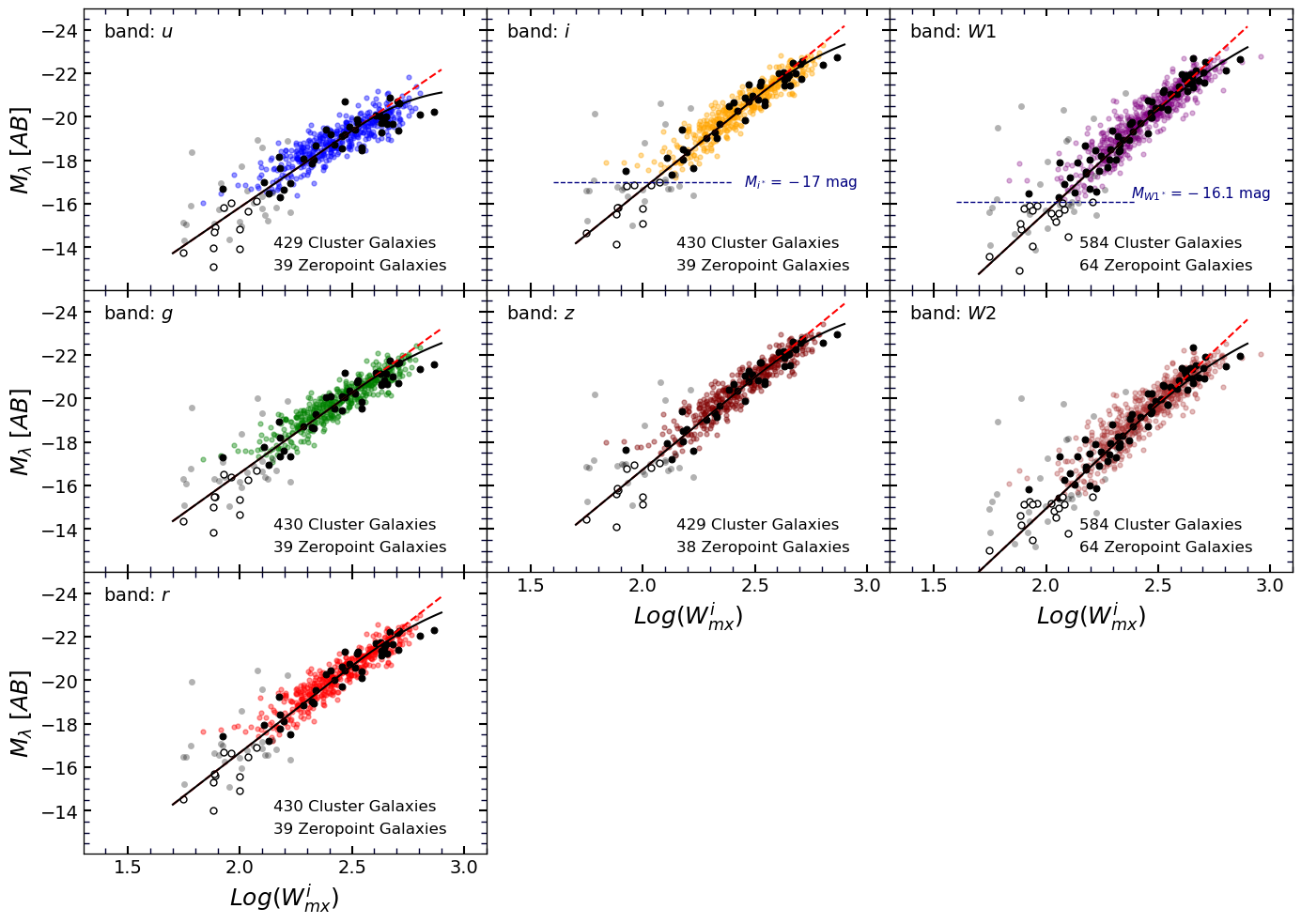}
\caption{Similar to Fig \ref{fig:TF_all}. The linear ITFR is replaced with quadratic functions at the bright ends. 
At the {\it u, g, r, i} and {\it z} optical bands, the departure from the linear relation begins at log$(W^i_{mx})=2.5$. At the {\it W1} and {\it W2} infrared bands, the break point is at log$(W^i_{mx})=2.4$. In each panel, the dashed line displays the continuation of the linear ITFR.
}
\label{fig:Curved_TFR}
\end{figure*}

\subsection{Cluster Distance Bias} \label{sec:bias} 

While the formalism of the ITFR effectively nulls the selection Malmquist bias, another bias remains because of the nonuniform population of the luminosity function of a cluster. Here our concern is restricted to only the gas-rich spiral populations, which, from compilations combining Virgo, Ursa Major, and Fornax samples, can be described by Schechter functions \citep{1976ApJ...203..297S} with bright end exponential cutoffs characterized by $M^{\star}_I=-23.0$ and $M^{\star}_{W1}=-22.0$ and faint-end power-law slopes $\alpha=-1.0$ \citep{2012ApJ...749...78T, 2014ApJ...792..129N}.  

The bias arises when the faint limit in the sampling of a cluster approaches the exponential cutoff magnitude. In such instances, there are more fainter galaxies available to scatter upward into the sample than brighter galaxies to scatter downward out of the sample.  The upward-scattered systems will tend to lie above an unbiased relation, resulting in a fit to the cluster that is slightly too close.  Since the Schechter faint-end slope parameter $\alpha=-1.0$ implies equal numbers per magnitude bin at fainter magnitudes, for relatively nearby clusters, there is little bias because there is a roughly equal probability of upward and downward scatter. See \citet{2014ApJ...792..129N} for further elaboration. 

We determine the amplitudes of the bias in each passband with simulations. However, before discussing that detail, we address another concern.  Simulations can be used to study the joint effects discussed in this section and the next.

\subsection{TFR Curvature} \label{sec:curvature}

A close look at Figure \ref{fig:TF_all} reveals that at the bright end of the TFR, almost all galaxies are below the fitted straight lines, implying that the linear TFR might not perfectly capture the general trend of data points. The departure of bright galaxies from the linear TFR is larger at longer wavelengths. Here we treat the effect of curvature as an additional bias that is addressed together with the effect of magnitude cutoff bias. 

Curvature of the near-IR TFR has been considered in previous studies using {\it H}-band luminosities \citep[and references therein]{1986ApJ...302..536A}. \citet{2000ApJ...529..698S} have used quadratic functions for the {\it BVRIH}$_{-0.5}$ bands and showed that the curvature term increases with wavelength, which is consistent with what we observe in Figure \ref{fig:TF_all}. \citet{2007MNRAS.381.1463N} described the effect with linear fits with a break between two slopes, noting systems with declining rotation curves at large radii dominate the flatter portion at the bright end and that formulating a ``baryonic" relation \citep{2005ApJ...632..859M} helps to minimize the change in slopes. \citet{2014ApJ...792..129N} accounted for curvature with the empirical strategy of employing quadratic functions for the TFR without adjustments of magnitudes or line widths. 

In this study, we accept the fitted ITFRs everywhere except for the bright end, where we model the deviations of data points from the linear relation using quadratic functions. Our attempt to fit an inverted quadratic relation failed due to the dramatic increase of scatter about the resulting TFR. The same issue was also reported by \citet{2014ApJ...792..129N}. Here, to fit the quadratic relation, we take the direct fitting approach and let the regression process minimize residuals along the magnitude axis. To keep the Malmquist bias at a minimum when fitting the quadratic curves, we still use the same template we constructed in the slope calibration process (see \S \ref{sec:tfr_slope}). The uncertainties of the magnitudes are much smaller than that of line widths; thus, we project the line widths error bars onto the magnitude axis based on the slope of the linear TFR. 
             
In our analysis, we examined different locations for the break point, where there is the transition from linear to quadratic forms. At optical bands, taking the break point as a free parameter of the fit results in unrealistic fitted curves, with harsh deviations from the linear relation beginning at log$(W^i_{mx})\sim2.6$ and a turnover after reaching to a maximum within the range of the data. Rather, at optical bands, we empirically explored break points to obtain a result that curves gently from the linear relations while leaving enough data points for the fitting process. We adopted log$(W^i_{mx})=2.5$ for the onset of deviation. At infrared bands, by contrast, the break point is reasonably found from minimization in the fitting process, giving  log$(W^i_{mx})=2.4$. The following quadratic relation formulates the deviation from the linear TFR after the break point
\begin{equation}
\label{Eq:TFRcurve}
    \Delta M_{\lambda} = A_2 X^2+A_1 X+A_0~,
\end{equation}
where $X=\rm{log}(W^i_{mx})-2.5$. 
Our fitting procedure requires that the quadratic curve and the linear TFR share the same slope at the break point, which reduces the number of independent parameters.  
Table \ref{tab:TFRcurve_params} lists $A_2$ and other essential parameters in the calculation of $A_1$ and $A_0$. 
Figure~\ref{fig:Curved_TFR} displays the resulting curved relations that apply brightward of the break points, with the previously determined linear ITFRs faintward of the break points. The rms scatter of the magnitude offsets from the curved relations is reported in the last column of Table~\ref{tab:TFRcurve_params}.  Scatter is not substantially reduced compared to the linear ITFRs because a large fraction of the luminosity-line width correlation is still modeled linearly, the deviation of the curved region from linear is modest, and there are not that many galaxies affected by the curvature at the bright end. 

\begin{table}[t]
\scriptsize
\tabletypesize{\tiny}
\setlength{\tabcolsep}{0.15cm}
\centering
\caption{Parameters of the curved ITFRs. 
$Slope$ and $ZP$ are parameters of the linear TFR.
} 
\label{tab:TFRcurve_params}
\begin{tabular}{cccccc}
\tablewidth{0pt}
\hline \hline
band & $Slope$ & $ZP$ & log$(W^i_{mx})$ & $A_2$ & rms \\
& & & break point & & [mag] \\
\hline 
\decimals
{\it u}  & -7.03$\pm$0.17 & -19.27$\pm$0.13 & 2.5 & 6.59$\pm$1.10 & 0.56 \\
{\it g}  & -7.37$\pm$0.13 & -20.15$\pm$0.11 & 2.5 & 4.18$\pm$0.90 & 0.48 \\
{\it r}  & -7.96$\pm$0.13 & -20.57$\pm$0.10 & 2.5 & 4.56$\pm$0.89 & 0.48 \\
{\it i}  & -8.32$\pm$0.13 & -20.80$\pm$0.10 & 2.5 & 5.34$\pm$0.91 & 0.48 \\
{\it z}  & -8.46$\pm$0.13 & -20.89$\pm$0.10 & 2.5 & 5.81$\pm$0.91 & 0.48 \\
{\it W1} & -9.47$\pm$0.14 & -20.36$\pm$0.07 & 2.4 & 3.81$\pm$0.42 & 0.56 \\
{\it W2} & -9.66$\pm$0.15 & -19.76$\pm$0.08 & 2.4 & 4.42$\pm$0.43 & 0.61 \\
\hline
\end{tabular}
\end{table}

\begin{figure*}[ht]
\centering
\includegraphics[width=0.45\linewidth]{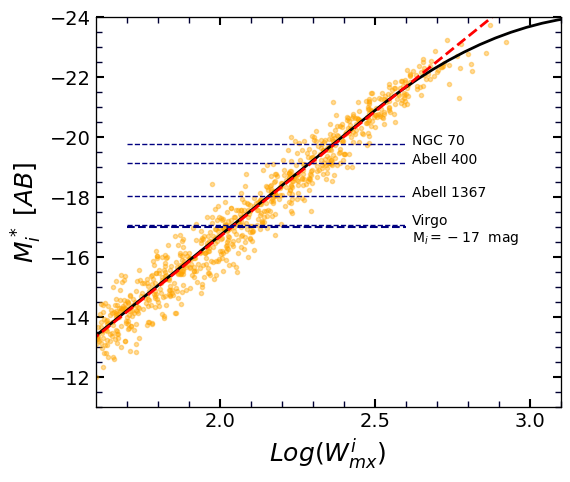}
\includegraphics[width=0.45\linewidth]{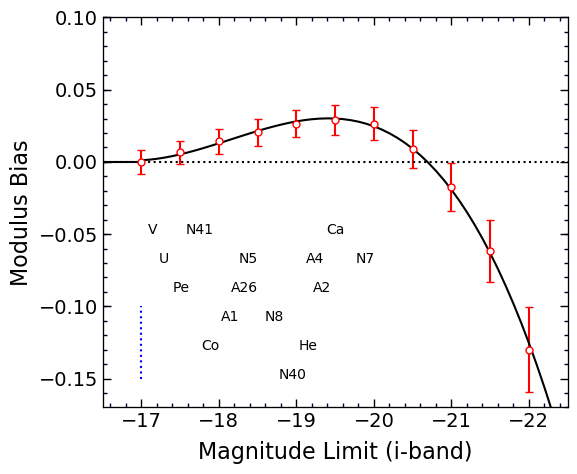}
\includegraphics[width=0.45\linewidth]{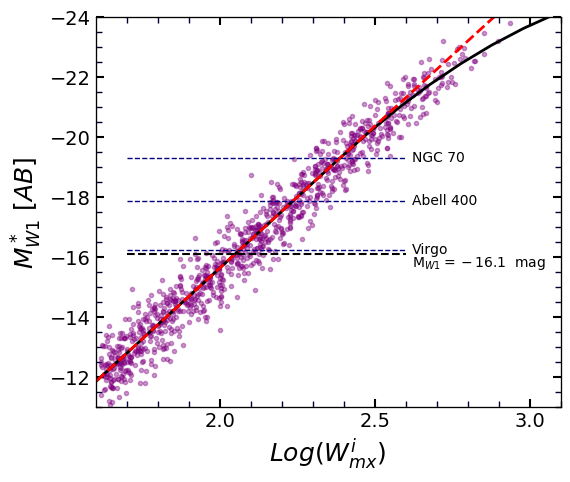}
\includegraphics[width=0.45\linewidth]{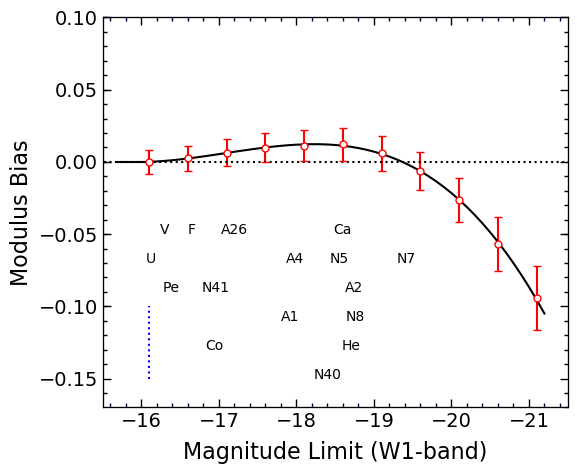}
\caption{{\bf Left:} 1000 simulated galaxies along the curved TFR drawing from the appropriate Schechter luminosity function and scattered along the magnitude axis based on the rms scatter model presented in Figure \ref{fig:RMScurves}. Horizontal dashed lines represent the magnitude limit of our sample at the {\it i} and {\it W1} bands. Dashed lines display the faint-end magnitude cutoffs of Virgo, Abell 1367, Abell 400 and NGC 70 clusters as examples. {\bf Right:} distance bias as a function of the faint-end magnitude limit of simulated TFRs. Cluster codes, placed horizontally, locate the faint-end magnitude coverage of calibrator clusters. Vertical dotted lines are drawn at the location of the magnitude cutoff of our samples at $M_i=-17$ and $M_{W1}=-16.1$ mag.}
\label{fig:Slpoe_calib}
\end{figure*}

\begin{table*}[t]
\scriptsize
\setlength{\tabcolsep}{0.15cm}
\centering
\caption{Parameters of the third-degree polynomial function, $b = 0.01\times \sum_{n=0}^{3} B_n M_{lim}^n$, that fits the distance modulus bias at different wave bands. The last two columns are the fitted parameters of luminosity function as described by the Schechter function. } 
\label{tab:pca_w1}
\begin{tabular}{cl|cccc|cc}
\tablewidth{0pt}
\hline \hline
 Band & TFR & \multicolumn{4}{c|}{Modulus Bias Function} &  \multicolumn{2}{c}{LF Parameters} \\
  & Code & $B_3$ & $B_2$ & $B_1$ & $B_0$ & $M^{\star}$  & $\alpha$  \\
 \hline 
\decimals
{\it u}	 & TF$_{c}$	& 1.08$\pm$0.19 & 3.73$\pm$0.95 & -0.87$\pm$1.23 & 0.16$\pm$0.38 &  -19.9$\pm$0.3	 & -1.0$\pm$0.3	 \\
{\it g}	 & TF$_{g}$	& 0.41$\pm$0.03 & 1.57$\pm$0.21 & -0.00$\pm$0.35 & 0.12$\pm$0.15 &  -21.3$\pm$0.5	 & -1.0$\pm$0.2	 \\
{\it r}	 & TF$_{r}$	& 0.48$\pm$0.04 & 1.82$\pm$0.24 & 0.37$\pm$0.43 & 0.18$\pm$0.19 & -21.8$\pm$0.2	 & -1.0$\pm$0.1	 \\
{\it i}	 & TF$_{i}$	 & 0.37$\pm$0.02 & 1.26$\pm$0.14 & -0.28$\pm$0.25 & 0.11$\pm$0.12 & -22.0$\pm$0.1	 & -1.0$\pm$0.1	 \\
{\it z}	 & TF$_{z}$	 & 0.34$\pm$0.02 & 1.14$\pm$0.15 & -0.30$\pm$0.26 & 0.09$\pm$0.11 &  -22.1$\pm$0.2	 & -1.0$\pm$0.1	 \\
{\it W1}	 & TF$_{W1}$ & 0.21$\pm$0.01 & 0.61$\pm$0.05 & -0.21$\pm$0.10 & 0.00$\pm$0.04 &  -21.9$\pm$0.1	 & -1.0$\pm$0.1	 \\
{\it W2}	 & TF$_{W2}$ & 0.25$\pm$0.02 & 0.98$\pm$0.15 & 0.40$\pm$0.28 & 0.12$\pm$0.14 & -21.3$\pm$0.2	 & -1.0$\pm$0.1	 \\
\hline
\end{tabular}
\end{table*}

\subsection{Simulations} \label{sec:simulation}

We follow a similar bias analysis as explained in \citet{2012ApJ...749...78T, 2013ApJ...765...94S}, and \citet{2014ApJ...792..129N} with one important difference.  Those earlier studies only considered the bias discussed in \S\ref{sec:bias} while now we also consider the bias due to curvature, the topic of \S\ref{sec:curvature}. These two biases act in opposite senses, and we study their effects jointly with simulations.

We approximate the luminosity function of TFR-applicable galaxies with the Schechter function \citep{1976ApJ...203..297S}.  To determine the faint-end slope, $\alpha$, we fit to the combination of Virgo-Ursa Major-Fornax cluster samples.  The bright-end cutoff parameter $M^{\star}$ is poorly constrained by only these three clusters, so, to improve statistics, we add clusters by distance groups, with $\alpha$ fixed by the nearest three clusters and magnitude ranges for the determination of $M^{\star}$ increasingly constrained with distance.
The fitted Schechter parameters at the different bands are given in Table \ref{tab:pca_w1}. 

The process of generating synthetic samples is given by the following recipe. (1) At each band, we generate a synthetic sample of galaxies with absolute magnitudes drawn from the Schechter luminosity function. (2) We assign a synthetic line width to each simulated galaxy based on the curved luminosity-line width correlations presented in \S \ref{sec:curvature}. (3) The magnitude of each galaxy is then statistically dispersed assuming a normal distribution with standard deviations taken from the scatter models similar to that presented in Figure~\ref{fig:RMScurves}, applied here to the curved luminosity-line width relations.
The left panels of Figure~\ref{fig:Slpoe_calib} illustrate the ensemble of 1000 simulated galaxies at the {\it i} and {\it W1} bands. 

To calculate the amplitude of bias, for any given faint-end magnitude limit, $M_{lim}$, we draw 10,000 galaxies randomly with simulated magnitudes brighter than the cutoff magnitude. 
In our analysis, the relative distance moduli of clusters are determined through fitting the linear luminosity-line width correlations; therefore, we follow the same recipe to measure the magnitudes of the random ensemble.
The line widths of randomly selected galaxies determine the assignments of absolute magnitudes from the model TFR. The average deviation of these measured magnitudes from input magnitudes is the bias, $b$.

The right panels of Figure~\ref{fig:Slpoe_calib} plot the bias as a function of magnitude limits. For each magnitude cutoff, we repeat the calculations for 100 different random ensembles and plot the averages and 1$\sigma$ standard deviations of the results. Data points are normalized to zero at the magnitude limits $M_i=-17$ and $M_{W1}=-16.1$ mag, where bias for a complete sample goes to zero.
In this plot, black solid curves are third-degree polynomial functions that fit the modulus bias, of the form $b = 0.01\times \sum_{n=0}^{3} B_n M_{lim}^n$, where $B_n$ is the coefficient of the polynomial function. 
Table~\ref{tab:pca_w1} lists the coefficients of the fitted polynomials. 

It is seen in Figure~\ref{fig:Slpoe_calib} that the combination of the two biases, those discussed in \S \ref{sec:bias} and \S \ref{sec:curvature} respectively, give results significantly different from those derived by \citet{2014ApJ...792..129N} and earlier because, while the bias due to scatter on the Schechter function is monotonically increasingly positive with distance, the curvature bias is monotonically increasingly negative with distance.  With increasing distance, the former bias is initially dominant, but then the latter takes over.  Over the range of relevance for our calibrator clusters, the bias remains positive but is subdued, reaching just $b\sim0.04$ at the {\it i} band and $b\sim0.02$ at the {\it W1} band.

\section{TFR with Additional Parameters} \label{sec:tfr_addTerms}  

There have been suggestions to reduce the dispersion in the TFR by including additional parameters such as type, color, or \hi-to-stellar content \citep{1985ApJ...289...81R, 1997AJ....113...53G, 2018MNRAS.479.3373M}.  In this spirit,
\citet{2006ApJ...653..861M} fit the TFR at the {\it I} band for early- and late-type spirals separately.
It has been known that the TFR is steeper at longer wavelengths \citep{1982ApJ...257..527T}.  This property must introduce a color term that affects scatter, producing systematic differences between passbands.

Larger spirals rotate faster, as manifested in larger \hi line widths. 
Larger galaxies are composed of older and redder stellar populations owing to their rapid star formation histories. Consequently, spirals with larger \hi line widths (bigger galaxies) are in general redder, as shown in Figure~\ref{fig:Color_LogW}, causing the fast-rotating end of the TFR to rise up relative to the slow-rotating end when progressing to longer wave bands. 
This effect is illustrated in Figure \ref{fig:TFallinONE}. The TFR progressively steepens from the optical {\it u} band to the infrared {\it W2} band. 

Consider two spirals that are different in color but are the same in line width. Toward infrared bands, the redder galaxy rises in luminosity relative to the bluer galaxy.  This effect induces an intrinsic scatter about the TFR that is a function of wave band and the color distribution of the sample galaxies.

The need for a color term in the infrared was clearly documented by \citet{2013ApJ...765...94S} at $3.6$ $\mu m$ using Spitzer IRAC images and by \citet{2014ApJ...792..129N} in their study of the TFR at the WISE {\it W1} and {\it W2} bands. In each of these cases, the scatter at infrared bands was reduced by $\sim0.1$~mag by the introduction of a color adjustment term.  Empirically, color terms can effectively reduce scatter in the TFR at blueward and infrared bands, becoming minimally applicable between the {\it i} and {\it z} bands.

\begin{figure}[t]
\centering
\includegraphics[width=0.9\linewidth]{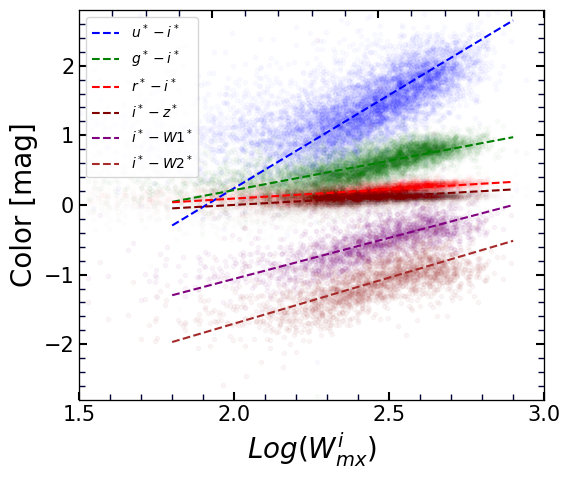}
\caption{Color terms vs. \hi line width. Each point displays a galaxy. Points are colored based on the color indices they represent. Dashed lines fit the data points of corresponding colors using the least-squares regression through minimizing the residuals along the vertical axis. The parameters of the fits are presented in Table \ref{table:omega_params}.
}
\label{fig:Color_LogW}
\end{figure}

\begin{figure}[t]
\centering
\includegraphics[width=0.9\linewidth]{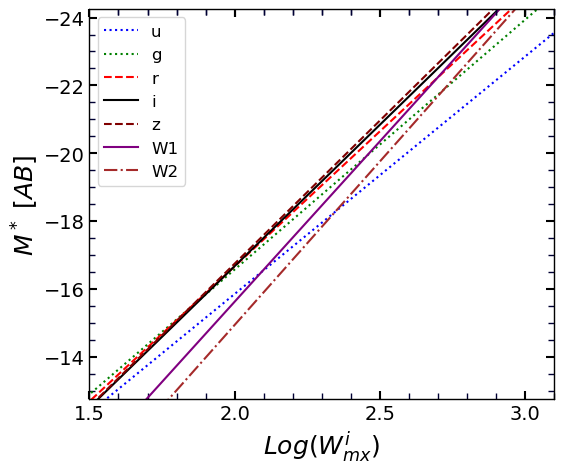}
\caption{The ITFR fits in different optical and infrared wave bands. The power-law lines are those displayed in Figure \ref{fig:TF_all}.}
\label{fig:TFallinONE}
\end{figure}

Here we evaluate correlations between TFR residuals and the various distance-independent observables that are available in our study. 
The features that we use to probe the type, morphology, and other general characteristics of spirals include (1) color indices calculated from optical and infrared magnitudes, (2) mean surface brightnesses of galaxies at different wave bands as probes of galaxy morphology, and (3) pseudocolors derived from the difference between the \hi magnitude calculated from Eq. \ref{Eq:m21} and optical/infrared magnitudes, $C_{21\lambda}=m_{21}-m_{\lambda}$, as an estimate of the ratio of the galaxy \hi content to its stellar mass.

These observables are not all independent of each other. 
In \S \ref{sec:observables-linewidth}, we explore the correlations between these observables and the \hi line width. 
The ability of quantifiable features to lower the scatter about the TFR is investigated in \S \ref{sec:scatter-observables}.

\subsection{Distance-independent Observables versus \hi Line Width} \label{sec:observables-linewidth} 

 In this section we perform our analysis using a sample of $\sim$10,000 spirals with high-quality photometry measurements.  Distance-independent characteristics are evaluated in reference to the \hi line width, the main distance-independent parameter of the TFR. We model the general relations between \hi line width and other observable features using power-laws that are formulated as
\begin{equation}
\label{Eq:omega}
    \Omega=\omega_0({\rm log}W_{mx}^i-2.5)+\omega_1~.
\end{equation}
where $\Omega$ can be any of the investigated features, such as color terms, surface brightness, or $C_{21\lambda}$ parameter. Here $\omega_0$ and $\omega_1$ are the slope and zero-point of the linear relations. We minimize residuals along the $\Omega$-axis in the least-squares regression to fit the linear relation.

In the absence of the galaxy SED, color indices are effective substitutes that provide useful information on the type, morphology, stellar populations, and general characteristics of galaxies.
Our spirals have measured magnitudes in seven different optical/infrared bands. Twenty-one color indices can be defined using these magnitudes. 
In this work, we only consider six independent color terms with reference to the {\it i} band. Our choice of {\it i} band is based on the relatively high image quality at that wave band; changing the reference passband does not change our conclusions.

The color terms with respect to the {\it i} band are seen in Figure \ref{fig:Color_LogW}. 
 Bigger spirals consist of redder/older stars and populate the larger line widths region of the diagram.  These relations steepen as the span of the wave bands that define the color term increases.

The correlations between line width and $C_{21\lambda}$ at different passbands are plotted in Fig \ref{fig:C21_LogW}. 
Spirals with more positive $C_{21\lambda}$ values (pseudoredder) have less \hi gas relative to their stellar content. The
$C_{21\lambda}$ increases with line width since the efficient star formation in bigger spirals consumes larger fractions of their \hi content to form stars. 
The linear relations steepen as the $C_{21\lambda}$ pseudocolor is based on longer wave bands. 
 For a given galaxy, the $21$cm magnitude is a fixed parameter in the calculation of $C_{21\lambda}$ so redder colors at larger line widths translate to steeper $C_{21\lambda}$ slopes toward the infrared. 

\begin{figure}[t]
\centering
\includegraphics[width=0.7\linewidth]{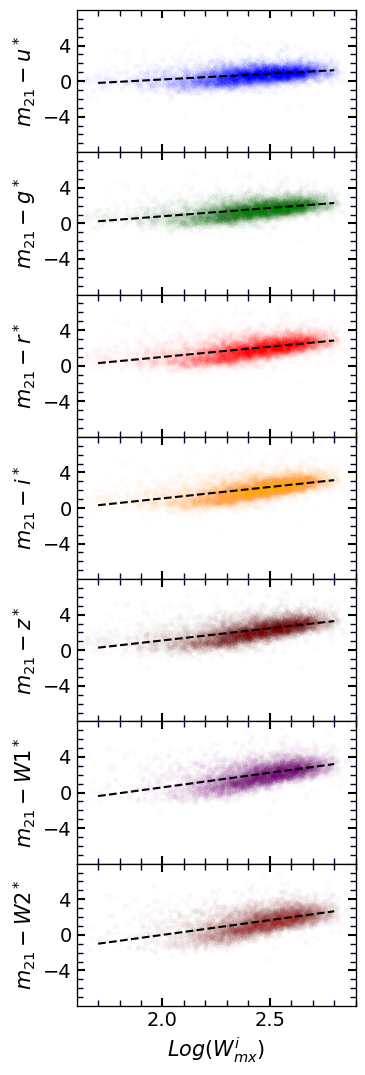}
\caption{Similar to Figure~\ref{fig:Color_LogW} but for $m_{21}-m_{\lambda}$ versus \hi line width, where $\lambda$ represents different optical/infrared bands. The unit of the vertical axis is magnitude. The parameters of the black dashed lines are presented in Table \ref{table:omega_params}.}
\label{fig:C21_LogW}
\end{figure}

\begin{figure}[t]
\centering
\includegraphics[width=0.7\linewidth]{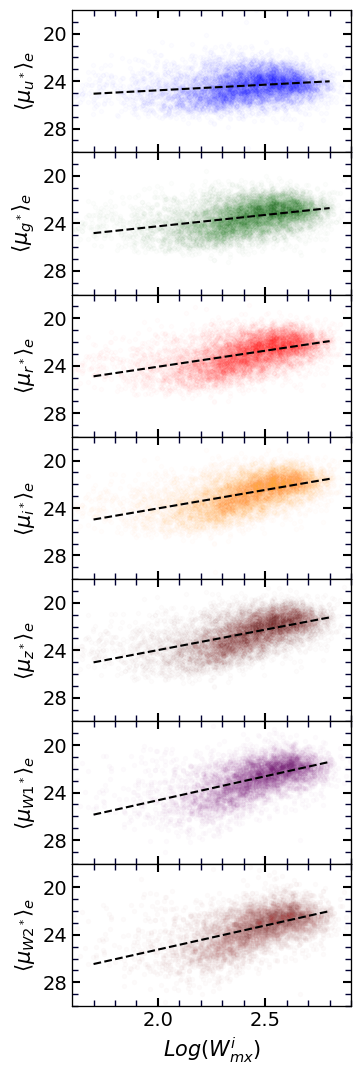}
\caption{Similar to Figure \ref{fig:Color_LogW} but for mean effective surface brightness, $\langle\mu_{\lambda^*}\rangle^{(i)}_e$, versus \hi line width at different wave bands. The unit of the vertical axis is mag arcsec$^{-2}$. The parameters of the black dashed fitted lines are presented in Table~\ref{table:omega_params}.
}
\label{fig:Mu50_LogW}
\end{figure}

Figure~\ref{fig:Mu50_LogW} plots the mean surface brightness within the half-light radius versus \hi line width. Surface brightness is corrected for the effect of inclination using Eq.~\ref{Eq:Mu50_i}. More massive spirals with larger \hi line widths tend toward higher surface brightness.

Table \ref{table:omega_params} contains the parameters of the linear fits that describe the correlations between various distance-dependant observables and \hi line widths.

\begin{table}
\setlength{\tabcolsep}{0.2cm}
\begin{center}
\caption{Parameters of the linear correlations between various distance independent observables and \hi line widths as expressed by Eq.~\ref{Eq:omega}. The parameter rms is the root mean square of the dispersion in data about the fitted lines along the axis of the corresponding parameter. $C$ in the last column is the correlation coefficient and quantifies the strength of the relations.} 
\label{table:omega_params}
\begin{tabular}{rlcccccc}
\hline \hline
\# & $\Omega^{\dag}$ & $\omega_0$ & $\omega_1$ & $rms$ & $C (\%)$ \\
(1) & (2) & (3) & (4) & (5) & (6) \\
\hline 
1 & $u^*-i^*$ &  2.68 & 1.58 & 0.87 & 33 \\
2 & $g^*-i^*$ &  0.84 & 0.64 & 0.18 & 66 \\
3 & $r^*-i^*$ &  0.27 & 0.23 & 0.07 & 61 \\
4 & $i^*-z^*$ &  0.25 & 0.12 & 0.07 & 47 \\
5 & $i^*-W1^*$ & 1.18 & -0.47 & 0.23 & 66 \\
6 & $i^*-W2^*$ & 1.32 & -1.04 & 0.28 & 60 \\
\hline 
7 & $m_{21}-u^*$ & 1.30 & 0.84 & 0.74 & 36 \\
8 & $m_{21}-g^*$ & 1.86 & 1.73 & 0.79 & 46 \\
9 & $m_{21}-r^*$ & 2.31 & 2.14 & 0.86 & 51 \\
10 & $m_{21}-i^*$ & 2.56 & 2.36 & 0.89 & 54 \\
11 & $m_{21}-z^*$ & 2.72 & 2.47 & 0.91 & 56 \\
12 & $m_{21}-W1^*$ & 3.27 & 2.22 & 1.00 & 53 \\
13 & $m_{21}-W2^*$ & 3.32 & 1.65 & 1.02 & 53 \\
\hline 
14 & $\langle\mu_{u^*}\rangle_e$ & -0.95 & 24.30 & 1.37 & 15 \\
15 & $\langle\mu_{g^*}\rangle_e$ & -1.92 & 23.28 & 1.08 & 37 \\
16 & $\langle\mu_{r^*}\rangle_e$ & -2.70 & 22.74 & 1.13 & 47 \\
17 & $\langle\mu_{i^*}\rangle_e$ & -3.12 & 22.48 & 1.14 & 52 \\
18 & $\langle\mu_{z^*}\rangle_e$ & -3.45 & 22.26 & 1.16 & 55 \\
19 & $\langle\mu_{W1^*}\rangle_e$ & -4.05 & 22.63 & 1.22 & 54 \\
20 & $\langle\mu_{W2^*}\rangle_e$ & -4.06 & 23.21 & 1.41 & 49 \\
\hline
\end{tabular}
\end{center}
\footnotesize{
\begin{itemize}
    \item[$\dag$] $\Omega_{1-13}$ are in mag.
    \item[$\dag$] $\Omega_{14-20}$ are in mag arcsec$^{-2}$.
\end{itemize}
}
\end{table}

\subsection{Linking TFR Dispersion to Distance-independent Observables} \label{sec:scatter-observables} 

It is shown that the scatter in the TFR can be reduced through introducing additional parameters. \citet{2013ApJ...765...94S} and \citet{2014ApJ...792..129N} found significant correlations between deviations from the infrared TFRs and their optical-infrared color terms. In this section, we explore a range of ways to reduce scatter about the TFR at different wave bands.

The deviation of a spiral galaxy in the slope calibrator sample from the ITFR is given by Eq.~\ref{Eq:mu_lambda}. 
These residuals from the ITFR are plotted against observables, and the trends of linear fits are evaluated with correlation coefficients. The strongest correlations are used to adjust magnitudes, resulting in modified ITFRs with smaller scatter. In this process, we consider two groups of observables. The first group consists of all 20 parameters  listed in the second column of Table~\ref{table:omega_params}. The second group considers the residuals of the distance-independent parameters from their linear relations with respect to \hi line widths
\begin{equation}
\label{Eq:Domega}
    \Delta \Omega^i = \Omega_{obs}^i - \Omega~,
\end{equation}
where $\Omega_{obs}^i$ represents any of the individual distance-independent observed characteristics and $\Omega$ is calculated from the line width following Eq.~\ref{Eq:omega} and the parameters in Table~\ref{table:omega_params}.

\begin{table}
\setlength{\tabcolsep}{0.1cm}
\scriptsize
\begin{center}
\caption{The parameters of the linear relation  between deviations of magnitudes from the unadjusted TFR as expressed in Eq. \ref{Eq:phi_model}. $C$ in the last column denotes the correlation strength between deviations and the corresponding observable feature, $\Theta$.
} 
\label{table:deviations_observables}
\begin{tabular}{c|l| ccr }
\hline \hline
Deviation & Correction & $\Phi_{Slope}$ & $\Phi_{ZP}$ & $C $ \\
from TFR & Parameter ($\Theta$) & & & $(\%)$ \\
(1) & (2) & (3) & (4) & (5) \\
\hline 
$u^*-$Mean   & $u^*-i^*$  & 0.76$\pm$0.05 & -1.13$\pm$0.07 & 61 \\
  Correlation$^{\dag}$   & $u^*-W1^*$  & 0.50$\pm$0.03 & -0.48$\pm$0.04 & 58 \\
                         & $\Delta({\langle\mu_{u^*}\rangle_e})$ & -0.13$\pm$0.02 & -0.21$\pm$0.04 & 21 \\
                         & $\Delta({m_{21}-u^*})$ & 0.17$\pm$0.03 & -0.29$\pm$0.05 & 27 \\
\hline
$g^*-$Mean    & $g^*-i^*$  & 0.73$\pm$0.09 & -0.45$\pm$0.06 & 35 \\
  Correlation            & $g^*-W1^*$  & 0.35$\pm$0.05 & -0.04$\pm$0.02 & 32 \\
                         & $i^*-W1^*$  & 0.49$\pm$0.08 & 0.26$\pm$0.05 & 25 \\
\hline
$r^*-$Mean     & $g^*-i^*$  & 0.44$\pm$0.09 & -0.27$\pm$0.06 & 21 \\
Correlation    & $i^*-W1^*$  & 0.34$\pm$0.08 & 0.18$\pm$0.05 & 16 \\
               & $\Delta({i^*-W1^*})$  & -0.14$\pm$0.11 & 0.00$\pm$0.02 & 22 \\
               & $\Delta({\langle\mu_{r^*}\rangle_e})$  & 0.06$\pm$0.02 & -0.01$\pm$0.02 & 23 \\
               & $\Delta({m_{21}-r^*})$ & -0.05$\pm$0.02 & 0.01$\pm$0.02 & 14 \\
\hline
$i^*-$Mean     & $g^*-i^*$  & 0.29$\pm$0.10 & -0.18$\pm$0.06 & 14 \\
Correlation    & $i^*-W1^*$  & 0.28$\pm$0.08 & 0.14$\pm$0.05 & 12 \\
               & $\Delta({g^*-i^*})$  & -0.31$\pm$0.14 & 0.00$\pm$0.02 & 19 \\ 
               & $\Delta({i^*-W1^*})$  & -0.24$\pm$0.11 & 0.00$\pm$0.02 & 26 \\
               & $\Delta({\langle\mu_{i^*}\rangle_e})$  & 0.06$\pm$0.02 & -0.02$\pm$0.02 & 24 \\
               & $\Delta({m_{21}-i^*})$ & -0.06$\pm$0.02 & 0.01$\pm$0.02 & 18 \\
\hline
$z^*-$Mean     & $g^*-i^*$  & 0.21$\pm$0.10 & -0.14$\pm$0.06 & 11 \\
Correlation    & $i^*-W1^*$  & 0.23$\pm$0.08 & 0.11$\pm$0.05 & 10 \\
               & $\Delta({g^*-i^*})$  & -0.39$\pm$0.13 & 0.00$\pm$0.02 & 21 \\ 
               & $\Delta({i^*-W1^*})$  & -0.31$\pm$0.11 & -0.01$\pm$0.02 & 29 \\
               & $\Delta({\langle\mu_{z^*}\rangle_e})$  & 0.07$\pm$0.02 & -0.02$\pm$0.02 & 26 \\
               & $\Delta({m_{21}-z^*})$ & -0.08$\pm$0.02 & 0.01$\pm$0.02 & 21 \\

\hline
$W1^*-$Mean   & $i^*-W1^*$  & -0.28$\pm$0.09 & -0.15$\pm$0.05 & 13  \\
  Correlation           & $\Delta({i^*-W1^*})$  & -1.08$\pm$0.11 & -0.04$\pm$0.02 & 53 \\
                         & $\Delta({\langle\mu_{W1^*}\rangle_e})$  & 0.16$\pm$0.02 & -0.05$\pm$0.02 & 38 \\
                         & $\Delta({m_{21}-W1^*})$  & -0.10$\pm$0.02 & 0.00$\pm$0.02 & 28 \\
\hline
$W2^*-$Mean   & $i^*-W2^*$  & -0.39$\pm$0.08 & -0.44$\pm$0.09 & 22 \\
 Correlation             & $\Delta({i^*-W2^*})$  & -1.09$\pm$0.09 & -0.07$\pm$0.02 & 59 \\
                         & $\Delta({\langle\mu_{W2^*}\rangle_e})$  & 0.19$\pm$0.02 & -0.07$\pm$0.02 & 40 \\
                         & $\Delta({m_{21}-W2^*})$  & -0.12$\pm$0.02 & -0.01$\pm$0.02 & 29 \\
\hline
\end{tabular}
\end{center}
\footnotesize{
${\dag}$ Mean correlation is given by the unadjusted ITFR at each band, TF$_{\lambda}$, where $\lambda=$ {\it u, g, r, i, z, W1} and {\it W2}, as formulated by Eq. \ref{Eq:TFR} with optimized parameters listed in Table \ref{tab:allITFRs}.
}
\end{table}

We express the general correlations between deviation from the ITFR and galaxy observables using the linear relation 
\begin{equation}
\label{Eq:phi_model}
    \Delta M_\lambda^{(\Theta)} = \Phi_{Slope}~ \Theta + \Phi_{ZP}~,
\end{equation}
where $ \Delta M_\lambda^{(\Theta)}$ is the output of the model for deviations from the ITFR at the wave band $\lambda$, and $\Theta$ is any of the observable features, $\Omega_{obs}$ or $\Delta \Omega$. Here $\Phi_{Slope}$ and $\Phi_{ZP}$ are the slope and zero-point of the linear relation. 
Table \ref{table:deviations_observables} lists the parameters of the correlations between various observables, $\Omega_{obs}$ or $\Delta \Omega$, and deviations from the mean unadjusted ITFRs (see Figure \ref{fig:TF_all}). To save space, this list only includes observables with the strongest correlations and a set of important features; these include mean effective surface brightness at each band, $\langle\mu_{\lambda^*}\rangle_e$, their deviations from the linear relations with line width as described by Eq.~\ref{Eq:omega} $\Delta({\langle\mu_{\lambda^*}\rangle_e})$, and the $m_{21}-m^*_{\lambda}$ pseudocolors.

An observable $\Theta$ that is highly correlated to the ITFR residuals, $\Delta M$, offers a modified ITFR with smaller dispersion. The modified relation is constructed based on the adjusted apparent magnitudes that are given as
\begin{equation}
\label{Eq:pseudomag}
    C_{\lambda}(\Theta) = m^*_{\lambda} - \Delta M_\lambda^{(\Theta)} ~,
\end{equation}
where $\Delta M_\lambda^{(\Theta)}$ is calculated based on the observable parameter, $\Theta$, following Eq.~\ref{Eq:phi_model}.  
The same algorithm as explained in \S \ref{sec:tfr_calib} is pursued to determine the slope and zero-point of the revised ITFR that has the form $\mathcal{M}^*_{C_\lambda(\Theta)}=Slope \big({\rm log}W^i_{mx}-2.5 \big)+ZP$. In Table \ref{tab:revised_ITFR}, we present the parameters of our ITFR fits for unadjusted and adjusted magnitudes. In this table,
the unadjusted relations are denoted by $TF_\lambda$ codes. We give specific attention to interesting cases where either the rms scatter of the relation is significantly reduced after adjusting magnitudes or the correction parameter is highly correlated with the dispersion of the original relation according to Table \ref{table:deviations_observables}. These cases are attributed unique codes for further discussion. 

Following the procedure discussed in \S \ref{sec:curvature}, we fit a second-degree polynomial to the bright end of the luminosity-line width correlations for each of the adjusted and unadjusted cases. The last two columns of Table \ref{tab:revised_ITFR} list the parameters of the fitted curves as formulated in Eq. \ref{Eq:TFRcurve}.

The distance modulus of a spiral galaxy with the absolute pseudomagnitude, $\mathcal{M}^*_{C_\lambda(\Theta)}$, that is given by the modified ITFR is derived as
\begin{equation}
\label{Eq:}
    \mu_{C_\lambda(\Theta)} = C_{\lambda}(\Theta) - \mathcal{M}^*_{C_\lambda(\Theta)}~.
\end{equation}
where $C_{\lambda}(\Theta)$ is the galaxy apparent pseudo-magnitude that is determined by Eq. \ref{Eq:pseudomag}.

\begin{figure}[t]
\centering
\includegraphics[width=0.9\linewidth]{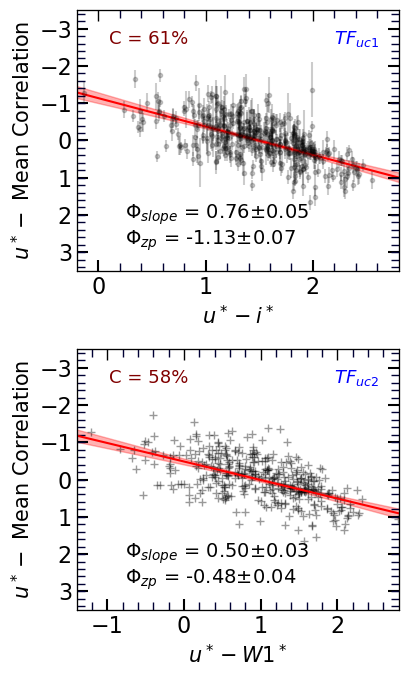}
\caption{Deviations from the mean unadjusted ITFRs at the {\it u} band vs. color. Each black point represents a galaxy.  Dominant errors are along the vertical axis. The best linear relation is derived by least-squares regression incorporating the vertical uncertainties. Vertical error bars are displayed in the top panel as an example.  The 95\% confidence envelope is displayed about each fitted line. In the top left corner of each panel, $C$ is the correlation coefficient of the plotted parameters as listed in Table \ref{table:deviations_observables}. In the top right corner is the code for the corresponding corrected ITFR.}
\label{fig:u_color_correlation}
\end{figure}

\begin{figure}[t]
\centering
\includegraphics[width=0.83\linewidth]{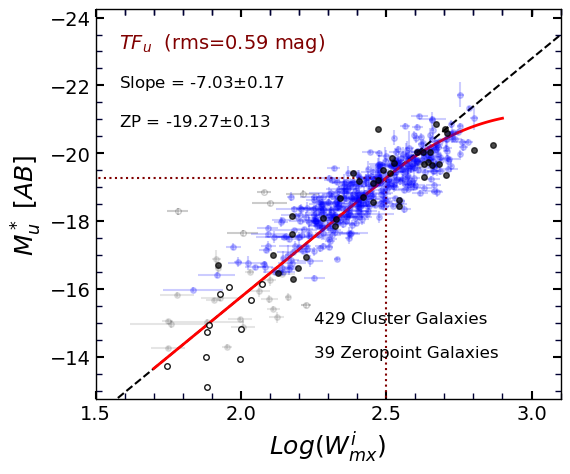}
\includegraphics[width=0.83\linewidth]{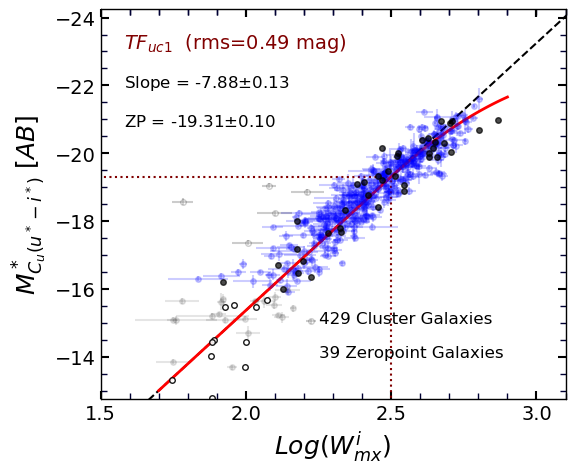}
\includegraphics[width=0.83\linewidth]{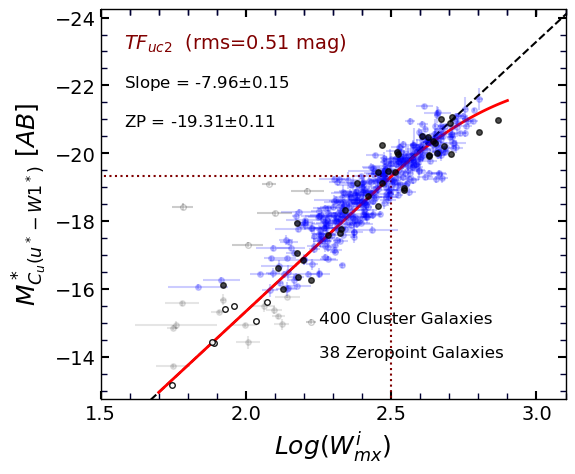}
\caption{Unadjusted and revised ITFR relations. {\bf Top:} unadjusted ITFR at the {\it u} band ($TF_u$). This panel is the same as the top left panel of Figure \ref{fig:TF_all}. The rms of vertical residuals about the relation is $0.59$ mag. {\bf Middle:} modified ITFR after adjusting magnitudes using the $u^*-i^*$ color ($TF_{uc1}$) with an rms scatter of $0.49$ mag. {\bf Bottom:} The modified ITFR after adjusting magnitudes based on the $u^*-W1^*$ color ($TF_{uc2}$) with an rms scatter of $0.51$ mag. 
}
\label{fig:MCu}
\end{figure}

\subsection{Modified ITFRs at {\it u} Band} \label{sec:modified_uband}

Among the correction parameters presented in the {\it u}-band section of Table \ref{table:deviations_observables}, the $u^*-i^*$ and $u^*-W1^*$ colors are highly correlated with deviations from the mean {\it u}-band ITFR with the correlation factors of 61\% and 58\%, respectively. These correlations are illustrated in Figure \ref{fig:u_color_correlation} together with the best-fit linear relations described by Eq. \ref{Eq:phi_model}. The parameters of the fitted red lines are derived following the least-squares optimization along the vertical axis. Our fitting procedure ignores the horizontal errors, because they are significantly smaller than the vertical uncertainties and relatively the same for all spirals. These two linear relations are 
\begin{subequations}
	\begin{align}
	\begin{split}
		 \Delta M_u^{(u^*-i^*)} = &- (1.13 \pm 0.07) \\
			&+ (0.76 \pm 0.05)(u^*-i^*),
	\label{eq:u_redfit:a}
	\end{split} \\
	\begin{split}
		 \Delta M_u^{(u^*-W1^*)} = & - (0.48 \pm 0.04) \\
			&+ (0.50 \pm 0.03)(u^*-W1^*),
	\label{eq:u_redfit:b}
	\end{split}
	\end{align}
	\label{eq:u_redfit}
\end{subequations}
where the parameter $\Theta$ in Eq.~\ref{Eq:phi_model} is replaced by $(u^*-i^*)$ and $(u^*-W1^*)$ color terms. 
The revised ITFR relations are then derived based on the {\it u}-band pseudomagnitudes, $C_u(\Theta)$, through
\begin{subequations}
	\begin{align}
	     C_u(u^*-i^*) &=& u^* &- \Delta M_u^{(u^*-i^*)}~, \\
		 C_u(u^*-W1^*) &=& u^* &- \Delta M_u^{(u^*-W1^*)}~.
	\end{align}
	\label{eq:u_pseudo}
\end{subequations}
The corresponding modified ITFRs have the form
\begin{subequations}
	\begin{align}
	\begin{split}
		\mathcal{M}^*_{C_u(u^*-i^*)} = &- (19.31 \pm 0.10) \\
			&- (7.88 \pm 0.13)({\rm log}W^i_{mx} - 2.5), 
	\label{eq:TF_uc1}
	\end{split} \\
	\begin{split}
		\mathcal{M}^*_{C_u(u^*-W1^*)} = &- (19.31 \pm 0.11) \\
			&- (7.96 \pm 0.15)({\rm log}W^i_{mx} - 2.5).
	\label{eq:TF_uc2}
	\end{split}
	\end{align}
	\label{eq:TF_u}
\end{subequations}

We assign the $TF_{uc1}$ and $TF_{uc2}$ codes to the revised ITFRs presented in Eq.~\ref{eq:TF_uc1} and \ref{eq:TF_uc2}, respectively. These relations are illustrated in the middle and bottom panels of Figure~\ref{fig:MCu}.
For comparison, the top panel of Figure~\ref{fig:MCu} shows the unadjusted ITFR constructed using {\it u} magnitudes ($TF_u$). In this figure, the partially curved relations are illustrated in solid red, with the linear part below the break point at ${\rm log}W^i_{mx}=2.5$ and the polynomial fits to data points brightward of the break point. Dashed black lines are the continuations of the linear ITFRs.
 To evaluate the performance of the models, we compare the scatters about the linear magnitude-line width relations.
The rms of the residuals from the unadjusted $TF_u$ of $0.59$ mag is significantly reduced to $0.49$ and $0.51$ mag for $TF_{uc1}$ and $TF_{uc2}$, respectively. 
The $TF_{uc2}$ model requires the availability of WISE photometry that is only present for 400 out of 429 galaxies with {\it u}-band photometry. However, we do not expect that the lack of 29 galaxies from the $TF_{uc2}$ sample has a significant influence on our final conclusion. 

The {\it u} band is sensitive to populations of blue stars and star forming regions, whereas the majority of the baryonic mass in spiral galaxies resides in older stellar populations that dominate at longer wave bands.  Color terms reduce scatter by providing information about older stellar populations that are subdominant in the {\it u}-band imaging.

\begin{figure}[t]
\centering
\includegraphics[width=0.9\linewidth]{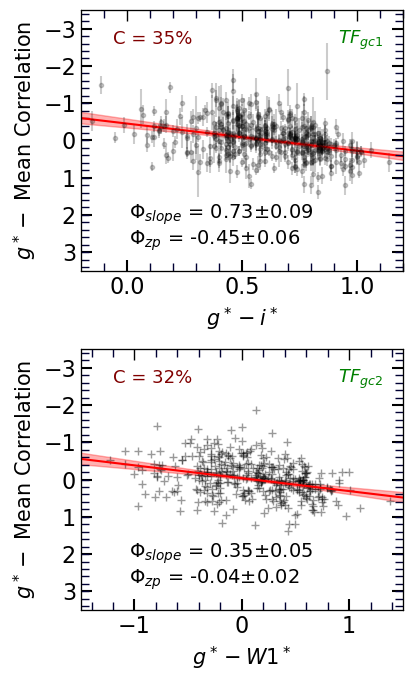}
\caption{Same as Figure~\ref{fig:u_color_correlation} but for deviations from the mean unadjusted ITFRs at {\it g}-band versus color.}
\label{fig:g_color_correlation}
\end{figure}

\begin{figure}[t]
\centering
\includegraphics[width=0.83\linewidth]{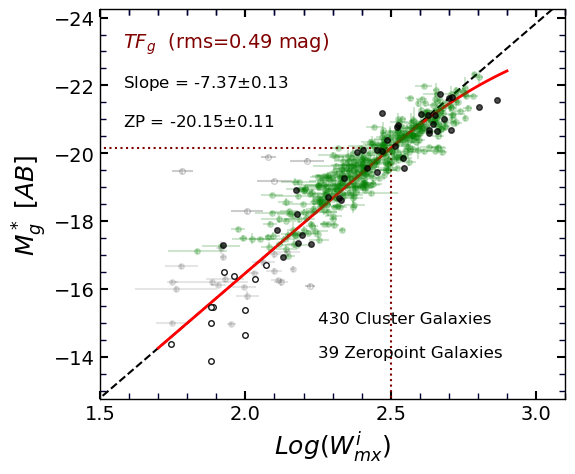}
\includegraphics[width=0.83\linewidth]{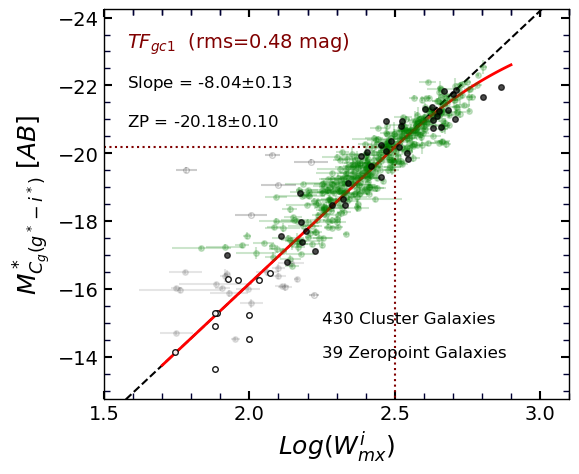}
\includegraphics[width=0.83\linewidth]{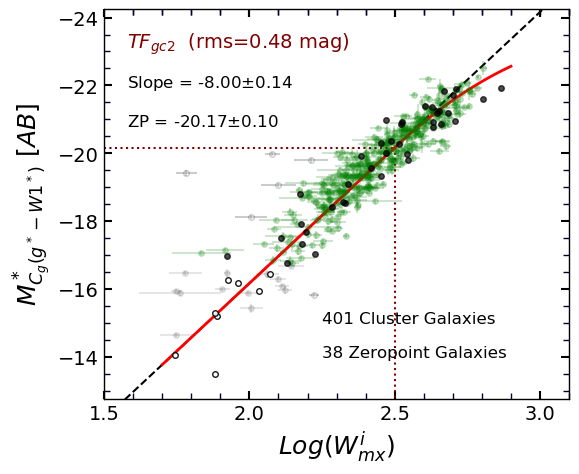}
\caption{Same as Figure~\ref{fig:MCu} for {\it g}-band.
}
\label{fig:MCg}
\end{figure}

\subsection{Modified ITFRs at {\it g} Band} \label{sec:modified_gband}

Resembling the {\it u} band, it is seen in Table~\ref{table:deviations_observables} that deviations from the mean unadjusted ITFRs at the {\it g} band are most correlated with $g^*-i^*$ and $g^*-W1^*$ color indices, with correlation factors of 35\% and 32\%, respectively. 
These correlations are shown in Figure~\ref{fig:u_color_correlation}. The parameters of the two linear fits required to adjust the {\it g}-band magnitudes are presented in Table~\ref{table:deviations_observables}. The corresponding adjusted ITFRs based on the {\it g}-band pseudo-magnitudes are given as
\begin{subequations}
	\begin{align}
	\begin{split}
		\mathcal{M}^*_{C_g(g^*-i^*)} = &- (20.18 \pm 0.10) \\
			&- (8.04 \pm 0.13)({\rm log}W^i_{mx} - 2.5), 
	\label{eq:TF_gc1}
	\end{split} \\
	\begin{split}
		\mathcal{M}^*_{C_g(g^*-W1^*)} = &- (20.17 \pm 0.10) \\
			&- (8.00 \pm 0.14)({\rm log}W^i_{mx} - 2.5).
	\label{eq:TF_gc2}
	\end{split}
	\end{align}
	\label{eq:TF_g}
\end{subequations}
The modified ITFR models that are represented by Eq.~\ref{eq:TF_gc1} and \ref{eq:TF_gc2} are labelled with $TF_{gc1}$ and $TF_{gc2}$ codes (also see Table~\ref{tab:revised_ITFR}).

\begin{figure}[t]
\centering
\includegraphics[width=0.9\linewidth]{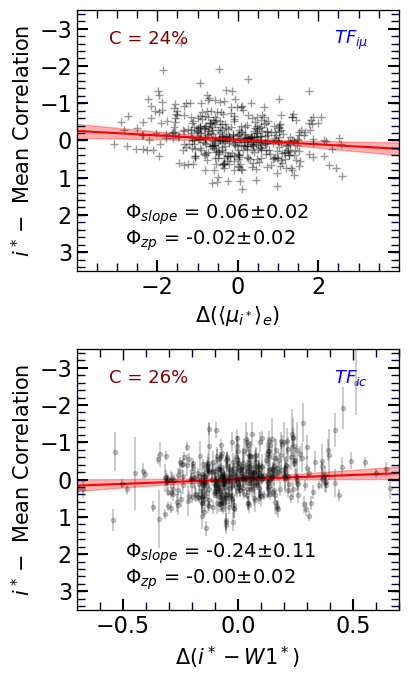}
\caption{Deviations from the mean ITFR at the {\it i} band as functions of surface brightness and color. }
\label{fig:i_color_correlation}
\end{figure}

\begin{figure}[t]
\centering
\includegraphics[width=0.83\linewidth]{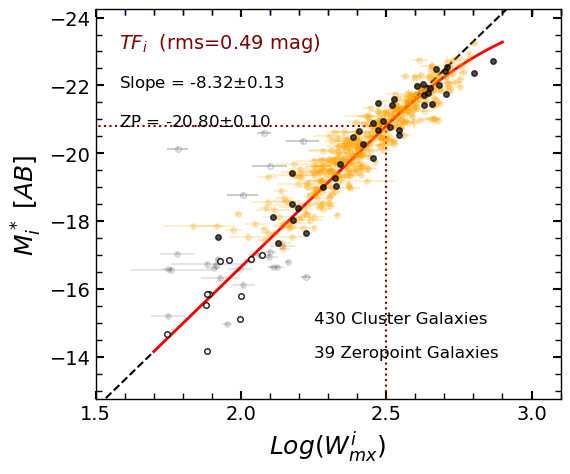}
\includegraphics[width=0.83\linewidth]{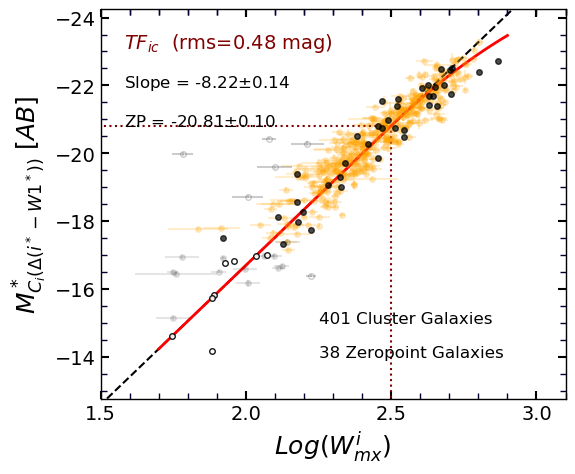}
\includegraphics[width=0.83\linewidth]{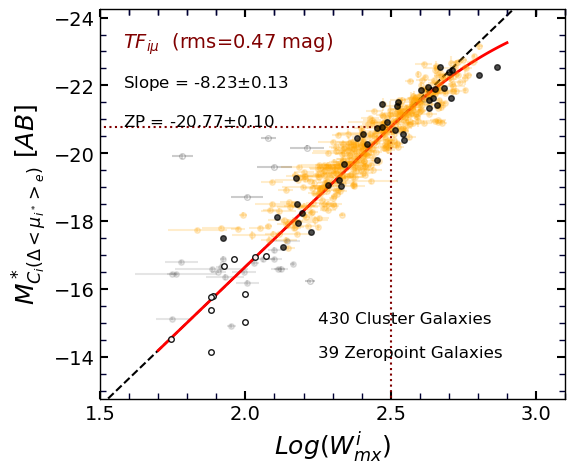}
\caption{Same as Figure~\ref{fig:MCu} for the {\it i} band.
}
\label{fig:MCi}
\end{figure}

\subsection{Modified ITFRs at {\it r, i}, and {\it z} Bands} \label{sec:modified_rizband}

The rms scatter in unadjusted ITFRs at the {\it r, i} and {\it z} bands is similar. 
Exploring tables~\ref{table:deviations_observables} and \ref{tab:revised_ITFR} jointly reveals the observables with the most significant correlations to ITFR residuals, hence most capable of reducing the scatter in adjusted ITFR relations, to be $\Delta(i^*-W1^*)$ and $\Delta(\langle\mu_{\lambda^*}\rangle_e)$.  The correlations with these observables in the case of the {\it i} band are shown in Figure~\ref{fig:i_color_correlation}. The adjusted ITFRs based on these models are labeled with the codes $TF_{\lambda c}$ and  $TF_{\lambda \mu}$, respectively, where $\lambda$ is replaced with {\it r}, {\it i}, and {\it z}. 

All of the correlations at the {\it r}, {\it i} and {\it z} bands are weak. Comparing the lower panels of Figures.~\ref{fig:u_color_correlation}, \ref{fig:g_color_correlation}, and \ref{fig:i_color_correlation}, the slopes of the correlations with colors $\lambda^{\star}-W1^{\star}$ transition from positive to negative, flattening to zero slope between {\it z} and {\it W1} bands. Adjustments coupled to color parameters do little to improve the TFR scatter (see Table \ref{tab:revised_ITFR} and Figure \ref{fig:MCi} in the case of {\it i}-band).  Likewise, correlation coefficients related to surface brightness or \hi pseudocolors are small. We favor {\it not} making third-parameter adjustments to the TFR at the {\it r}, {\it i} and {\it z} bands, since adding parameters adds uncertainties.

\subsection{Modified ITFRs at infrared {\it W1} and {\it W2} bands} \label{sec:modified_W12}

The effectiveness of the optical-infrared colors to reduce the scatters of the unadjusted ITFRs at infrared bands have already been noticed in previous studies. For example, \citet{2013ApJ...765...94S} observed a significant correlation between the {\it I}-IRAC [3.6] color and deviations from the mean ITFR constructed at the infrared IRAC [3.6] band. Inspired by their work, \citet{2014ApJ...792..129N} applied the same methodology to reduce the scatter about the WISE {\it W1} and {\it W2} ITFRs. 

\begin{figure}[t]
\centering
\includegraphics[width=0.9\linewidth]{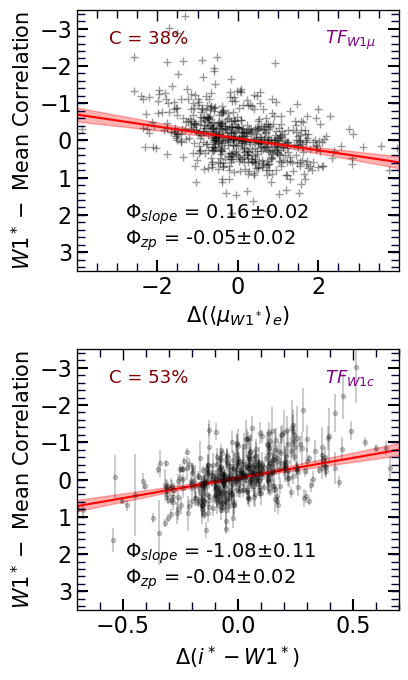}
\caption{Same as Figure \ref{fig:i_color_correlation} but for deviations from the mean ITFR at the infrared {\it W1-} band. }
\label{fig:W1_color_correlation}
\end{figure}

\begin{figure}[t]
\centering
\includegraphics[width=0.83\linewidth]{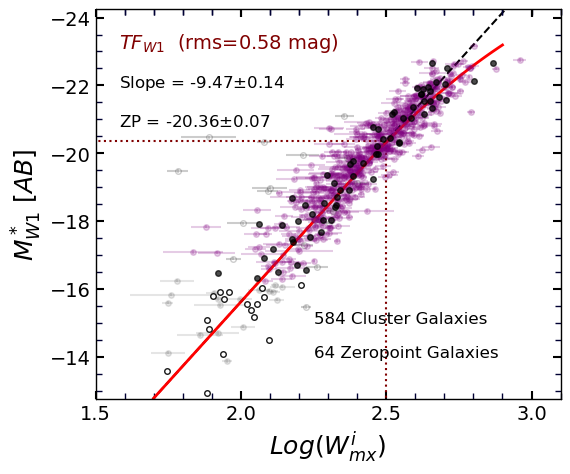}
\includegraphics[width=0.83\linewidth]{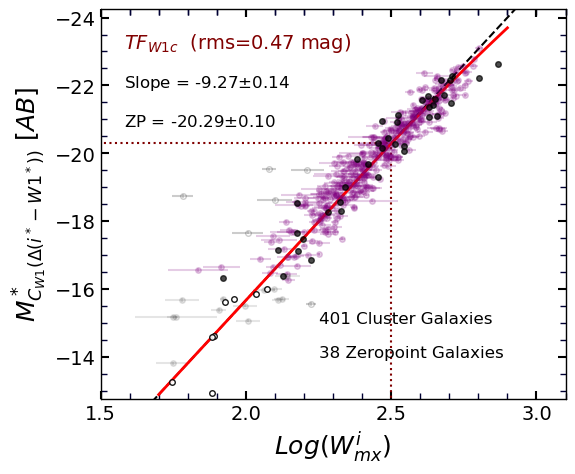}
\includegraphics[width=0.83\linewidth]{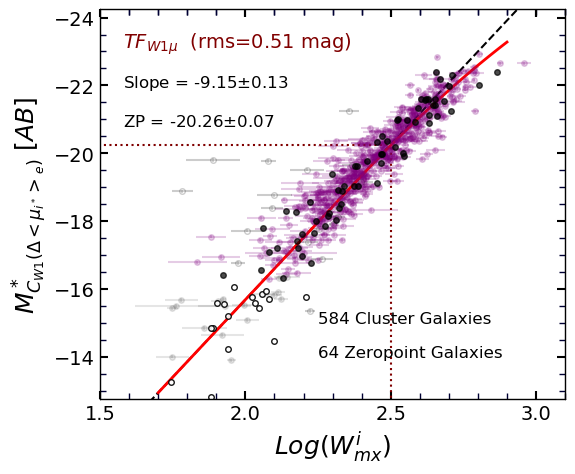}
\caption{Same as Figure~\ref{fig:MCu} for the {\it W1} band. Red solid lines display the partially curved ITFRs that deviate from the linear relation for galaxies with \hi line widths larger than log$(W^i_{mx})=2.4$.
}
\label{fig:MCw1}
\end{figure}

Giving consideration to a similar color correlation with the present data, Table~\ref{table:deviations_observables} lists the linear dependencies of unadjusted {\it W1} and {\it W2} ITFRs residuals on the $i^*-W1^*$ and $i^*-W2^*$ parameters. We find similar trends as previous studies; however, the correlations are not very significant and cannot usefully reduce the scatter of the ITFR.\footnote{Similarly, \citet{2017MNRAS.469.2387P} found that the color term in a bivariate solution for the luminosity-line width relation at the IRAC [3.6] band did not rise to a level of significance.} Instead, we consider 
$\Delta(i^*-W1^*)$ and $\Delta(i^*-W2^*)$ color terms that are in higher correlation with the ITFR residuals, with the correlation factors of 53\% and 59\%, respectively. In addition, the surface brightness seems to be capable of reducing the scatter efficiently. 

Figure \ref{fig:W1_color_correlation} displays the deviations from the {\it W1}-band ITFR versus $\Delta(i^*-W1^*)$ and $\Delta(\langle\mu_{W1^*}\rangle_e)$. In both cases, the slopes and zero-points of the fitted linear relations are significant. Analogous relations have been derived for the {\it W2} band. The middle and bottom panels of Figure~\ref{fig:MCw1} plot the revised luminosity-line width relations after applying corrections based on the $\Delta(i^*-W1^*)$ and $\Delta(\langle\mu_{W1^*}\rangle_e)$ parameters, labeled $TF_{W1c}$ and $TF_{W1 \mu}$, respectively. In addition, the adjustments using additional parameters reduce the curvature of the best fit after the break point at log$(W^i_{mx})=2.4$. This reduction is in agreement with the results of \citet{2014ApJ...792..129N}, where no curvature for the relation of the color-adjusted magnitude with line width is apparent. The larger correlation factor of $\Delta(i^*-W1^*)$ (53\%) compared to that of $\Delta(\langle\mu_{W1^*}\rangle_e)$ (38\%) translates to greater reduction in the scatter about the adjusted TF model. The $0.58$ mag rms scatter of $TF_{W1}$ is lowered to $0.47$ mag and $0.51$ mag for the $TF_{W1c}$ and $TF_{W1 \mu}$ cases, respectively.

Ever since \citet{1979ApJ...229....1A} it has been anticipated that minimal scatter in the TFR would be achieved at infrared bands.  Indeed, \citet{2017MNRAS.469.2387P} explored correlations in 12 bands from the far ultraviolet to $4.5\mu$m and found the tightest fit at $3.6\mu$m. With extensive modeling, adjustments can be made to account for radiation from dust associated with star forming regions that can contribute $10-30\%$ of light at $3.4-3.6\mu$m \citep{2015ApJS..219....5Q}. In any event, the minimization in scatter found by \citet{2017MNRAS.469.2387P} and similarly by \citet{2016ApJ...816L..14L} refers to the scatter orthogonal to mean bivariate fits.  The scatter relevant for the acquisition of distances is that in magnitudes and is aggravated by the steepening of slopes toward the infrared.

\begin{table*}[t]
\scriptsize
\setlength{\tabcolsep}{0.2cm}
\centering
\caption{TFR Parameters Before and After Corrections} 
\label{tab:revised_ITFR}
\begin{tabular}{cc|ll| ccc |ccc| cc}
\tablewidth{0pt}
\hline \hline
 \multirow{2}{*}{Band} & \multirow{2}{*}{Sample} & Correction & TFR & \multicolumn{3}{c|}{Universal Slope} &  \multicolumn{3}{c|}{Zero-point} &  \multicolumn{2}{c}{Curvature} \\
 \cline{5-7}  \cline{8-10}  \cline{11-12}
    & &  Parameter & Code & Ngal & Slope &  rms &  Ngal & Mag  & rms & Break Point & $A_2$ \\
\hline 
\decimals
{\it u}   &   OP  &   \dots  &  TF$_{u}$  & 429 &   -7.03$\pm$0.17 &  0.59 & 39 &   -19.27$\pm$0.13   & 0.75   & log$(W^i_{mx})$=2.5  &  6.59$\pm$1.10 \\
{\it u}   &   OP  &  $u^*-i^*$  &  TF$_{uc1}$   & 429 &   -7.88$\pm$0.13 &  0.49 & 39 &   -19.31$\pm$0.10   & 0.60   & 2.5  & 5.03$\pm$0.90  \\
{\it u}   &   OP+IR  &  $u^*-W1^*$  &  TF$_{uc2}$  & 400 &   -7.96$\pm$0.15  &  0.51 & 38 & -19.31$\pm$0.11  & 0.61   & 2.5  & 5.90$\pm$0.91 \\
{\it u}   &   OP  &  $\Delta(\langle\mu_{u^*}\rangle_e)$  &    & 429 &   -7.35$\pm$0.16  & 0.59 & 39 &  -19.33$\pm$0.12  & 0.60   & 2.5  &  6.55$\pm$1.10 \\
{\it u}   &   OP  &  $\Delta(m_{21}-u^*)$  &    & 435 &   -7.58$\pm$0.17 & 0.59 & 39 &   -19.43$\pm$0.13   & 0.75   & 2.5  &  6.58$\pm$1.12 \\
\hline 
{\it g}   &   OP  &   \dots  &  TF$_{g}$  & 430 &   -7.37$\pm$0.13  &  0.49 & 39 &   -20.15$\pm$0.11   & 0.62   & 2.5  & 4.18$\pm$0.90  \\
{\it g}   &   OP  &  $g^*-i^*$  &  TF$_{gc1}$  & 430 &   -8.04$\pm$0.13  &  0.48 & 39 &   -20.18$\pm$0.10   & 0.59   & 2.5  & 4.91$\pm$0.89  \\
{\it g}   &   OP+IR  &  $g^*-W1^*$  & TF$_{gc2}$    & 401 &   -8.00$\pm$0.14  &  0.48 & 38 &  -20.17$\pm$0.10   & 0.59   & 2.5  & 5.01$\pm$0.89  \\
{\it g}   &   OP+IR  &  $i^*-W1^*$  &    & 401 &   -7.85$\pm$0.14  &  0.48 & 38 &  -20.17$\pm$0.11   & 0.61   &  2.5 & 5.12$\pm$0.88  \\
\hline
{\it r}  &  OP  &  \dots & TF$_{r}$ &  430  &  -7.96$\pm$0.13  &  0.49  & 39  &  -20.57$\pm$0.10  &  0.59   &  2.5 & 4.56$\pm$0.89 \\
{\it r}  &   OP  &  $g^*-i^*$  &   & 430 &   -8.41$\pm$0.13  &  0.49 & 39 &   -20.59$\pm$0.10   & 0.58   & 2.5  & 5.23$\pm$0.91 \\
{\it r}   &   OP+IR  &  $i^*-W1^*$  &    & 401 &   -8.28$\pm$0.14  &  0.49 & 38 &  -20.58$\pm$0.10  & 0.58   & 2.5 &  5.13$\pm$0.90 \\
{\it r}   &   OP+IR  &  $\Delta(i^*-W1^*)$  &  TF$_{rc}$  & 401 &   -7.89$\pm$0.14  &  0.48 & 38 &  -20.57$\pm$0.10 & 0.58   & 2.5  & 3.59$\pm$0.92 \\
{\it r}   &   OP  &    $\Delta(\langle\mu_{r^*}\rangle_e)$  &   TF$_{r\mu}$  & 430 &  -7.89$\pm$0.13  &  0.47 & 39 &   -20.55$\pm$0.10  & 0.59   & 2.5  & 4.34$\pm$0.90 \\
{\it r}   &   OP  &  $\Delta(m_{21}-r^*)$  &    & 430 &   -7.81$\pm$0.13 &  0.47 & 39 &   -20.53$\pm$0.10   & 0.59   & 2.5  & 4.35$\pm$0.87  \\
\hline
{\it i}   &   OP  &   \dots  &  TF$_{i}$   & 430 &   -8.32$\pm$0.13  &  0.49 & 39 & -20.80$\pm$0.10 & 0.59   & 2.5  & 5.34$\pm$0.91 \\
{\it i}  &   OP  &  $g^*-i^*$  &   & 430 &   -8.64$\pm$0.14  &  0.49 & 39 &   -20.81$\pm$0.10   & 0.59   & 2.5  & 5.87$\pm$0.94 \\
{\it i}   &   OP+IR  &  $i^*-W1^*$  &    & 401 &   -8.58$\pm$0.14  &  0.49 & 38 &  -20.81$\pm$0.10  & 0.58   & 2.5  & 5.71$\pm$0.94 \\
{\it i}   &   OP  &  $\Delta(g^*-i^*)$  &   & 430 &  -8.22$\pm$0.13 &0.49 & 39 &  -20.80$\pm$0.10   & 0.59   & 2.5  & 4.68$\pm$0.89 \\
{\it i}   &   OP+IR  &  $\Delta(i^*-W1^*)$  &  TF$_{ic}$  & 401 &   -8.22$\pm$0.14  &  0.48 & 39 &  -20.81$\pm$0.10 & 0.57   & 2.5  & 3.91$\pm$0.94  \\
{\it i}   &   OP  &    $\Delta(\langle\mu_{i^*}\rangle_e)$  &   TF$_{i\mu}$  & 430 &  -8.23$\pm$0.13  &  0.47 & 39 &   -20.77$\pm$0.10  & 0.59   & 2.5  &  5.03$\pm$0.92 \\
{\it i}   &   OP  &  $\Delta(m_{21}-i^*)$  &    & 435 &   -8.13$\pm$0.13 &  0.48 & 39 &   -20.75$\pm$0.10   & 0.59   & 2.5  &  5.01$\pm$0.89 \\
\hline
{\it z}  &  OP  &  \dots & TF$_{z}$ &  429  &  -8.46$\pm$0.13  &   0.50  & 38  &  -20.89$\pm$0.10  &  0.57   & 2.5  &  5.81$\pm$0.91 \\
{\it z}  &   OP  &  $g^*-i^*$  &   & 429 &   -8.69$\pm$0.13  &  0.49 & 38 &   -20.88$\pm$0.10   & 0.59   & 2.5  & 6.21$\pm$0.93 \\
{\it z}   &   OP+IR  &  $i^*-W1^*$  &    & 400 &   -8.60$\pm$0.14  &  0.50 & 37 &  -20.88$\pm$0.10  & 0.57   & 2.5  &  5.82$\pm$0.94 \\
{\it z}   &   OP  &  $\Delta(g^*-i^*)$  &   & 429 &  -8.33$\pm$0.13 &  0.48 & 38 &  -20.88$\pm$0.10   & 0.56   & 2.5  & 4.96$\pm$0.89  \\
{\it z}   &   OP+IR  &  $\Delta(i^*-W1^*)$  &  TF$_{zc}$  & 400 &   -8.29$\pm$0.14  &  0.47 & 37 &  -20.88$\pm$0.10 & 0.55   & 2.5  & 3.77$\pm$0.94  \\
{\it z}   &   OP  &    $\Delta(\langle\mu_{z^*}\rangle_e)$  &   TF$_{z\mu}$  & 429 &  -8.35$\pm$0.13  &  0.47 & 38 &  -20.86$\pm$0.10  & 0.57   &  2.5 & 3.88$\pm$0.95 \\
{\it z}   &   OP  &  $\Delta(m_{21}-z^*)$  &    & 429 &   -8.22$\pm$0.13 &  0.47 & 38 &   -20.81$\pm$0.10   & 0.57   & 2.5  & 5.34$\pm$0.88 \\
\hline 
{\it W1}   &   IR  &   \dots  &  TF$_{W1}$  & 584 &   -9.47$\pm$0.14  &  0.58 & 64 &   -20.36$\pm$0.07   & 0.60   & 2.4  & 3.81$\pm$0.42 \\
{\it W1}   &   OP+IR  &  $i^*-W1^*$  &    & 401 &   -9.21$\pm$0.15  &  0.54 & 38 &   -20.30$\pm$0.10   & 0.60   &  2.4 &  3.41$\pm$0.44 \\
{\it W1}   &   OP+IR  &  $\Delta(i^*-W1^*)$  &  TF$_{W1c}$   & 401 &   -9.27$\pm$0.14  &  0.47 & 38 &   -20.29$\pm$0.10   & 0.56   & 2.4  & 1.22$\pm$0.40 \\
{\it W1}   &   IR  &  $\Delta(\langle\mu_{W1^*}\rangle_e)$  &  TF$_{W1\mu }$   & 584 &  -9.15$\pm$0.13  &  0.51 & 64 &   -20.26$\pm$0.07  & 0.57   &  2.4 & 2.55$\pm$0.40 \\
{\it W1}   &   IR  &  $\Delta(m_{21}-W1^*)$  &   & 584 &  -9.12$\pm$0.13 &  0.53 & 64 &  -20.29$\pm$0.08  & 0.58   & 2.4  & 3.35$\pm$0.40 \\
\hline 
{\it W2}   &   IR  &   \dots  &  TF$_{W2}$   & 584 &   -9.66$\pm$0.15  &  0.62 & 64 &   -19.76$\pm$0.08   & 0.65   & 2.4  & 4.42$\pm$0.43  \\
{\it W2}   &   OP+IR  &   $i^*-W2^*$  &    & 401 &   -9.14$\pm$0.15  &  0.55 & 38 &   -19.70$\pm$0.11   & 0.63   & 2.4 & 3.47$\pm$0.43  \\
{\it W2}   &   OP+IR  &   $\Delta(i^*-W2^*)$  &  TF$_{W2c}$  & 401 &   -9.40$\pm$0.14  &  0.47 & 38 &   -19.70$\pm$0.10   & 0.56   & 2.4  & 1.13$\pm$0.41  \\
{\it W2}   &   IR  &   $\Delta(\langle\mu_{W2^*}\rangle_e)$  &  TF$_{W2\mu}$   & 584 &   -9.18$\pm$0.13  &  0.52 & 64 &   -19.62$\pm$0.07   & 0.60   & 2.4  &  2.67$\pm$0.39 \\
{\it W2}   &   IR  &   $\Delta(m_{21}-W1^*)$  &   & 584 &   -9.22$\pm$0.14  &  0.56 & 64 &   -19.67$\pm$0.08   & 0.62   & 2.4  &  3.79$\pm$0.41 \\
\hline \hline 
\end{tabular}
\end{table*}

\subsection{The Baryonic Tully-Fisher Relation}
\label{sec:BTFR}

\citet{2005ApJ...632..859M} pointed out that the fractional representation of interstellar gas mass increases progressively toward galaxies of lower stellar mass.  He suggested that luminosity be replaced by a linear combination of luminosity and the \hi flux as a measure of the galactic baryon content, $M_b$.
\begin{equation}
    M_b = 1.33M_{HI} + \Upsilon_{\star} L_{\lambda}
\end{equation}
where $\Upsilon_{\star}$ is the stellar mass-to-light ratio ($\sim0.5 M_{\odot}/L_{\odot}$ at $\lambda = 3.6\mu$m) and the multiplier 1.33 on $M_{HI}$ accounts for the unseen contribution of helium \citep{2019MNRAS.484.3267L}.

The baryonic formulation tends to flatten the TFR by raising the low line width end relative to the high line width end and may somewhat mitigate curvature \citep{2007MNRAS.381.1463N}.  Nevertheless, we have not given the baryonic TFR close attention.  The neutral gas only becomes a substantial component of the baryonic mass of a galaxy in the vicinity of our faint cutoff at $M_i=-17$.  In particular, though, we are not confident of the reliability of the \hi flux parameters in our \hi catalogs at the Extragalactic Distance Database.  \citet{2016MNRAS.463.4052P} has noted the basis of our concerns.  In the compilation of the ADHI catalog, great care was given to the accumulation of line widths on a coherent system, but \hi fluxes have not received such attention.  Contributions come from a variety of telescopes with beams of different sizes that may or may not capture most of the \hi flux.  Inhomogeneities in flux measurements might depend on the telescope, and hence portions of the sky, and on target sizes, and hence distance.

\section{Cluster Radial Velocities} \label{sec:ClusterRadial}

The mean radial velocity of each cluster is determined by averaging the velocities of all known constituents (not just those with distance measurements) using the biweight statistics discussed in \citet{1990AJ....100...32B}.  For clusters beyond 3,500~\kms\ memberships are based on the 2MASS galaxy group catalog of \citet{2015AJ....149..171T} while within 3,500 km s$^{-1}$, we draw on the group catalog of \citet{2017ApJ...843...16K}. The average radial velocities of clusters are then shifted to the cosmic microwave background (CMB) rest frame $V_{cmb}$, and further adjusted for cosmological curvature $V_{mod}=fV_{cmb}$ where 
\begin{equation}
\label{hi}
f = 1 + \frac{1}{2} [1 - q_0]z - \frac{1}{6} [1 - q_0 -3 q_0^2 + j_0]z^2~,
\end{equation}
the jerk parameter $j_0 = 1$, and the acceleration parameter $q_0 = {\frac{1}{2}} ( \Omega_m -2 \Omega_{\Lambda} ) = -0.595$ assuming $\Omega_m=0.27$, $\Omega_m+\Omega_{\Lambda}=1$.
These cosmological adjustments are negligible for the nearby clusters, reaching in the extreme 3\% for the Hercules cluster. The second column of Table \ref{Tab:cluster_H0} lists these adjusted radial velocities, $V_{mod}$, and their uncertainties. 

\begin{figure}[t]
\centering
\includegraphics[width=0.90\linewidth]{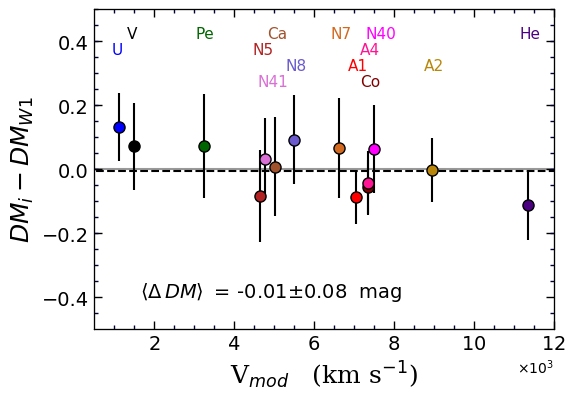}
\caption{ 
Differences between the measured cluster distance moduli at the {\it i} and {\it W1} bands vs. their radial velocities, $V_{mod}$. The cluster codes are horizontally placed based on the corresponding $V_{mod}$ values. The horizontal axis is the cosmologically corrected radial velocity of clusters in the CMB rest frame. The dashed horizontal line displays the average of the $DM$ offset, $\langle \Delta DM \rangle$, calculated by incorporating the vertical error bars. 
}
\label{fig:DM_i_w1}
\end{figure}

\begin{figure}[t]
\centering
\includegraphics[width=0.90\linewidth]{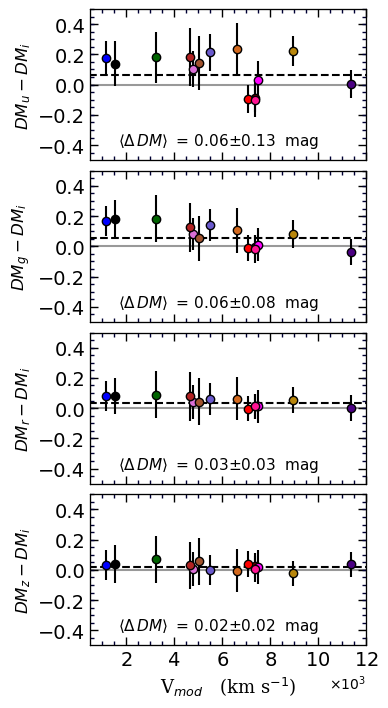}
\caption{Offset of cluster distance moduli, $DM$, measured at the optical {\it u}, {\it g}, {\it r} and {\it z} bands from those measured at the {\it i} band using the unadjusted TFR relations. The color scheme and other details are the same as in Figure~\ref{fig:DM_i_w1}.
}
\label{fig:OP_DM_i}
\end{figure}

\begin{figure*}[ht]
\centering
\includegraphics[width=0.9\linewidth]{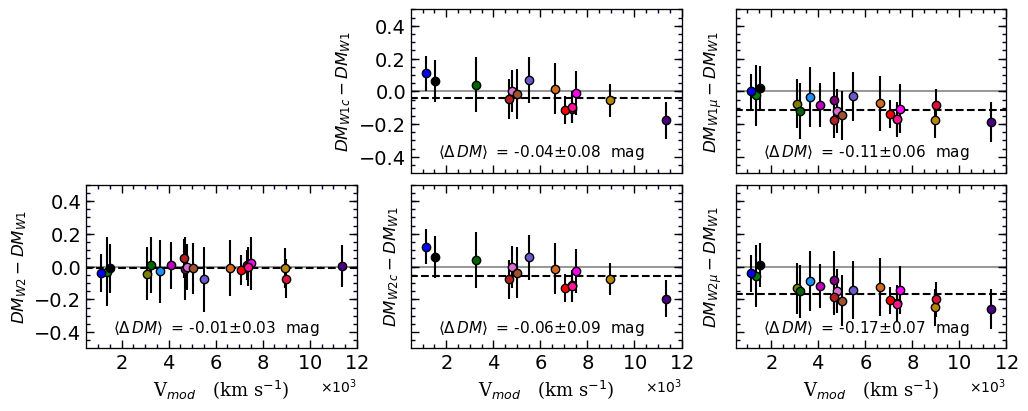}
\caption{Unaltered distance moduli at the {\it W1} band are taken as the reference of comparison in all panels.  The bottom panels show comparisons with {\it W2}, with the panel at left involving unaltered moduli.  The middle panels give color-adjusted moduli differences with respect to unadjusted {\it W1} moduli, with the {\it W1} case on top and the {\it W2} case on the bottom.  The panels at the right are similar but now with surface brightness corrections. The The color scheme and other details are the same as in Figure \ref{fig:DM_i_w1}.
}
\label{fig:IR_DM_W1}
\end{figure*}

\section{Distances of clusters} \label{sec:ClusterDM.png}

A product of the TFR calibration is the distance moduli of the galaxy clusters used to form the TFR template. 
The raw distance moduli of clusters, $DM_o$, are listed in column (6) of Table~\ref{Tab:cluster_H0} for different TFR models listed in column (4). Column (7) contains $b$, the luminosity function scatter bias combined with the effect of TFR curvature that is calculated following the discussion presented in \S \ref{sec:bias}. Column (8) contains any other adjustment bias that will be discussed later in this section, $\Delta DM$. Column (9) tabulates the corrected distance moduli, given as $DM_c=DM_o+b+\Delta DM$, that are converted to distances in units of Mpc and presented in column (10).

Initially, we set $\Delta DM=0$, and we plot the differences of the bias-corrected moduli as a function of the cluster radial velocities to test the consistency of the measurements at different passbands.
At optical bands, 14 galaxy clusters contributed in our calibration process, while in the infrared, 20 clusters are available. 
First, we check the consistency of our distance measurements between the optical and infrared bands using the {\it i} and {\it W1} bands as proxies, and then we evaluate the agreements within the optical and infrared bands separately. 

Figure \ref{fig:DM_i_w1} plots the differences of the distance moduli measured at the {\it i} and {\it W1} band for the 14 clusters in common. No significant offset is apparent. We notice positive offsets for the nearest three clusters, Virgo, Ursa Major, and Fornax, and a negative offset for the most distant Hercules cluster. After further exploration, we find that the Virgo galaxies are slightly redder and the Hercules galaxies are bluer compared to the average color distribution of the template galaxies used in the calibration process. Redder galaxies appear to be more luminous at longer wavelengths, and therefore they are located at smaller distances. The opposite effect is true for galaxies that are in general bluer than average. This subtle effect was seen earlier \citep{2000ApJ...533..744T, 2014ApJ...792..129N} with the very nearby clusters, in the first case with entirely distinct optical and infrared photometry and in the second case with distinct optical photometry.

We look for offsets of the cluster moduli, measured using the unadjusted TFRs (TF$_\lambda$ models), at optical passbands relative to that at the {\it i} band in Figure~\ref{fig:OP_DM_i}. It is seen that systematic differences in moduli are highest for the {\it u} band and decrease toward longer wavelengths. These offsets are not statistically significant; nonetheless, we offer adjustments at the {\it u} and {\it g} bands so moduli in these bands are on the same scale as the {\it i} band. In our final analysis, though, we do {\it not} make use of {\it u} and {\it g} moduli. By contrast, we see in Figure~\ref{fig:OP_DM_i} that the moduli at {\it r}, {\it i} and {\it z} are in close agreement without any modifications. Going forward, we only use the {\it r}, {\it i} and {\it z} relations without relative bandpass adjustments to measure distances with optical photometry.

For evaluation of the infrared distances, we choose the {\it W1} band as reference. 
In order to measure distances at this infrared band, we apply the adjusted ITFRs that have smaller scatters. 
As discussed in \S \ref{sec:modified_rizband}, both $\Delta(i^*-W1^*)$ and $\Delta(\langle\mu_{W1^*}\rangle_e)$ parameters are capable of reducing the scatter. However, it is more practical for our purposes to use the corrections based on the {\it W1} surface brightness because we sometimes lack optical photometry.

\begin{figure}[t]
\centering
\includegraphics[width=0.85\linewidth]{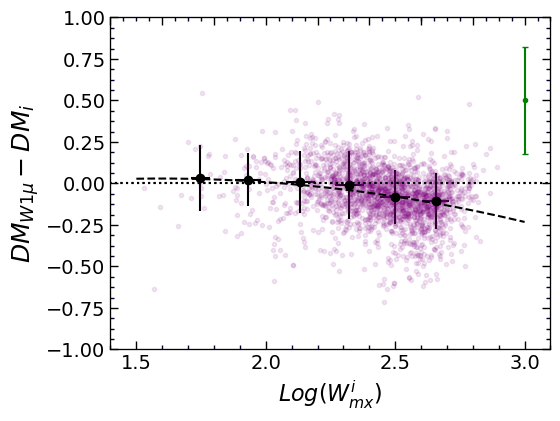}
\includegraphics[width=0.85\linewidth]{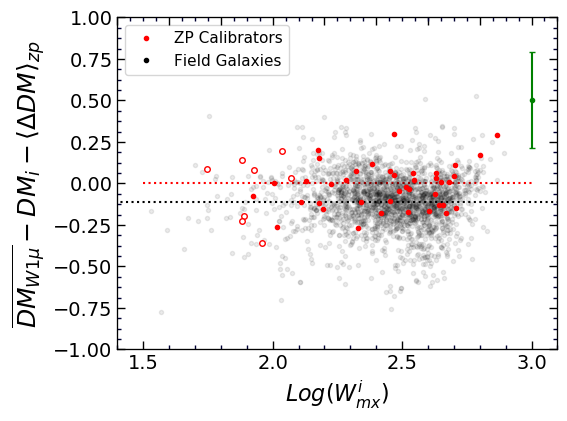}
\caption{ {\bf Top:} differences in distance moduli of field galaxies, $DM$, between the $TF_{W1 \mu}$ and $TF_{i}$ cases. Each purple point represents a galaxy, with black points showing averages in horizontal bins of $0.2$. The black dashed line is the best quadratic fitted function on all data points, minimizing vertical residuals, with expression $DM_{W1 \mu} - DM_i = (-0.13\pm0.06) X^2+(0.42\pm0.28) X + (-0.31\pm0.33)$, where $X={\rm log}W^i_{mx}$. {\bf Bottom:} offset of distance moduli after correcting $DM_{W1 \mu}$ for the line width-dependent bias presented in top panel, $\overline{DM_{W1 \mu}} - DM_i$. The zero-point of the plot is set at $\langle \Delta DM \rangle_{zp}=0.11$ mag, the average of the moduli differences of the zero-point calibrators, displayed as red filled points. Gray points represent field galaxies, while red open points are rejected faint calibrators. The black dotted line at $-0.11$ is at the average moduli offset of field galaxies. Green error bars at the top right give typical differences for field galaxies.
}
\label{fig:dDM_logW}
\end{figure}

\begin{figure}[t]
\centering
\includegraphics[width=0.85\linewidth]{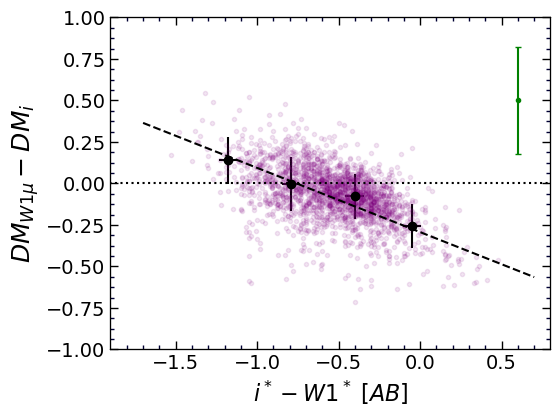}
\includegraphics[width=0.85\linewidth]{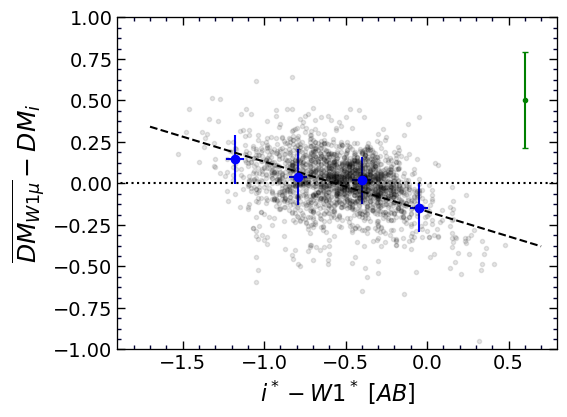}
\caption{ {\bf Top:} distance moduli offset of the field galaxies versus $i^*-W1^*$ color. Other details are the same as the top panel of Figure \ref{fig:dDM_logW}. {\bf Bottom:} same as the top panel but with the bias-corrected $\overline{DM_{W1 \mu }}$.
}
\label{fig:dDM_SBW1i_iW1}
\end{figure}

Figure \ref{fig:IR_DM_W1} displays the differences of the cluster moduli at adjusted and unadjusted infrared bands with reference to the {\it W1} band. The left panel of this plot shows an excellent consistency between the {\it W1} and {\it W2} results. The middle panels of Figure~\ref{fig:IR_DM_W1} show that the color-adjusted models, $TF_{W1c}$ and $TF_{W2c}$, result in slightly smaller distance moduli compared to that measured using the unadjusted models. 
The right panels of Figure~\ref{fig:IR_DM_W1} reveal more significant offsets for the $TF_{W1 \mu}$ and $TF_{W2\mu}$ moduli, calling for adjustments to the distance moduli of $DM_{W1 \mu}+0.11$ and $DM_{W2 \mu}+0.17$. These offsets are listed as $\Delta DM$ in column (8) of Table~\ref{tab_tfr_params} and are included to form the adjusted moduli, $DM_c$.

These discrepancies arise from the still-modest size of the zero-point calibrator sample that does not fairly represent the global distribution of galaxies. The bias is evaluated with improved statistics using the field galaxies of the impending {\it Cosmicflows-4} distance compilation. We take as a baseline $DM_{i}$ rather than $DM_{W1}$ because, as shown in Figure \ref{fig:DM_i_w1}, cluster distance moduli measured at the {\it i} and {\it W1} bands are reasonably consistent, and with this choice, we can test for other discrepancies between measured moduli at optical and infrared passbands.

The top panels of Figure~\ref{fig:dDM_logW} and Figure~\ref{fig:dDM_SBW1i_iW1} plot the $DM_{W1 \mu} - DM_i$ of the field galaxies versus their line widths and $i^*-W1^*$ colors, respectively. It would be preferable to use the $i^*-W1^*$ color term to redress the observed trends because the ITFR already uses line widths to measure distances. However, the $i^*-W1^*$ color does not provide a practical solution if optical photometry is missing. The correlation between color terms and line width shown in Figure~\ref{fig:Color_LogW}, provides a route to the bias correction that is available for all of our samples. 

The black dashed curve in the top panel of Figure~\ref{fig:dDM_logW} shows our best fit to the offsets of moduli as a quadratic function of line width, shifting the $DM_{W1 \mu}$ values to the same level as $DM_{i}$. The offsets of the modified distance moduli, $\overline{DM_{W1 \mu}} - DM_i$, are plotted in the bottom panels of Figures \ref{fig:dDM_logW} and \ref{fig:dDM_SBW1i_iW1}. 

In the bottom panel of Figure~\ref{fig:dDM_logW} the red filled points represent the zero-point calibrators after adjustments following the description in the caption for the top panel of this figure.  The gray points illustrate the related adjusted values for the full field sample. On average, the field sample manifests a vertical drift of $0.11\pm0.08$ mag in the negative direction.  The value of this net offset is in agreement with the moduli offset in the top right panel of Figure~\ref{fig:IR_DM_W1}.  Evidently, the pattern of surface brightness corrections imposed on the zero-point calibrators does not tightly follow that of the field galaxies.  It is expected that this discrepancy would be neutralized by adopting a larger/fairer sample of zero-point calibrators.

Comparing the two panels of Figure~\ref{fig:dDM_logW}, we infer that our quadratic formulation based on line width significantly but imperfectly improves the consistency of the distance measurements between $TF_{W1 \mu}$ and $TF_{i}$ alternatives.  In this paper, our focus is on galaxy clusters that contain spirals with a variety of properties such that, on average, wavelength discrepancies are expected to be moderate.  In applications to the full field sample, discussed in a follow-up paper, concerns for bias in distances to individual galaxies due to wavelength-dependent color variations will require attention.

\begin{figure}[t]
\centering
\includegraphics[width=0.9\linewidth]{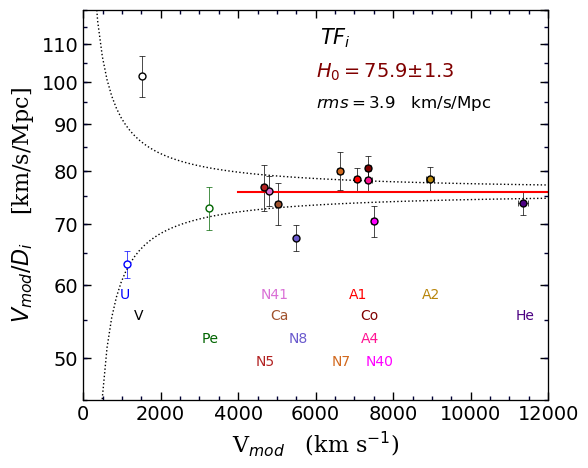}
\includegraphics[width=0.9\linewidth]{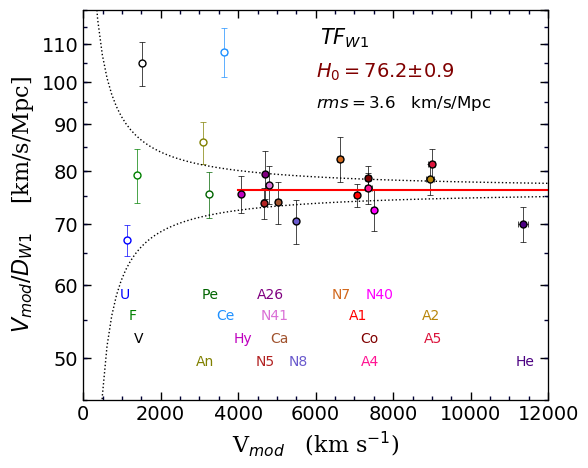}
\caption{Hubble parameter as a function of systemic velocity at the SDSS {\it i} (top) and WISE {\it W1} (bottom) bands. 
Dotted curves display a $\pm 200$ km s$^{-1}$ envelope in velocities. Red horizontal lines lies at the log average of the Hubble parameter of clusters beyond 4000~\kms, those represented by filled circles.   Corresponding codes are horizontally placed by velocity. Clusters that do not contribute to the averaging process are denoted by open circles.
}
\label{fig:H0_i_W1}
\end{figure}

\section{\hnut~ from clusters} \label{sec:HubbleParam}

For each cluster, $j$, the Hubble parameter is calculated as $H_j=V_{mod, j}/D_j$. The values of the Hubble parameters based on different TFR models are listed in column (11) of Table~\ref{Tab:cluster_H0}.  
Figure~\ref{fig:H0_i_W1} provides plots of the resulting $H_j$ given the distance moduli measured at the {\it i} and {\it W1} bands using unadjusted TFR fits. Large scatters are observed for clusters within 4,000~\kms\ due to substantial peculiar motions within our host supercluster complex. In Figure~\ref{fig:H0_i_W1}, dotted lines show an error envelope of $\pm 200$ km s$^{-1}$ to illustrate how the Hubble parameter is influenced by peculiar velocities as a function of distance. We find a Hubble constant of $H_0=75.9\pm1.3$~\kms Mpc$^{-1}$ based on the unadjusted TFR analysis at the {\it i} band with an rms scatter of $3.9$~\kms Mpc$^{-1}$. For the unadjusted TFR at the {\it W1} band, $H_0=76.2\pm0.9$~\kms Mpc$^{-1}$ with an rms scatter of $3.6$~\kms Mpc$^{-1}$.

\begin{figure*}[t]
\centering
\includegraphics[width=0.80\linewidth]{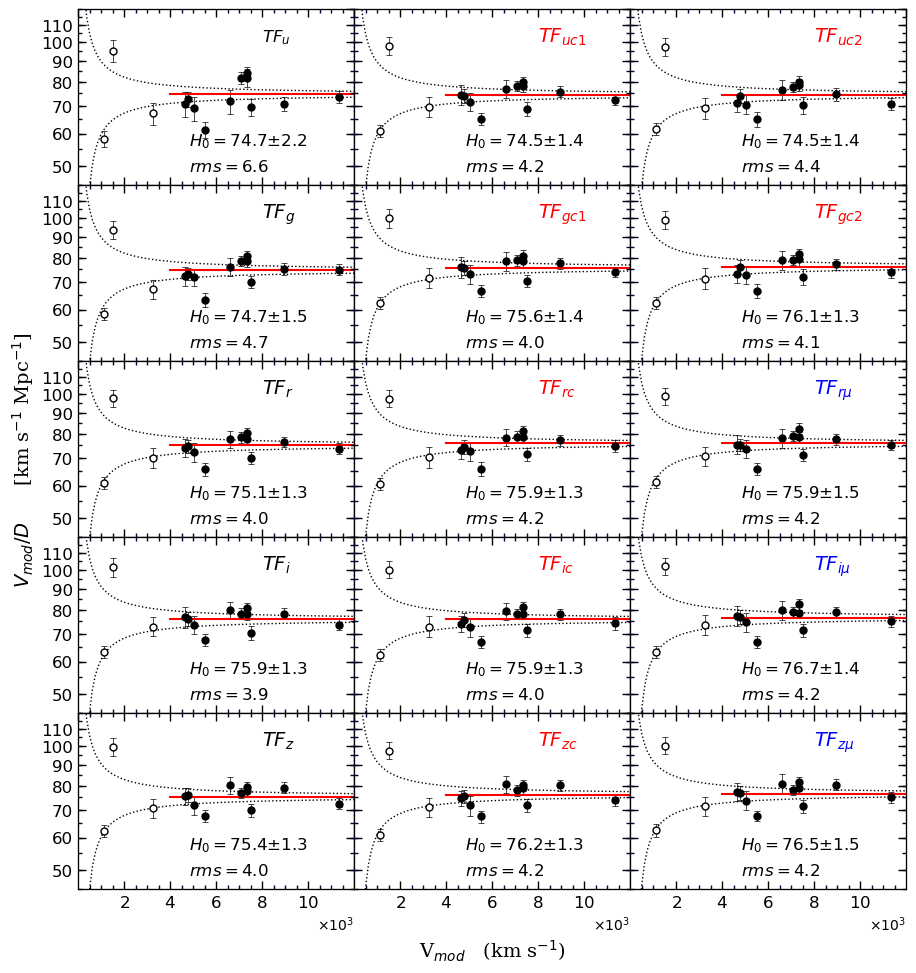}
\caption{Same as Figure \ref{fig:H0_i_W1} for the optical TFR models presented in Table \ref{tab:revised_ITFR}. The code of each model is displayed on the top right corner of each panel.
}
\label{fig:H0_OP}
\end{figure*}

\begin{figure*}[t]
\centering
\includegraphics[width=0.80\linewidth]{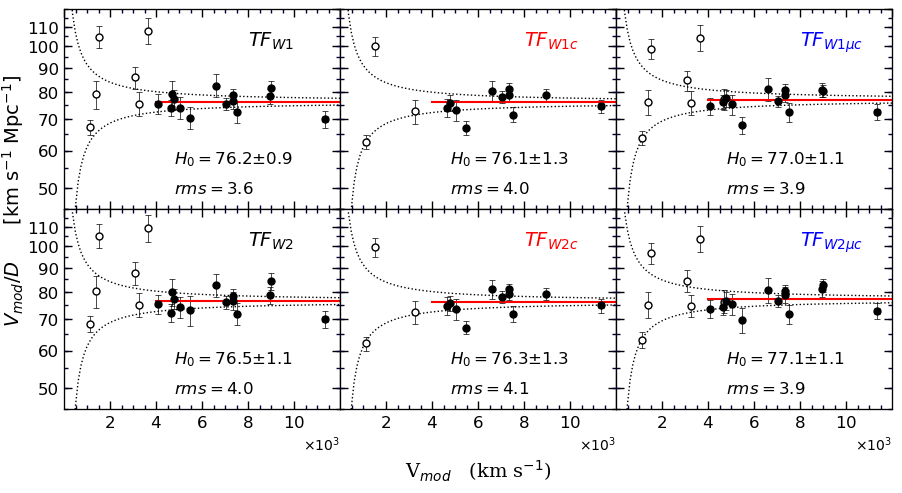}
\caption{Same as Figure \ref{fig:H0_OP} for the infrared ITFR models.}
\label{fig:H0_IR}
\end{figure*}

Figure~\ref{fig:H0_OP} and \ref{fig:H0_IR} present Hubble parameter derivations from selected formulations labeled in the fourth column of Table~\ref{tab:revised_ITFR}. The cluster moduli determined at optical bands are in good agreement with each other; hence, there is no motivation for adjustments to cluster moduli measured at optical passbands. However, the cluster moduli derived from the adjusted models at infrared bands, $DM_{W1c}$, $DM_{W2c}$, $DM_{W1 \mu}$ and $DM_{W2 \mu}$, must receive offsets to be on the same scale as the measured moduli at optical bands. We apply the offset values shown in Figure~\ref{fig:IR_DM_W1} to the moduli prior to the calculation of the Hubble parameters. 

The Hubble constant values given in Figure~\ref{fig:H0_OP} and \ref{fig:H0_IR} are consistent within the claimed statistical uncertainties. The rms scatter of the Hubble parameters is larger than the statistical uncertainties owing to cluster peculiar velocities.
It may well be that our calibration clusters do not extend to distances such that peculiar velocities are negligible.  
In addition, any systematic errors in the distances of the zero-point calibrators systematically affect our measurements of \hnut.  
The main purpose of this study is not to determine the absolute value of $H_0$ but rather to provide calibrations of the TFR methodology for measuring relative galaxy distances that are consistent across passbands and parameterizations.

\section{Conclusions} \label{sec:conclusions}

Every bit of information used in the calibration of the luminosity-line width relations of spirals might introduce biases and uncertainties in the final derived relations. 
The main purpose of this calibration program is to determine the peculiar flow of galaxies; thus, any unaccounted biases might generate artificial flows suggesting unrealistic cosmological structures. 
In this program, our main focus is to obtain distances at multiple passbands that are in good agreement with each other and are on the same scale. 

The quality of photometry data, the \hi line widths, and measurements of inclinations are all important inputs.  Corrections for dust extinction can be important toward blueward bands, and our formalism to measure dust attenuation \citep{2019ApJ...884...82K} has been constructed based on the same catalog of spirals used for the current study. 

The TFR calibrating process has two major steps: definition of slopes at each passband and then establishment of zero-points. 
The derivations of slopes of the TFR from an ensemble of cluster galaxies is a robust process, but small variations in the properties of galaxies in each cluster influence that cluster's measured distance relative to the other clusters in the ensemble.
For instance, clusters with, on average, redder galaxies are located at farther distances at redward bands.

Zero-point calibrators are the foundation of our calibration process, and any systematic in their distances propagates through the entire process. We separately examined the applicability of two sets of distance calibrators, one using Population I Cepheid variables and the other using Population II red giant branch stars. We find consistency of scales between the two calibrations as manifested in implied values of $H_0$ using the scale for Cepheids given by \cite{2019ApJ...876...85R} and the scale for TRGB luminosities given by \citet{2007ApJ...661..815R}. The TRGB scale advocated by \citet{2019ApJ...882...34F} gives inconsistency with the Cepheid scale. This alternate scaling does caution us to the plausible existence of zero-point systematics at the level of 3\%.  

We find good agreement in TFR distances measured at the SDSS {\it r}, {\it i}, and {\it z} bands and WISE {\it W1} and {\it W2} bands. The SDSS {\it u} and {\it g} bands are less useful for determining TFR distances, but the broad span of wavelength coverage they provide offers insight into obscuration and color term issues. Color or surface brightness correlations do not significantly reduced TFR scatter at the optical bands, so they are not used. However, infrared TFR scatter can be substantially reduced with either optical to infrared color or surface brightness terms. The color correlations are strongest but practically not so useful, since the availability of optical information negates the need for an infrared-based measurement. A surface brightness-adjustment can be constructed with only infrared photometric information. However, the surface brightness adjusted TFR at the {\it W1} and {\it W2} bands must be compensated for offsets evidently due to statistical differences between the surface brightness properties of the zero-point calibrators and the full field sample. The offset at the level of 5\%  is a cautionary example of possible systematics.

The TFR scatter varies with luminosity, such that it is smaller at the bright end and increases toward fainter galaxies. The scatter at the slow-rotating, faint-end of the TFR can be reduced by accounting for substantial contributions of dark baryons in the form of neutral gas \citep{2000ApJ...533L..99M}.  We have not explored this added complexity.  In the redshift regime $3,000-10,000$~\kms\ where the numbers in our field sample peak, our multiband measurements will typically render distance with 20\% accuracy.

There is a slight curvature at the bright end of the TFR, an effect that has been noticed by others.  The departure from linearity in the relation between magnitude and log line width results in a small bias.  Another bias results from asymmetric scatter due to magnitude uncertainties in the regime of the Schechter function exponential cutoff.  These two biases act in opposite senses, and simulations demonstrate that they roughly but not entirely cancel.

Measurements of distances of clusters at $V_{mod}>4000$~\kms give a preliminary determination of the Hubble constant, \hnut.  The sample is small, 11 optical and 14 infrared, and peculiar velocities may not be negligible.  In any event, our measurements of \hnut\ at all passbands based on the adjusted and unadjusted TFR models are consistent. We find $H_0 = 75.9\pm1.3$ and $H_0=76.2\pm0.9$~\kmsMpc\ for the unadjusted TFR at the {\it i} and {\it W1} bands. With these values, a 3\% systematic offset in the distance zero-point translates to a $\pm2.3$~\kms Mpc$^{-1}$ systematic error in the estimated Hubble constant.

In this study, we only considered biases in the distance measurements of the clusters, where constituent galaxies are treated as an ensemble. There can be additional biases in measuring the distances of individual field galaxies. This issue is important, especially if different passbands are utilized in different parts of the sky. Even if color terms do not lower the scatter of the calibrated relations at the {\it r}, {\it i} and {\it z} bands, they might still help cancel systematics with application to field galaxies. In a subsequent study, we will investigate the possible ways to reduce biases in the distance measurements of field spirals and present formulations for getting consistent distances at optical and infrared bands based on the calibrations provided in this paper.

\section*{Acknowledgments}
\acknowledgments
We are pleased to acknowledge the citizen participation in scientific research of undergraduate students at the University of Hawaii; members of amateur astronomy clubs in France at the Plan\'etarium de Vaulx-en-Velin, Association Clair d'\'etoiles et Brin d'jardin, Soci\'et\'e astronomique de Lyon, Club d'astronomie Lyon Amp\`ere, Club d'astronomie des monts du lyonnais, and Club d'astronomie de Dijon; and friends who helped us with measuring inclinations of spiral galaxies in our sample.

Support for E.K. and R.B.T. was provided by NASA through grant No. 88NSSC18K0424
and for G.A. and R.B.T. through grants from the Space Telescope Science Institute. 
H.C. acknowledges support from Institut Universitaire de France.

This research has made use of the NASA/IPAC Extragalactic Database,\footnote{\url{http://ned.ipac.caltech.edu/}} which is operated by the Jet Propulsion Laboratory, California Institute of Technology, under contract with the National Aeronautics and Space Administration.

\clearpage
\startlongtable
\begin{deluxetable*}{ll|c|l|c|ccc|cc|c}
\tablewidth{0in}
\tabletypesize{\scriptsize}
\setlength{\tabcolsep}{0.15cm}
\tablecaption{\label{Tab:cluster_H0}
Cluster Distances and $H_0$ Data\label{tab_tfr_params}}

\tablehead{
  &  &  &  & & & & & & & 
\\
Cluster &  V$_{mod}$ &  TFR &  TFR  &  Ngal &  DM$_{o}$ &  Bias & $\Delta DM$ & DM$_c$  & D &  V$_{mod}$/D \\ 
  & \kms & Band & Model & & mag & mag & mag & mag & Mpc & \kms Mpc$^{-1}$ \\
  (1) & (2) & (3) & (4) & (5) & (6) & (7) & (8) & (9) & (10) & (11) 
 }
\startdata
Virgo  & 1516$\pm$45 & u & TF$_{u}$ & 23 & 31.01 & 0.000 &  & 31.01$\pm$0.12 & 15.92$\pm$0.88 & 95.21$\pm$5.97 \\
 & & u & TF$_{uc1}$ & 23 & 30.95 & 0.000 &  & 30.95$\pm$0.09 & 15.49$\pm$0.64 & 97.88$\pm$4.99 \\
 & & u & TF$_{uc2}$ & 23 & 30.96 & 0.000 &  & 30.96$\pm$0.09 & 15.56$\pm$0.64 & 97.43$\pm$4.97 \\
 & & g & TF$_{g}$ & 24 & 31.05 & 0.001 &  & 31.05$\pm$0.09 & 16.23$\pm$0.67 & 93.42$\pm$4.76 \\
 & & g & TF$_{gc1}$ & 24 & 30.98 & 0.000 &  & 30.98$\pm$0.09 & 15.70$\pm$0.65 & 96.54$\pm$4.92 \\
 & & g & TF$_{gc2}$ & 24 & 31.00 & 0.000 &  & 31.00$\pm$0.09 & 15.85$\pm$0.66 & 95.65$\pm$4.88 \\
 & & r & TF$_{r}$ & 24 & 30.95 & 0.002 &  & 30.95$\pm$0.08 & 15.50$\pm$0.57 & 97.81$\pm$4.63 \\
 & & r & TF$_{rc}$ & 24 & 30.96 & 0.001 &  & 30.96$\pm$0.08 & 15.56$\pm$0.57 & 97.41$\pm$4.61 \\
 & & r & TF$_{r \mu}$ & 24 & 30.93 & 0.001 &  & 30.93$\pm$0.08 & 15.35$\pm$0.57 & 98.75$\pm$4.67 \\
 & & i & TF$_{i}$ & 24 & 30.87 & 0.001 &  & 30.87$\pm$0.09 & 14.94$\pm$0.62 & 101.49$\pm$5.17 \\
 & & i & TF$_{ic}$ & 24 & 30.90 & 0.001 &  & 30.90$\pm$0.08 & 15.14$\pm$0.56 & 100.11$\pm$4.74 \\
 & & i & TF$_{i \mu}$ & 24 & 30.86 & 0.001 &  & 30.86$\pm$0.08 & 14.87$\pm$0.55 & 101.97$\pm$4.82 \\
 & & z & TF$_{z}$ & 23 & 30.91 & 0.002 &  & 30.91$\pm$0.09 & 15.22$\pm$0.63 & 99.63$\pm$5.08 \\
 & & z & TF$_{zc}$ & 23 & 30.96 & 0.001 &  & 30.96$\pm$0.08 & 15.57$\pm$0.57 & 97.40$\pm$4.61 \\
 & & z & TF$_{z \mu}$ & 23 & 30.90 & 0.000 &  & 30.90$\pm$0.08 & 15.14$\pm$0.56 & 100.15$\pm$4.74 \\
 & & W1 & TF$_{W1}$ & 24 & 30.80 & 0.000 &  & 30.80$\pm$0.10 & 14.46$\pm$0.67 & 104.86$\pm$5.75 \\
 & & W1 & TF$_{W1c}$ & 24 & 30.86 & 0.001 & 0.04 & 30.90$\pm$0.08 & 15.14$\pm$0.56 & 100.11$\pm$4.74 \\
 & & W1 & TF$_{W1 \mu}$ & 24 & 30.82 & -0.001 & 0.11 & 30.93$\pm$0.09 & 15.34$\pm$0.64 & 98.83$\pm$5.04 \\
 & & W2 & TF$_{W2}$ & 24 & 30.79 & 0.001 &  & 30.79$\pm$0.11 & 14.39$\pm$0.73 & 105.33$\pm$6.18 \\
 & & W2 & TF$_{W2c}$ & 24 & 30.86 & 0.000 & 0.06 & 30.92$\pm$0.08 & 15.28$\pm$0.56 & 99.23$\pm$4.69 \\
 & & W2 & TF$_{W2 \mu}$ & 24 & 30.81 & -0.001 & 0.17 & 30.98$\pm$0.09 & 15.70$\pm$0.65 & 96.58$\pm$4.92 \\
\hline
 Ursa Major  & 1141$\pm$13 & u & TF$_{u}$ & 36 & 31.46 & 0.000 &  & 31.46$\pm$0.09 & 19.59$\pm$0.81 & 58.25$\pm$2.50 \\
 & & u & TF$_{uc1}$ & 36 & 31.36 & 0.000 &  & 31.36$\pm$0.07 & 18.71$\pm$0.60 & 60.99$\pm$2.09 \\
 & & u & TF$_{uc2}$ & 36 & 31.34 & 0.000 &  & 31.34$\pm$0.07 & 18.54$\pm$0.60 & 61.56$\pm$2.10 \\
 & & g & TF$_{g}$ & 36 & 31.45 & 0.003 &  & 31.45$\pm$0.07 & 19.52$\pm$0.63 & 58.45$\pm$2.00 \\
 & & g & TF$_{gc1}$ & 36 & 31.38 & 0.002 &  & 31.38$\pm$0.07 & 18.90$\pm$0.61 & 60.37$\pm$2.06 \\
 & & g & TF$_{gc2}$ & 36 & 31.38 & 0.001 &  & 31.38$\pm$0.07 & 18.89$\pm$0.61 & 60.40$\pm$2.07 \\
 & & r & TF$_{r}$ & 36 & 31.36 & 0.002 &  & 31.36$\pm$0.07 & 18.73$\pm$0.60 & 60.93$\pm$2.08 \\
 & & r & TF$_{rc}$ & 36 & 31.38 & -0.001 &  & 31.38$\pm$0.07 & 18.88$\pm$0.61 & 60.45$\pm$2.07 \\
 & & r & TF$_{r \mu}$ & 36 & 31.35 & 0.000 &  & 31.35$\pm$0.07 & 18.62$\pm$0.60 & 61.27$\pm$2.09 \\
 & & i & TF$_{i}$ & 36 & 31.28 & 0.002 &  & 31.28$\pm$0.07 & 18.05$\pm$0.58 & 63.21$\pm$2.16 \\
 & & i & TF$_{ic}$ & 36 & 31.31 & 0.003 &  & 31.31$\pm$0.07 & 18.30$\pm$0.59 & 62.33$\pm$2.13 \\
 & & i & TF$_{i \mu}$ & 36 & 31.28 & 0.002 &  & 31.28$\pm$0.07 & 18.05$\pm$0.58 & 63.23$\pm$2.16 \\
 & & z & TF$_{z}$ & 36 & 31.31 & 0.003 &  & 31.31$\pm$0.07 & 18.31$\pm$0.59 & 62.31$\pm$2.13 \\
 & & z & TF$_{zc}$ & 36 & 31.36 & 0.002 &  & 31.36$\pm$0.07 & 18.72$\pm$0.60 & 60.95$\pm$2.08 \\
 & & z & TF$_{z \mu}$ & 36 & 31.31 & -0.002 &  & 31.31$\pm$0.07 & 18.26$\pm$0.59 & 62.48$\pm$2.14 \\
 & & W1 & TF$_{W1}$ & 36 & 31.15 & 0.000 &  & 31.15$\pm$0.08 & 16.98$\pm$0.63 & 67.19$\pm$2.59 \\
 & & W1 & TF$_{W1c}$ & 36 & 31.26 & 0.000 & 0.04 & 31.30$\pm$0.07 & 18.20$\pm$0.59 & 62.69$\pm$2.14 \\
 & & W1 & TF$_{W1 \mu}$ & 36 & 31.15 & 0.000 & 0.11 & 31.26$\pm$0.07 & 17.86$\pm$0.58 & 63.87$\pm$2.18 \\
 & & W2 & TF$_{W2}$ & 36 & 31.11 & 0.001 &  & 31.11$\pm$0.08 & 16.68$\pm$0.61 & 68.41$\pm$2.64 \\
 & & W2 & TF$_{W2c}$ & 36 & 31.27 & 0.000 & 0.06 & 31.33$\pm$0.07 & 18.45$\pm$0.59 & 61.83$\pm$2.11 \\
 & & W2 & TF$_{W2 \mu}$ & 36 & 31.11 & -0.000 & 0.17 & 31.28$\pm$0.08 & 18.03$\pm$0.66 & 63.30$\pm$2.44 \\
\hline
 Fornax  & 1383$\pm$32 & W1 & TF$_{W1}$ & 17 & 31.21 & 0.002 &  & 31.21$\pm$0.14 & 17.48$\pm$1.13 & 79.13$\pm$5.42 \\
 & & W1 & TF$_{W1 \mu}$ & 17 & 31.19 & -0.003 & 0.11 & 31.30$\pm$0.12 & 18.17$\pm$1.00 & 76.11$\pm$4.56 \\
 & & W2 & TF$_{W2}$ & 17 & 31.18 & 0.001 &  & 31.18$\pm$0.16 & 17.23$\pm$1.27 & 80.27$\pm$6.20 \\
 & & W2 & TF$_{W2 \mu}$ & 17 & 31.16 & -0.005 & 0.17 & 31.32$\pm$0.13 & 18.40$\pm$1.10 & 75.15$\pm$4.82 \\
\hline
 Pegasus  & 3249$\pm$61 & u & TF$_{u}$ & 24 & 33.42 & 0.005 &  & 33.42$\pm$0.13 & 48.41$\pm$2.90 & 67.12$\pm$4.21 \\
 & & u & TF$_{uc1}$ & 24 & 33.34 & 0.003 &  & 33.34$\pm$0.11 & 46.62$\pm$2.36 & 69.69$\pm$3.77 \\
 & & u & TF$_{uc2}$ & 22 & 33.36 & 0.003 &  & 33.36$\pm$0.12 & 47.06$\pm$2.60 & 69.03$\pm$4.03 \\
 & & g & TF$_{g}$ & 24 & 33.42 & 0.005 &  & 33.43$\pm$0.11 & 48.42$\pm$2.45 & 67.10$\pm$3.63 \\
 & & g & TF$_{gc1}$ & 24 & 33.35 & 0.005 &  & 33.36$\pm$0.11 & 46.89$\pm$2.38 & 69.29$\pm$3.74 \\
 & & g & TF$_{gc2}$ & 22 & 33.36 & 0.003 &  & 33.36$\pm$0.12 & 47.06$\pm$2.60 & 69.04$\pm$4.03 \\
 & & r & TF$_{r}$ & 24 & 33.33 & 0.003 &  & 33.33$\pm$0.11 & 46.42$\pm$2.35 & 69.99$\pm$3.78 \\
 & & r & TF$_{rc}$ & 22 & 33.32 & -0.001 &  & 33.32$\pm$0.12 & 46.12$\pm$2.55 & 70.45$\pm$4.11 \\
 & & r & TF$_{r \mu}$ & 24 & 33.31 & 0.000 &  & 33.31$\pm$0.11 & 45.93$\pm$2.33 & 70.74$\pm$3.82 \\
 & & i & TF$_{i}$ & 24 & 33.24 & 0.004 &  & 33.24$\pm$0.11 & 44.54$\pm$2.26 & 72.94$\pm$3.94 \\
 & & i & TF$_{ic}$ & 22 & 33.24 & 0.005 &  & 33.24$\pm$0.12 & 44.56$\pm$2.46 & 72.91$\pm$4.26 \\
 & & i & TF$_{i \mu}$ & 24 & 33.22 & 0.003 &  & 33.22$\pm$0.11 & 44.12$\pm$2.24 & 73.64$\pm$3.98 \\
 & & z & TF$_{z}$ & 24 & 33.31 & 0.005 &  & 33.31$\pm$0.11 & 46.02$\pm$2.33 & 70.59$\pm$3.81 \\
 & & z & TF$_{zc}$ & 22 & 33.30 & 0.002 &  & 33.30$\pm$0.11 & 45.75$\pm$2.32 & 71.02$\pm$3.84 \\
 & & z & TF$_{z \mu}$ & 24 & 33.29 & -0.003 &  & 33.29$\pm$0.11 & 45.44$\pm$2.30 & 71.50$\pm$3.86 \\
 & & W1 & TF$_{W1}$ & 23 & 33.17 & 0.001 &  & 33.17$\pm$0.12 & 43.06$\pm$2.38 & 75.45$\pm$4.40 \\
 & & W1 & TF$_{W1c}$ & 22 & 33.21 & 0.001 & 0.04 & 33.25$\pm$0.12 & 44.69$\pm$2.47 & 72.69$\pm$4.24 \\
 & & W1 & TF$_{W1 \mu}$ & 23 & 33.05 & -0.000 & 0.11 & 33.16$\pm$0.12 & 42.85$\pm$2.37 & 75.82$\pm$4.43 \\
 & & W2 & TF$_{W2}$ & 23 & 33.18 & 0.001 &  & 33.18$\pm$0.12 & 43.27$\pm$2.39 & 75.09$\pm$4.38 \\
 & & W2 & TF$_{W2c}$ & 22 & 33.21 & 0.000 & 0.06 & 33.27$\pm$0.12 & 45.09$\pm$2.49 & 72.06$\pm$4.21 \\
 & & W2 & TF$_{W2 \mu}$ & 23 & 33.02 & 0.000 & 0.17 & 33.19$\pm$0.11 & 43.46$\pm$2.20 & 74.76$\pm$4.04 \\
\hline
 Centaurus  & 3645$\pm$56 & W1 & TF$_{W1}$ & 22 & 32.63 & 0.012 &  & 32.64$\pm$0.13 & 33.76$\pm$2.02 & 107.96$\pm$6.67 \\
 & & W1 & TF$_{W1 \mu}$ & 22 & 32.62 & -0.013 & 0.11 & 32.72$\pm$0.13 & 34.94$\pm$2.09 & 104.33$\pm$6.45 \\
 & & W2 & TF$_{W2}$ & 22 & 32.60 & 0.013 &  & 32.61$\pm$0.14 & 33.31$\pm$2.15 & 109.44$\pm$7.25 \\
 & & W2 & TF$_{W2 \mu}$ & 22 & 32.58 & -0.024 & 0.17 & 32.73$\pm$0.13 & 35.10$\pm$2.10 & 103.86$\pm$6.42 \\
\hline
 Antlia  & 3103$\pm$53 & W1 & TF$_{W1}$ & 17 & 32.78 & 0.009 &  & 32.79$\pm$0.11 & 36.12$\pm$1.83 & 85.90$\pm$4.59 \\
 & & W1 & TF$_{W1 \mu}$ & 17 & 32.72 & -0.007 & 0.11 & 32.82$\pm$0.10 & 36.69$\pm$1.69 & 84.58$\pm$4.15 \\
 & & W2 & TF$_{W2}$ & 17 & 32.74 & 0.006 &  & 32.75$\pm$0.12 & 35.42$\pm$1.96 & 87.60$\pm$5.07 \\
 & & W2 & TF$_{W2 \mu}$ & 17 & 32.67 & -0.013 & 0.17 & 32.83$\pm$0.11 & 36.77$\pm$1.86 & 84.39$\pm$4.51 \\
\hline
 Hydra  & 4084$\pm$44 & W1 & TF$_{W1}$ & 44 & 33.66 & 0.008 &  & 33.67$\pm$0.10 & 54.14$\pm$2.49 & 75.43$\pm$3.57 \\
 & & W1 & TF$_{W1 \mu}$ & 44 & 33.59 & -0.007 & 0.11 & 33.69$\pm$0.09 & 54.79$\pm$2.27 & 74.54$\pm$3.19 \\
 & & W2 & TF$_{W2}$ & 44 & 33.67 & 0.004 &  & 33.67$\pm$0.10 & 54.31$\pm$2.50 & 75.19$\pm$3.56 \\
 & & W2 & TF$_{W2 \mu}$ & 44 & 33.56 & -0.010 & 0.17 & 33.72$\pm$0.09 & 55.47$\pm$2.30 & 73.63$\pm$3.15 \\
\hline
 Abell 262  & 4684$\pm$50 & W1 & TF$_{W1}$ & 54 & 33.85 & 0.006 &  & 33.86$\pm$0.13 & 59.03$\pm$3.53 & 79.34$\pm$4.83 \\
 & & W1 & TF$_{W1 \mu}$ & 54 & 33.81 & -0.005 & 0.11 & 33.91$\pm$0.11 & 60.67$\pm$3.07 & 77.21$\pm$4.00 \\
 & & W2 & TF$_{W2}$ & 54 & 33.84 & 0.003 &  & 33.84$\pm$0.14 & 58.70$\pm$3.78 & 79.79$\pm$5.21 \\
 & & W2 & TF$_{W2 \mu}$ & 54 & 33.78 & -0.009 & 0.17 & 33.94$\pm$0.12 & 61.41$\pm$3.39 & 76.27$\pm$4.29 \\
\hline
 NGC 507  & 4660$\pm$64 & u & TF$_{u}$ & 20 & 34.07 & 0.027 &  & 34.10$\pm$0.15 & 65.98$\pm$4.56 & 70.63$\pm$4.97 \\
 & & u & TF$_{uc1}$ & 20 & 33.97 & 0.014 &  & 33.98$\pm$0.12 & 62.64$\pm$3.46 & 74.40$\pm$4.24 \\
 & & u & TF$_{uc2}$ & 19 & 34.06 & 0.018 &  & 34.08$\pm$0.11 & 65.41$\pm$3.31 & 71.25$\pm$3.74 \\
 & & g & TF$_{g}$ & 20 & 34.03 & 0.014 &  & 34.04$\pm$0.11 & 64.40$\pm$3.26 & 72.36$\pm$3.80 \\
 & & g & TF$_{gc1}$ & 20 & 33.99 & 0.019 &  & 34.01$\pm$0.12 & 63.37$\pm$3.50 & 73.54$\pm$4.19 \\
 & & g & TF$_{gc2}$ & 19 & 34.07 & 0.019 &  & 34.09$\pm$0.10 & 65.72$\pm$3.03 & 70.90$\pm$3.41 \\
 & & r & TF$_{r}$ & 20 & 33.98 & 0.014 &  & 33.99$\pm$0.11 & 62.93$\pm$3.19 & 74.05$\pm$3.89 \\
 & & r & TF$_{rc}$ & 19 & 34.02 & 0.003 &  & 34.02$\pm$0.10 & 63.75$\pm$2.94 & 73.09$\pm$3.51 \\
 & & r & TF$_{r \mu}$ & 20 & 33.95 & 0.005 &  & 33.95$\pm$0.11 & 61.80$\pm$3.13 & 75.41$\pm$3.96 \\
 & & i & TF$_{i}$ & 20 & 33.90 & 0.017 &  & 33.92$\pm$0.12 & 60.73$\pm$3.36 & 76.73$\pm$4.37 \\
 & & i & TF$_{ic}$ & 19 & 33.97 & 0.021 &  & 33.99$\pm$0.09 & 62.84$\pm$2.60 & 74.15$\pm$3.24 \\
 & & i & TF$_{i \mu}$ & 20 & 33.88 & 0.017 &  & 33.90$\pm$0.12 & 60.17$\pm$3.32 & 77.45$\pm$4.41 \\
 & & z & TF$_{z}$ & 20 & 33.93 & 0.018 &  & 33.95$\pm$0.10 & 61.60$\pm$2.84 & 75.65$\pm$3.64 \\
 & & z & TF$_{zc}$ & 19 & 33.97 & 0.005 &  & 33.97$\pm$0.09 & 62.37$\pm$2.59 & 74.71$\pm$3.26 \\
 & & z & TF$_{z \mu}$ & 20 & 33.90 & -0.003 &  & 33.90$\pm$0.10 & 60.17$\pm$2.77 & 77.45$\pm$3.72 \\
 & & W1 & TF$_{W1}$ & 22 & 33.99 & 0.012 &  & 34.00$\pm$0.08 & 63.15$\pm$2.33 & 73.79$\pm$2.90 \\
 & & W1 & TF$_{W1c}$ & 19 & 33.93 & 0.025 & 0.04 & 33.99$\pm$0.09 & 62.94$\pm$2.61 & 74.04$\pm$3.23 \\
 & & W1 & TF$_{W1 \mu}$ & 22 & 33.84 & -0.015 & 0.11 & 33.94$\pm$0.07 & 61.25$\pm$1.97 & 76.08$\pm$2.67 \\
 & & W2 & TF$_{W2}$ & 22 & 34.04 & 0.013 &  & 34.05$\pm$0.09 & 64.65$\pm$2.68 & 72.08$\pm$3.15 \\
 & & W2 & TF$_{W2c}$ & 19 & 33.92 & 0.004 & 0.06 & 33.98$\pm$0.09 & 62.65$\pm$2.60 & 74.39$\pm$3.25 \\
 & & W2 & TF$_{W2 \mu}$ & 22 & 33.84 & -0.023 & 0.17 & 33.99$\pm$0.08 & 62.73$\pm$2.31 & 74.28$\pm$2.92 \\
\hline
 NGC 410  & 4792$\pm$53 & u & TF$_{u}$ & 33 & 34.09 & 0.008 &  & 34.10$\pm$0.09 & 66.02$\pm$2.74 & 72.58$\pm$3.11 \\
 & & u & TF$_{uc1}$ & 33 & 34.05 & 0.006 &  & 34.06$\pm$0.08 & 64.74$\pm$2.39 & 74.02$\pm$2.85 \\
 & & u & TF$_{uc2}$ & 31 & 34.05 & 0.006 &  & 34.06$\pm$0.08 & 64.76$\pm$2.39 & 74.00$\pm$2.85 \\
 & & g & TF$_{g}$ & 33 & 34.07 & 0.010 &  & 34.08$\pm$0.07 & 65.45$\pm$2.11 & 73.21$\pm$2.50 \\
 & & g & TF$_{gc1}$ & 33 & 34.06 & 0.013 &  & 34.07$\pm$0.08 & 65.26$\pm$2.40 & 73.43$\pm$2.82 \\
 & & g & TF$_{gc2}$ & 31 & 34.06 & 0.010 &  & 34.07$\pm$0.08 & 65.16$\pm$2.40 & 73.54$\pm$2.83 \\
 & & r & TF$_{r}$ & 33 & 34.03 & 0.007 &  & 34.04$\pm$0.08 & 64.19$\pm$2.36 & 74.65$\pm$2.87 \\
 & & r & TF$_{rc}$ & 31 & 34.04 & -0.000 &  & 34.04$\pm$0.08 & 64.26$\pm$2.37 & 74.57$\pm$2.87 \\
 & & r & TF$_{r \mu}$ & 33 & 34.02 & 0.002 &  & 34.02$\pm$0.07 & 63.73$\pm$2.05 & 75.19$\pm$2.56 \\
 & & i & TF$_{i}$ & 33 & 33.99 & 0.006 &  & 34.00$\pm$0.08 & 62.98$\pm$2.32 & 76.08$\pm$2.93 \\
 & & i & TF$_{ic}$ & 31 & 34.00 & 0.007 &  & 34.01$\pm$0.08 & 63.31$\pm$2.33 & 75.69$\pm$2.91 \\
 & & i & TF$_{i \mu}$ & 33 & 33.97 & 0.005 &  & 33.98$\pm$0.08 & 62.38$\pm$2.30 & 76.82$\pm$2.95 \\
 & & z & TF$_{z}$ & 33 & 33.99 & 0.009 &  & 34.00$\pm$0.08 & 63.08$\pm$2.32 & 75.97$\pm$2.92 \\
 & & z & TF$_{zc}$ & 31 & 34.01 & 0.003 &  & 34.01$\pm$0.07 & 63.48$\pm$2.05 & 75.49$\pm$2.57 \\
 & & z & TF$_{z \mu}$ & 33 & 33.98 & -0.004 &  & 33.98$\pm$0.08 & 62.41$\pm$2.30 & 76.79$\pm$2.95 \\
 & & W1 & TF$_{W1}$ & 31 & 33.96 & 0.004 &  & 33.96$\pm$0.10 & 62.05$\pm$2.86 & 77.23$\pm$3.66 \\
 & & W1 & TF$_{W1c}$ & 31 & 33.96 & 0.005 & 0.04 & 34.00$\pm$0.08 & 63.24$\pm$2.33 & 75.78$\pm$2.91 \\
 & & W1 & TF$_{W1 \mu}$ & 31 & 33.85 & -0.004 & 0.11 & 33.96$\pm$0.09 & 61.84$\pm$2.56 & 77.49$\pm$3.32 \\
 & & W2 & TF$_{W2}$ & 31 & 33.96 & 0.003 &  & 33.96$\pm$0.10 & 62.02$\pm$2.86 & 77.27$\pm$3.66 \\
 & & W2 & TF$_{W2c}$ & 31 & 33.96 & 0.001 & 0.06 & 34.02$\pm$0.08 & 63.71$\pm$2.35 & 75.22$\pm$2.89 \\
 & & W2 & TF$_{W2 \mu}$ & 31 & 33.82 & -0.007 & 0.17 & 33.98$\pm$0.09 & 62.61$\pm$2.60 & 76.54$\pm$3.28 \\
\hline
 Cancer  & 5025$\pm$71 & u & TF$_{u}$ & 18 & 34.24 & 0.071 &  & 34.31$\pm$0.14 & 72.80$\pm$4.69 & 69.02$\pm$4.56 \\
 & & u & TF$_{uc1}$ & 18 & 34.19 & 0.040 &  & 34.23$\pm$0.11 & 70.15$\pm$3.55 & 71.63$\pm$3.77 \\
 & & u & TF$_{uc2}$ & 17 & 34.22 & 0.045 &  & 34.27$\pm$0.11 & 71.30$\pm$3.61 & 70.48$\pm$3.71 \\
 & & g & TF$_{g}$ & 18 & 34.19 & 0.035 &  & 34.23$\pm$0.10 & 69.99$\pm$3.22 & 71.80$\pm$3.46 \\
 & & g & TF$_{gc1}$ & 18 & 34.20 & 0.057 &  & 34.26$\pm$0.11 & 71.02$\pm$3.60 & 70.76$\pm$3.72 \\
 & & g & TF$_{gc2}$ & 17 & 34.21 & 0.056 &  & 34.27$\pm$0.10 & 71.31$\pm$3.28 & 70.46$\pm$3.39 \\
 & & r & TF$_{r}$ & 18 & 34.18 & 0.032 &  & 34.21$\pm$0.11 & 69.56$\pm$3.52 & 72.24$\pm$3.80 \\
 & & r & TF$_{rc}$ & 17 & 34.19 & 0.012 &  & 34.20$\pm$0.11 & 69.25$\pm$3.51 & 72.56$\pm$3.82 \\
 & & r & TF$_{r \mu}$ & 18 & 34.15 & 0.019 &  & 34.17$\pm$0.11 & 68.21$\pm$3.46 & 73.67$\pm$3.87 \\
 & & i & TF$_{i}$ & 18 & 34.14 & 0.030 &  & 34.17$\pm$0.11 & 68.24$\pm$3.46 & 73.64$\pm$3.87 \\
 & & i & TF$_{ic}$ & 17 & 34.16 & 0.038 &  & 34.20$\pm$0.11 & 69.13$\pm$3.50 & 72.69$\pm$3.82 \\
 & & i & TF$_{i \mu}$ & 18 & 34.11 & 0.032 &  & 34.14$\pm$0.11 & 67.35$\pm$3.41 & 74.61$\pm$3.92 \\
 & & z & TF$_{z}$ & 18 & 34.20 & 0.027 &  & 34.23$\pm$0.11 & 70.04$\pm$3.55 & 71.74$\pm$3.77 \\
 & & z & TF$_{zc}$ & 17 & 34.22 & 0.007 &  & 34.23$\pm$0.12 & 70.05$\pm$3.87 & 71.73$\pm$4.09 \\
 & & z & TF$_{z \mu}$ & 18 & 34.17 & -0.003 &  & 34.17$\pm$0.11 & 68.14$\pm$3.45 & 73.74$\pm$3.88 \\
 & & W1 & TF$_{W1}$ & 17 & 34.15 & 0.012 &  & 34.16$\pm$0.11 & 67.98$\pm$3.44 & 73.92$\pm$3.89 \\
 & & W1 & TF$_{W1c}$ & 17 & 34.12 & 0.026 & 0.04 & 34.19$\pm$0.11 & 68.73$\pm$3.48 & 73.11$\pm$3.85 \\
 & & W1 & TF$_{W1 \mu}$ & 17 & 34.03 & -0.015 & 0.11 & 34.12$\pm$0.10 & 66.82$\pm$3.08 & 75.20$\pm$3.62 \\
 & & W2 & TF$_{W2}$ & 17 & 34.14 & 0.013 &  & 34.15$\pm$0.11 & 67.70$\pm$3.43 & 74.22$\pm$3.90 \\
 & & W2 & TF$_{W2c}$ & 17 & 34.12 & 0.006 & 0.06 & 34.19$\pm$0.11 & 68.73$\pm$3.48 & 73.12$\pm$3.85 \\
 & & W2 & TF$_{W2 \mu}$ & 17 & 33.98 & -0.027 & 0.17 & 34.12$\pm$0.11 & 66.76$\pm$3.38 & 75.26$\pm$3.96 \\
\hline
 NGC 80  & 5499$\pm$42 & u & TF$_{u}$ & 14 & 34.72 & 0.046 &  & 34.77$\pm$0.10 & 89.77$\pm$4.13 & 61.25$\pm$2.86 \\
 & & u & TF$_{uc1}$ & 14 & 34.61 & 0.024 &  & 34.63$\pm$0.07 & 84.47$\pm$2.72 & 65.10$\pm$2.16 \\
 & & u & TF$_{uc2}$ & 13 & 34.61 & 0.026 &  & 34.64$\pm$0.09 & 84.57$\pm$3.51 & 65.02$\pm$2.74 \\
 & & g & TF$_{g}$ & 14 & 34.67 & 0.024 &  & 34.69$\pm$0.08 & 86.87$\pm$3.20 & 63.30$\pm$2.38 \\
 & & g & TF$_{gc1}$ & 14 & 34.62 & 0.035 &  & 34.66$\pm$0.07 & 85.32$\pm$2.75 & 64.45$\pm$2.14 \\
 & & g & TF$_{gc2}$ & 13 & 34.62 & 0.032 &  & 34.65$\pm$0.08 & 85.20$\pm$3.14 & 64.54$\pm$2.43 \\
 & & r & TF$_{r}$ & 14 & 34.59 & 0.023 &  & 34.61$\pm$0.08 & 83.67$\pm$3.08 & 65.72$\pm$2.47 \\
 & & r & TF$_{rc}$ & 13 & 34.60 & 0.006 &  & 34.61$\pm$0.08 & 83.42$\pm$3.07 & 65.92$\pm$2.48 \\
 & & r & TF$_{r \mu}$ & 14 & 34.60 & 0.010 &  & 34.61$\pm$0.07 & 83.58$\pm$2.69 & 65.80$\pm$2.18 \\
 & & i & TF$_{i}$ & 14 & 34.53 & 0.023 &  & 34.55$\pm$0.07 & 81.39$\pm$2.62 & 67.57$\pm$2.24 \\
 & & i & TF$_{ic}$ & 13 & 34.55 & 0.027 &  & 34.58$\pm$0.07 & 82.29$\pm$2.65 & 66.82$\pm$2.21 \\
 & & i & TF$_{i \mu}$ & 14 & 34.55 & 0.024 &  & 34.57$\pm$0.07 & 82.17$\pm$2.65 & 66.92$\pm$2.22 \\
 & & z & TF$_{z}$ & 14 & 34.53 & 0.022 &  & 34.55$\pm$0.07 & 81.36$\pm$2.62 & 67.59$\pm$2.24 \\
 & & z & TF$_{zc}$ & 13 & 34.55 & 0.006 &  & 34.56$\pm$0.07 & 81.50$\pm$2.63 & 67.47$\pm$2.24 \\
 & & z & TF$_{z \mu}$ & 14 & 34.55 & -0.002 &  & 34.55$\pm$0.06 & 81.20$\pm$2.24 & 67.72$\pm$1.94 \\
 & & W1 & TF$_{W1}$ & 13 & 34.45 & 0.011 &  & 34.46$\pm$0.12 & 78.02$\pm$4.31 & 70.49$\pm$3.93 \\
 & & W1 & TF$_{W1c}$ & 13 & 34.50 & 0.029 & 0.04 & 34.57$\pm$0.07 & 82.01$\pm$2.64 & 67.06$\pm$2.22 \\
 & & W1 & TF$_{W1 \mu}$ & 13 & 34.45 & -0.019 & 0.11 & 34.54$\pm$0.09 & 80.93$\pm$3.35 & 67.94$\pm$2.86 \\
 & & W2 & TF$_{W2}$ & 13 & 34.37 & 0.012 &  & 34.38$\pm$0.16 & 75.25$\pm$5.54 & 73.08$\pm$5.41 \\
 & & W2 & TF$_{W2c}$ & 13 & 34.51 & 0.007 & 0.06 & 34.58$\pm$0.07 & 82.29$\pm$2.65 & 66.82$\pm$2.21 \\
 & & W2 & TF$_{W2 \mu}$ & 13 & 34.35 & -0.033 & 0.17 & 34.49$\pm$0.13 & 78.95$\pm$4.73 & 69.66$\pm$4.20 \\
\hline
 NGC 70  & 6619$\pm$80 & u & TF$_{u}$ & 11 & 34.74 & 0.085 &  & 34.83$\pm$0.14 & 92.28$\pm$5.95 & 71.73$\pm$4.71 \\
 & & u & TF$_{uc1}$ & 11 & 34.62 & 0.051 &  & 34.67$\pm$0.11 & 85.96$\pm$4.35 & 77.00$\pm$4.01 \\
 & & u & TF$_{uc2}$ & 11 & 34.63 & 0.057 &  & 34.69$\pm$0.12 & 86.56$\pm$4.78 & 76.47$\pm$4.33 \\
 & & g & TF$_{g}$ & 11 & 34.66 & 0.036 &  & 34.70$\pm$0.11 & 86.93$\pm$4.40 & 76.14$\pm$3.97 \\
 & & g & TF$_{gc1}$ & 11 & 34.64 & 0.059 &  & 34.70$\pm$0.11 & 87.07$\pm$4.41 & 76.02$\pm$3.96 \\
 & & g & TF$_{gc2}$ & 11 & 34.63 & 0.059 &  & 34.69$\pm$0.11 & 86.64$\pm$4.39 & 76.40$\pm$3.98 \\
 & & r & TF$_{r}$ & 11 & 34.62 & 0.030 &  & 34.65$\pm$0.10 & 85.10$\pm$3.92 & 77.77$\pm$3.70 \\
 & & r & TF$_{rc}$ & 11 & 34.62 & 0.012 &  & 34.63$\pm$0.10 & 84.42$\pm$3.89 & 78.41$\pm$3.73 \\
 & & r & TF$_{r \mu}$ & 11 & 34.62 & 0.022 &  & 34.64$\pm$0.11 & 84.80$\pm$4.30 & 78.05$\pm$4.06 \\
 & & i & TF$_{i}$ & 11 & 34.56 & 0.028 &  & 34.59$\pm$0.10 & 82.73$\pm$3.81 & 80.01$\pm$3.81 \\
 & & i & TF$_{ic}$ & 11 & 34.56 & 0.040 &  & 34.60$\pm$0.10 & 83.18$\pm$3.83 & 79.57$\pm$3.79 \\
 & & i & TF$_{i \mu}$ & 11 & 34.56 & 0.031 &  & 34.59$\pm$0.11 & 82.82$\pm$4.20 & 79.92$\pm$4.16 \\
 & & z & TF$_{z}$ & 11 & 34.56 & 0.022 &  & 34.58$\pm$0.10 & 82.49$\pm$3.80 & 80.24$\pm$3.82 \\
 & & z & TF$_{zc}$ & 11 & 34.56 & 0.007 &  & 34.57$\pm$0.10 & 81.92$\pm$3.77 & 80.80$\pm$3.85 \\
 & & z & TF$_{z \mu}$ & 11 & 34.57 & -0.010 &  & 34.56$\pm$0.11 & 81.67$\pm$4.14 & 81.04$\pm$4.22 \\
 & & W1 & TF$_{W1}$ & 11 & 34.52 & 0.002 &  & 34.52$\pm$0.12 & 80.23$\pm$4.43 & 82.50$\pm$4.67 \\
 & & W1 & TF$_{W1c}$ & 11 & 34.50 & 0.037 & 0.04 & 34.58$\pm$0.10 & 82.31$\pm$3.79 & 80.42$\pm$3.83 \\
 & & W1 & TF$_{W1 \mu}$ & 11 & 34.48 & -0.031 & 0.11 & 34.56$\pm$0.12 & 81.61$\pm$4.51 & 81.11$\pm$4.59 \\
 & & W2 & TF$_{W2}$ & 11 & 34.51 & 0.004 &  & 34.51$\pm$0.12 & 79.96$\pm$4.42 & 82.78$\pm$4.68 \\
 & & W2 & TF$_{W2c}$ & 11 & 34.50 & 0.009 & 0.06 & 34.57$\pm$0.10 & 81.98$\pm$3.78 & 80.74$\pm$3.84 \\
 & & W2 & TF$_{W2 \mu}$ & 11 & 34.45 & -0.051 & 0.17 & 34.57$\pm$0.13 & 82.01$\pm$4.91 & 80.71$\pm$4.93 \\
\hline
 Abell 1367  & 7060$\pm$61 & u & TF$_{u}$ & 68 & 34.67 & 0.008 &  & 34.68$\pm$0.07 & 86.20$\pm$2.78 & 81.90$\pm$2.73 \\
 & & u & TF$_{uc1}$ & 68 & 34.77 & 0.008 &  & 34.78$\pm$0.06 & 90.28$\pm$2.49 & 78.21$\pm$2.26 \\
 & & u & TF$_{uc2}$ & 62 & 34.78 & 0.009 &  & 34.79$\pm$0.06 & 90.74$\pm$2.51 & 77.81$\pm$2.25 \\
 & & g & TF$_{g}$ & 68 & 34.75 & 0.012 &  & 34.76$\pm$0.06 & 89.61$\pm$2.48 & 78.78$\pm$2.28 \\
 & & g & TF$_{gc1}$ & 68 & 34.81 & 0.019 &  & 34.83$\pm$0.06 & 92.41$\pm$2.55 & 76.40$\pm$2.21 \\
 & & g & TF$_{gc2}$ & 62 & 34.81 & 0.015 &  & 34.83$\pm$0.06 & 92.26$\pm$2.55 & 76.52$\pm$2.22 \\
 & & r & TF$_{r}$ & 68 & 34.76 & 0.010 &  & 34.77$\pm$0.06 & 89.97$\pm$2.49 & 78.47$\pm$2.27 \\
 & & r & TF$_{rc}$ & 62 & 34.76 & 0.001 &  & 34.76$\pm$0.06 & 89.57$\pm$2.47 & 78.82$\pm$2.28 \\
 & & r & TF$_{r \mu}$ & 68 & 34.75 & 0.003 &  & 34.75$\pm$0.06 & 89.25$\pm$2.47 & 79.10$\pm$2.29 \\
 & & i & TF$_{i}$ & 68 & 34.76 & 0.013 &  & 34.77$\pm$0.06 & 90.08$\pm$2.49 & 78.37$\pm$2.27 \\
 & & i & TF$_{ic}$ & 62 & 34.76 & 0.015 &  & 34.78$\pm$0.06 & 90.18$\pm$2.49 & 78.29$\pm$2.27 \\
 & & i & TF$_{i \mu}$ & 68 & 34.74 & 0.013 &  & 34.75$\pm$0.06 & 89.23$\pm$2.47 & 79.12$\pm$2.29 \\
 & & z & TF$_{z}$ & 68 & 34.80 & 0.011 &  & 34.81$\pm$0.06 & 91.64$\pm$2.53 & 77.04$\pm$2.23 \\
 & & z & TF$_{zc}$ & 62 & 34.78 & 0.003 &  & 34.78$\pm$0.06 & 90.50$\pm$2.50 & 78.01$\pm$2.26 \\
 & & z & TF$_{z \mu}$ & 68 & 34.78 & -0.004 &  & 34.78$\pm$0.06 & 90.20$\pm$2.49 & 78.27$\pm$2.27 \\
 & & W1 & TF$_{W1}$ & 62 & 34.85 & 0.011 &  & 34.86$\pm$0.06 & 93.80$\pm$2.59 & 75.26$\pm$2.18 \\
 & & W1 & TF$_{W1c}$ & 62 & 34.73 & 0.015 & 0.04 & 34.79$\pm$0.06 & 90.58$\pm$2.50 & 77.94$\pm$2.26 \\
 & & W1 & TF$_{W1 \mu}$ & 62 & 34.73 & -0.009 & 0.11 & 34.83$\pm$0.06 & 92.50$\pm$2.56 & 76.32$\pm$2.21 \\
 & & W2 & TF$_{W2}$ & 62 & 34.83 & 0.010 &  & 34.84$\pm$0.07 & 92.88$\pm$2.99 & 76.01$\pm$2.54 \\
 & & W2 & TF$_{W2c}$ & 62 & 34.73 & 0.003 & 0.06 & 34.79$\pm$0.06 & 90.90$\pm$2.51 & 77.67$\pm$2.25 \\
 & & W2 & TF$_{W2 \mu}$ & 62 & 34.67 & -0.016 & 0.17 & 34.82$\pm$0.06 & 92.23$\pm$2.55 & 76.55$\pm$2.22 \\
\hline
 Coma  & 7352$\pm$70 & u & TF$_{u}$ & 79 & 34.68 & 0.027 &  & 34.71$\pm$0.07 & 87.40$\pm$2.82 & 84.12$\pm$2.83 \\
 & & u & TF$_{uc1}$ & 79 & 34.80 & 0.020 &  & 34.82$\pm$0.06 & 92.05$\pm$2.54 & 79.87$\pm$2.33 \\
 & & u & TF$_{uc2}$ & 75 & 34.79 & 0.022 &  & 34.81$\pm$0.06 & 91.70$\pm$2.53 & 80.17$\pm$2.34 \\
 & & g & TF$_{g}$ & 79 & 34.78 & 0.012 &  & 34.79$\pm$0.06 & 90.87$\pm$2.51 & 80.91$\pm$2.36 \\
 & & g & TF$_{gc1}$ & 79 & 34.84 & 0.019 &  & 34.86$\pm$0.06 & 93.70$\pm$2.59 & 78.46$\pm$2.29 \\
 & & g & TF$_{gc2}$ & 75 & 34.83 & 0.015 &  & 34.85$\pm$0.06 & 93.12$\pm$2.57 & 78.95$\pm$2.31 \\
 & & r & TF$_{r}$ & 79 & 34.80 & 0.010 &  & 34.81$\pm$0.06 & 91.64$\pm$2.53 & 80.23$\pm$2.34 \\
 & & r & TF$_{rc}$ & 75 & 34.78 & 0.001 &  & 34.78$\pm$0.06 & 90.39$\pm$2.50 & 81.34$\pm$2.38 \\
 & & r & TF$_{r \mu}$ & 79 & 34.75 & 0.003 &  & 34.75$\pm$0.06 & 89.24$\pm$2.47 & 82.39$\pm$2.41 \\
 & & i & TF$_{i}$ & 79 & 34.79 & 0.009 &  & 34.80$\pm$0.06 & 91.16$\pm$2.52 & 80.65$\pm$2.36 \\
 & & i & TF$_{ic}$ & 75 & 34.77 & 0.010 &  & 34.78$\pm$0.06 & 90.38$\pm$2.50 & 81.34$\pm$2.38 \\
 & & i & TF$_{i \mu}$ & 79 & 34.74 & 0.008 &  & 34.75$\pm$0.06 & 89.04$\pm$2.46 & 82.57$\pm$2.41 \\
 & & z & TF$_{z}$ & 79 & 34.82 & 0.008 &  & 34.83$\pm$0.06 & 92.38$\pm$2.55 & 79.58$\pm$2.33 \\
 & & z & TF$_{zc}$ & 75 & 34.80 & 0.003 &  & 34.80$\pm$0.06 & 91.31$\pm$2.52 & 80.52$\pm$2.35 \\
 & & z & TF$_{z \mu}$ & 79 & 34.77 & -0.004 &  & 34.77$\pm$0.06 & 89.80$\pm$2.48 & 81.87$\pm$2.39 \\
 & & W1 & TF$_{W1}$ & 75 & 34.85 & 0.004 &  & 34.85$\pm$0.06 & 93.49$\pm$2.58 & 78.64$\pm$2.30 \\
 & & W1 & TF$_{W1c}$ & 75 & 34.74 & 0.004 & 0.04 & 34.78$\pm$0.06 & 90.54$\pm$2.50 & 81.20$\pm$2.37 \\
 & & W1 & TF$_{W1 \mu}$ & 75 & 34.69 & -0.004 & 0.11 & 34.80$\pm$0.06 & 91.05$\pm$2.52 & 80.74$\pm$2.36 \\
 & & W2 & TF$_{W2}$ & 75 & 34.86 & 0.002 &  & 34.86$\pm$0.07 & 93.86$\pm$3.03 & 78.33$\pm$2.63 \\
 & & W2 & TF$_{W2c}$ & 75 & 34.74 & 0.001 & 0.06 & 34.80$\pm$0.06 & 91.23$\pm$2.52 & 80.59$\pm$2.36 \\
 & & W2 & TF$_{W2 \mu}$ & 75 & 34.64 & -0.005 & 0.17 & 34.80$\pm$0.06 & 91.40$\pm$2.53 & 80.44$\pm$2.35 \\
\hline
 Abell 400  & 7357$\pm$85 & u & TF$_{u}$ & 21 & 34.72 & 0.048 &  & 34.77$\pm$0.10 & 89.88$\pm$4.14 & 81.85$\pm$3.89 \\
 & & u & TF$_{uc1}$ & 21 & 34.84 & 0.032 &  & 34.87$\pm$0.06 & 94.26$\pm$2.60 & 78.05$\pm$2.34 \\
 & & u & TF$_{uc2}$ & 20 & 34.82 & 0.034 &  & 34.85$\pm$0.07 & 93.51$\pm$3.01 & 78.68$\pm$2.69 \\
 & & g & TF$_{g}$ & 21 & 34.82 & 0.031 &  & 34.85$\pm$0.07 & 93.36$\pm$3.01 & 78.81$\pm$2.70 \\
 & & g & TF$_{gc1}$ & 21 & 34.88 & 0.049 &  & 34.93$\pm$0.06 & 96.80$\pm$2.67 & 76.00$\pm$2.28 \\
 & & g & TF$_{gc2}$ & 20 & 34.86 & 0.047 &  & 34.91$\pm$0.07 & 95.81$\pm$3.09 & 76.79$\pm$2.63 \\
 & & r & TF$_{r}$ & 21 & 34.85 & 0.030 &  & 34.88$\pm$0.06 & 94.63$\pm$2.61 & 77.74$\pm$2.33 \\
 & & r & TF$_{rc}$ & 20 & 34.85 & 0.010 &  & 34.86$\pm$0.06 & 93.78$\pm$2.59 & 78.45$\pm$2.35 \\
 & & r & TF$_{r \mu}$ & 21 & 34.84 & 0.016 &  & 34.86$\pm$0.06 & 93.60$\pm$2.59 & 78.60$\pm$2.35 \\
 & & i & TF$_{i}$ & 21 & 34.84 & 0.029 &  & 34.87$\pm$0.06 & 94.15$\pm$2.60 & 78.14$\pm$2.34 \\
 & & i & TF$_{ic}$ & 20 & 34.83 & 0.035 &  & 34.86$\pm$0.06 & 93.97$\pm$2.60 & 78.29$\pm$2.34 \\
 & & i & TF$_{i \mu}$ & 21 & 34.82 & 0.030 &  & 34.85$\pm$0.06 & 93.33$\pm$2.58 & 78.83$\pm$2.36 \\
 & & z & TF$_{z}$ & 21 & 34.85 & 0.027 &  & 34.88$\pm$0.06 & 94.48$\pm$2.61 & 77.87$\pm$2.33 \\
 & & z & TF$_{zc}$ & 20 & 34.84 & 0.007 &  & 34.85$\pm$0.06 & 93.19$\pm$2.57 & 78.95$\pm$2.36 \\
 & & z & TF$_{z \mu}$ & 21 & 34.84 & -0.002 &  & 34.84$\pm$0.06 & 92.80$\pm$2.56 & 79.28$\pm$2.37 \\
 & & W1 & TF$_{W1}$ & 23 & 34.90 & 0.011 &  & 34.91$\pm$0.08 & 96.00$\pm$3.54 & 76.63$\pm$2.96 \\
 & & W1 & TF$_{W1c}$ & 20 & 34.80 & 0.016 & 0.04 & 34.86$\pm$0.06 & 93.60$\pm$2.59 & 78.60$\pm$2.35 \\
 & & W1 & TF$_{W1 \mu}$ & 23 & 34.75 & -0.009 & 0.11 & 34.85$\pm$0.07 & 93.35$\pm$3.01 & 78.81$\pm$2.70 \\
 & & W2 & TF$_{W2}$ & 23 & 34.90 & 0.011 &  & 34.91$\pm$0.09 & 95.97$\pm$3.98 & 76.66$\pm$3.30 \\
 & & W2 & TF$_{W2c}$ & 20 & 34.79 & 0.003 & 0.06 & 34.85$\pm$0.06 & 93.46$\pm$2.58 & 78.72$\pm$2.36 \\
 & & W2 & TF$_{W2 \mu}$ & 23 & 34.70 & -0.017 & 0.17 & 34.85$\pm$0.08 & 93.47$\pm$3.44 & 78.71$\pm$3.04 \\
\hline
 NGC 4065  & 7501$\pm$63 & u & TF$_{u}$ & 14 & 35.11 & 0.055 &  & 35.16$\pm$0.10 & 107.89$\pm$4.97 & 69.52$\pm$3.25 \\
 & & u & TF$_{uc1}$ & 14 & 35.15 & 0.033 &  & 35.18$\pm$0.08 & 108.80$\pm$4.01 & 68.95$\pm$2.61 \\
 & & u & TF$_{uc2}$ & 12 & 35.11 & 0.035 &  & 35.15$\pm$0.10 & 106.91$\pm$4.92 & 70.16$\pm$3.28 \\
 & & g & TF$_{g}$ & 14 & 35.12 & 0.028 &  & 35.15$\pm$0.07 & 107.04$\pm$3.45 & 70.08$\pm$2.33 \\
 & & g & TF$_{gc1}$ & 14 & 35.16 & 0.044 &  & 35.20$\pm$0.07 & 109.83$\pm$3.54 & 68.30$\pm$2.28 \\
 & & g & TF$_{gc2}$ & 12 & 35.12 & 0.040 &  & 35.16$\pm$0.09 & 107.65$\pm$4.46 & 69.68$\pm$2.95 \\
 & & r & TF$_{r}$ & 14 & 35.12 & 0.026 &  & 35.15$\pm$0.07 & 106.97$\pm$3.45 & 70.12$\pm$2.34 \\
 & & r & TF$_{rc}$ & 12 & 35.10 & 0.008 &  & 35.11$\pm$0.08 & 105.09$\pm$3.87 & 71.37$\pm$2.70 \\
 & & r & TF$_{r \mu}$ & 14 & 35.10 & 0.013 &  & 35.11$\pm$0.07 & 105.32$\pm$3.40 & 71.22$\pm$2.37 \\
 & & i & TF$_{i}$ & 14 & 35.11 & 0.025 &  & 35.14$\pm$0.08 & 106.43$\pm$3.92 & 70.48$\pm$2.66 \\
 & & i & TF$_{ic}$ & 12 & 35.08 & 0.029 &  & 35.11$\pm$0.08 & 105.16$\pm$3.87 & 71.33$\pm$2.70 \\
 & & i & TF$_{i \mu}$ & 14 & 35.08 & 0.026 &  & 35.11$\pm$0.08 & 104.99$\pm$3.87 & 71.45$\pm$2.70 \\
 & & z & TF$_{z}$ & 14 & 35.13 & 0.024 &  & 35.15$\pm$0.08 & 107.35$\pm$3.95 & 69.87$\pm$2.64 \\
 & & z & TF$_{zc}$ & 12 & 35.09 & 0.006 &  & 35.10$\pm$0.08 & 104.52$\pm$3.85 & 71.77$\pm$2.71 \\
 & & z & TF$_{z \mu}$ & 14 & 35.11 & -0.002 &  & 35.11$\pm$0.08 & 105.11$\pm$3.87 & 71.37$\pm$2.70 \\
 & & W1 & TF$_{W1}$ & 12 & 35.06 & 0.012 &  & 35.07$\pm$0.11 & 103.38$\pm$5.24 & 72.56$\pm$3.73 \\
 & & W1 & TF$_{W1c}$ & 12 & 35.04 & 0.022 & 0.04 & 35.10$\pm$0.08 & 104.82$\pm$3.86 & 71.56$\pm$2.70 \\
 & & W1 & TF$_{W1 \mu}$ & 12 & 34.98 & -0.013 & 0.11 & 35.08$\pm$0.10 & 103.61$\pm$4.77 & 72.40$\pm$3.39 \\
 & & W2 & TF$_{W2}$ & 12 & 35.08 & 0.012 &  & 35.09$\pm$0.12 & 104.33$\pm$5.77 & 71.90$\pm$4.02 \\
 & & W2 & TF$_{W2c}$ & 12 & 35.04 & 0.004 & 0.06 & 35.10$\pm$0.08 & 104.91$\pm$3.86 & 71.50$\pm$2.70 \\
 & & W2 & TF$_{W2 \mu}$ & 12 & 34.95 & -0.020 & 0.17 & 35.10$\pm$0.10 & 104.69$\pm$4.82 & 71.65$\pm$3.35 \\
\hline
 Abell 539  & 8995$\pm$87 & W1 & TF$_{W1}$ & 22 & 35.22 & -0.004 &  & 35.22$\pm$0.08 & 110.48$\pm$4.07 & 81.41$\pm$3.10 \\
 & & W1 & TF$_{W1 \mu}$ & 22 & 35.17 & -0.037 & 0.11 & 35.24$\pm$0.06 & 111.86$\pm$3.09 & 80.41$\pm$2.35 \\
 & & W2 & TF$_{W2}$ & 22 & 35.14 & 0.002 &  & 35.14$\pm$0.09 & 106.77$\pm$4.43 & 84.24$\pm$3.59 \\
 & & W2 & TF$_{W2 \mu}$ & 22 & 35.07 & -0.053 & 0.17 & 35.19$\pm$0.07 & 108.97$\pm$3.51 & 82.54$\pm$2.78 \\
\hline
 Abell 2634/66  & 8954$\pm$98 & u & TF$_{u}$ & 29 & 35.45 & 0.061 &  & 35.51$\pm$0.08 & 126.53$\pm$4.66 & 70.77$\pm$2.72 \\
 & & u & TF$_{uc1}$ & 29 & 35.33 & 0.032 &  & 35.36$\pm$0.06 & 118.13$\pm$3.26 & 75.80$\pm$2.25 \\
 & & u & TF$_{uc2}$ & 26 & 35.36 & 0.036 &  & 35.40$\pm$0.07 & 120.01$\pm$3.87 & 74.61$\pm$2.54 \\
 & & g & TF$_{g}$ & 29 & 35.34 & 0.033 &  & 35.37$\pm$0.07 & 118.75$\pm$3.83 & 75.40$\pm$2.57 \\
 & & g & TF$_{gc1}$ & 29 & 35.33 & 0.052 &  & 35.38$\pm$0.06 & 119.24$\pm$3.29 & 75.09$\pm$2.23 \\
 & & g & TF$_{gc2}$ & 26 & 35.34 & 0.051 &  & 35.39$\pm$0.06 & 119.72$\pm$3.31 & 74.79$\pm$2.22 \\
 & & r & TF$_{r}$ & 29 & 35.31 & 0.031 &  & 35.34$\pm$0.06 & 117.02$\pm$3.23 & 76.52$\pm$2.27 \\
 & & r & TF$_{rc}$ & 26 & 35.31 & 0.011 &  & 35.32$\pm$0.06 & 115.95$\pm$3.20 & 77.22$\pm$2.30 \\
 & & r & TF$_{r \mu}$ & 29 & 35.29 & 0.018 &  & 35.31$\pm$0.06 & 115.23$\pm$3.18 & 77.71$\pm$2.31 \\
 & & i & TF$_{i}$ & 29 & 35.26 & 0.030 &  & 35.29$\pm$0.06 & 114.27$\pm$3.16 & 78.36$\pm$2.33 \\
 & & i & TF$_{ic}$ & 26 & 35.26 & 0.036 &  & 35.30$\pm$0.06 & 114.62$\pm$3.17 & 78.12$\pm$2.32 \\
 & & i & TF$_{i \mu}$ & 29 & 35.24 & 0.031 &  & 35.27$\pm$0.06 & 113.29$\pm$3.13 & 79.04$\pm$2.35 \\
 & & z & TF$_{z}$ & 29 & 35.24 & 0.027 &  & 35.27$\pm$0.06 & 113.08$\pm$3.12 & 79.19$\pm$2.35 \\
 & & z & TF$_{zc}$ & 26 & 35.23 & 0.007 &  & 35.24$\pm$0.06 & 111.53$\pm$3.08 & 80.29$\pm$2.39 \\
 & & z & TF$_{z \mu}$ & 29 & 35.23 & -0.002 &  & 35.23$\pm$0.06 & 111.06$\pm$3.07 & 80.63$\pm$2.40 \\
 & & W1 & TF$_{W1}$ & 26 & 35.28 & 0.011 &  & 35.29$\pm$0.08 & 114.34$\pm$4.21 & 78.31$\pm$3.01 \\
 & & W1 & TF$_{W1c}$ & 26 & 35.21 & 0.027 & 0.04 & 35.28$\pm$0.06 & 113.62$\pm$3.14 & 78.81$\pm$2.34 \\
 & & W1 & TF$_{W1 \mu}$ & 26 & 35.13 & -0.017 & 0.11 & 35.22$\pm$0.07 & 110.82$\pm$3.57 & 80.80$\pm$2.75 \\
 & & W2 & TF$_{W2}$ & 26 & 35.27 & 0.013 &  & 35.28$\pm$0.09 & 113.92$\pm$4.72 & 78.60$\pm$3.37 \\
 & & W2 & TF$_{W2c}$ & 26 & 35.21 & 0.006 & 0.06 & 35.28$\pm$0.06 & 113.53$\pm$3.14 & 78.87$\pm$2.34 \\
 & & W2 & TF$_{W2 \mu}$ & 26 & 35.07 & -0.027 & 0.17 & 35.21$\pm$0.08 & 110.33$\pm$4.06 & 81.16$\pm$3.12 \\
\hline
 Abell 2151 (Hercules)  & 11353$\pm$121 & u & TF$_{u}$ & 39 & 35.86 & 0.083 &  & 35.94$\pm$0.07 & 154.40$\pm$4.98 & 73.53$\pm$2.50 \\
 & & u & TF$_{uc1}$ & 39 & 35.92 & 0.054 &  & 35.97$\pm$0.06 & 156.60$\pm$4.33 & 72.49$\pm$2.15 \\
 & & u & TF$_{uc2}$ & 33 & 35.97 & 0.061 &  & 36.03$\pm$0.07 & 160.74$\pm$5.18 & 70.63$\pm$2.40 \\
 & & g & TF$_{g}$ & 39 & 35.87 & 0.031 &  & 35.90$\pm$0.06 & 151.42$\pm$4.18 & 74.98$\pm$2.22 \\
 & & g & TF$_{gc1}$ & 39 & 35.95 & 0.050 &  & 36.00$\pm$0.06 & 158.50$\pm$4.38 & 71.63$\pm$2.12 \\
 & & g & TF$_{gc2}$ & 33 & 35.95 & 0.048 &  & 36.00$\pm$0.07 & 158.37$\pm$5.11 & 71.68$\pm$2.43 \\
 & & r & TF$_{r}$ & 39 & 35.91 & 0.029 &  & 35.94$\pm$0.06 & 154.11$\pm$4.26 & 73.67$\pm$2.18 \\
 & & r & TF$_{rc}$ & 33 & 35.90 & 0.010 &  & 35.91$\pm$0.07 & 152.03$\pm$4.90 & 74.68$\pm$2.54 \\
 & & r & TF$_{r \mu}$ & 39 & 35.88 & 0.015 &  & 35.89$\pm$0.06 & 151.00$\pm$4.17 & 75.18$\pm$2.23 \\
 & & i & TF$_{i}$ & 39 & 35.91 & 0.028 &  & 35.94$\pm$0.06 & 154.04$\pm$4.26 & 73.70$\pm$2.18 \\
 & & i & TF$_{ic}$ & 33 & 35.89 & 0.033 &  & 35.92$\pm$0.07 & 153.00$\pm$4.93 & 74.20$\pm$2.52 \\
 & & i & TF$_{i \mu}$ & 39 & 35.87 & 0.029 &  & 35.90$\pm$0.06 & 151.28$\pm$4.18 & 75.05$\pm$2.22 \\
 & & z & TF$_{z}$ & 39 & 35.95 & 0.025 &  & 35.97$\pm$0.06 & 156.67$\pm$4.33 & 72.47$\pm$2.15 \\
 & & z & TF$_{zc}$ & 33 & 35.92 & 0.006 &  & 35.93$\pm$0.07 & 153.20$\pm$4.94 & 74.11$\pm$2.52 \\
 & & z & TF$_{z \mu}$ & 39 & 35.90 & -0.002 &  & 35.90$\pm$0.06 & 151.23$\pm$4.18 & 75.07$\pm$2.22 \\
 & & W1 & TF$_{W1}$ & 33 & 36.04 & 0.011 &  & 36.05$\pm$0.09 & 162.28$\pm$6.73 & 69.96$\pm$2.99 \\
 & & W1 & TF$_{W1c}$ & 33 & 35.85 & 0.025 & 0.04 & 35.91$\pm$0.07 & 152.40$\pm$4.91 & 74.49$\pm$2.53 \\
 & & W1 & TF$_{W1 \mu}$ & 33 & 35.88 & -0.016 & 0.11 & 35.97$\pm$0.08 & 156.59$\pm$5.77 & 72.50$\pm$2.78 \\
 & & W2 & TF$_{W2}$ & 33 & 36.04 & 0.013 &  & 36.05$\pm$0.09 & 162.40$\pm$6.73 & 69.91$\pm$2.99 \\
 & & W2 & TF$_{W2c}$ & 33 & 35.85 & 0.005 & 0.06 & 35.92$\pm$0.07 & 152.42$\pm$4.91 & 74.49$\pm$2.53 \\
 & & W2 & TF$_{W2 \mu}$ & 33 & 35.82 & -0.027 & 0.17 & 35.96$\pm$0.08 & 155.83$\pm$5.74 & 72.85$\pm$2.79 \\
\enddata
\end{deluxetable*}

\bibliography{paper}

\end{document}